\documentclass[a4paper,fleqn,usenatbib]{mnras}

\usepackage{mathptmx}

\usepackage{ae,aecompl}
\usepackage{supertabular}
\usepackage[utf8]{inputenc}

\usepackage{graphicx}	
\usepackage{amsmath}	
\usepackage{amssymb}	
\usepackage{pdflscape}

\title[Stellar activity with LAMOST]{Stellar activity with LAMOST. III. Temporal variability pattern in Pleiades, Praesepe, and Hyades}

\author[X.-S. Fang et al.]{
Xiang-Song Fang,$^{1,2,3}$\thanks{E-mail: xsfang@bao.ac.cn}
Christian Moni Bidin,$^{3}$
Gang Zhao,$^{1,4}$\thanks{E-mail: gzhao@bao.ac.cn}
Li-Yun Zhang$^{5}$ 
\newauthor and Yerra Bharat Kumar$^{1}$\thanks{LAMOST Fellow}
\\
$^{1}$CAS Key Laboratory of Optical Astronomy, National Astronomical Observatories, Chinese Academy of Sciences, Beijing 100101, China\\
$^{2}$Chinese Academy of Sciences South America Center for Astronomy, National Astronomical Observatories, CAS, Beijing 100101, China\\
$^{3}$Instituto de Astronom\'{i}a, Universidad Cat\'{o}lica del Norte, Av. Angamos 0610 Antofagasta, Chile\\
$^{4}$School of Astronomy and Space Science, University of Chinese Academy of Sciences, Beijing 100049, China\\
$^{5}$College of Physics \& Guizhou Provincial Key Laboratory of Public Big Data, Guizhou University, Guiyang 550025, China \\
}

\date{Accepted 2020 May 15. Received 2020 May 13; in original form 2019 September 10}
\pubyear{2020}

\begin{document}
\label{firstpage}
\pagerange{\pageref{firstpage}--\pageref{lastpage}}
\maketitle

\begin{abstract}
We present the results from a systematic study of temporal variation of stellar activity in young late-type stars. We used multi-epoch LAMOST low-resolution spectra of over 300 member candidates in three young open clusters: Pleiades, Praesepe, and Hyades.  The spectral measurements of TiO band strength near 7050~\AA~(TiO2) and equivalent width of H$\alpha$ line (EW$_{\text{H}\alpha}$) are used as the tracers of cool-spot coverage and chromospheric emission strength, respectively. The analysis of time-variation patterns of these two tracers suggested there exist detectable variabilities in TiO2 and EW$_{\text{H}\alpha}$, and their time-scales are in the wide range from days to years. Results showed that more active stars, younger and fast rotators, tend to have larger activity variations. There is a tendency of anti-correlation between temporal variations in TiO2 and EW$_{\text{H}\alpha}$. Also, appreciable anti-correlation in the rotational phase between H$\alpha$ emission and K2 brightness is detected in some M dwarfs, indicating spatial co-location of the plages with cool starspots, however, cool stars do not always show such co-location features. Furthermore, spot coverage and H$\alpha$ emission were evident at all rotational phases of several M dwarfs, indicating a basal level of activity, perhaps due to many small and randomly located active regions in the atmosphere.
\end{abstract}




\section{Introduction}
Stellar activity is a ubiquitous phenomenon among late-type stars, which is believed to be closely linked to the generation and evolution of magnetic fields in stellar interiors and atmosphere. Stellar activity shows both short- and long-term variability. For instance, the rapid evolution of starspots \citep[e.g., the appearance of new spots and decay of old ones,][]{gile2017} produces the short-term variability in light curve shape in time-scales of days to weeks \citep[e.g.][]{rebu2016a,rebu2016b,rebu2017}. Short-term chromospheric variabilities were also detected in time-scales of days to weeks and months in young active stars, e.g. EK Dra \citep{jarv2007}, V889 Her \citep{jarv2008}, AP 149 \citep{sava2003,fang2010}, and a few Pleiades members \citep{sode1993}. Short-term variation pattern also includes sudden enhancements of brightness \citep[e.g.][]{hawl2014,dave2014}, chromospheric and coronal emissions \citep[e.g.][]{berd1999,cres2007}, which may be due to flares with time-scales of minutes to hours. On the other hand, long-term photometric brightness variations in the order of years have been detected in many late-type stars \citep[e.g.][]{alek2018,will2019}. The long-term chromospheric emission variability has been investigated by several projects \citep[e.g.][]{hall2007,gray2015,radi2018}, e.g., the Mount Wilson Observatory Ca~{\sc ii} HK project \citep{wils1978,bali1995,bali1998}; it is found that older stars tend to either vary in a smooth and cyclic fashion or have steady levels of the H \& K emission, but the young and active stars show irregular variations lacking smooth cycles. However, some younger and cooler stars like M-type stars still show cyclic activity behaviour, e.g., recently \citet{iban2019} detected a possible chromospheric activity cycle in the very young active dM1 star AU Microscopii. Studies also indicate the co-existence of long and short chromospheric activity cycles in some young solar-type stars \citep[e.g.,][and references therein]{metc2013,egel2015,bran2017}. 

Investigation of contemporaneous long-term photometric and chromospheric observations shows that young active stars become fainter when their Ca~{\sc ii} emission increases on year-to-year time-scale, while less active older stars become brighter \citep[e.g.][]{radi1998,lock2007,radi2018}, suggesting a change from spot-dominated to faculae-dominated activity with the star's age. On the other hand, both young and old stars tend to become fainter as their chromospheric emission increases in short-term time-scales such as day to day \citep[e.g.][]{radi1998,jarv2007,jarv2008}, indicating the cool starspots are co-located spatially with the chromospheric active regions (like in the Sun). However, such spatial correlation features might be more complex than expected, e.g., the chromospheric emission of two K dwarfs in Praesepe were found to remain relatively constant, lacking correlation with broadband flux in rotational phase \citep{morr2018}. 

As mentioned above, great efforts during the last decades have uncovered many mysteries in stellar activity variability, providing a good opportunity to understand the underlying physics and further the stellar dynamo mechanism. However, previous studies primarily focused on old solar-like stars (e.g. ages older than 1 Gyr), lack of attention on the young cooler stars such as KM-type stars.Thus, a full characterization of activity variability among cool stars including M-type stars is indeed needed. LAMOST \citep[Large Sky Area Multi-Object Fibre Spectroscopic Telescope,][]{cui+2012} has collected millions of stellar spectra for stars in the Milky Way and particularly has conducted multi-epoch observations for many stars, which allows us to initiate a systematic study of the time-variation of stellar activity among low mass stars in different population. In fact, we detected an evident chromospheric activity variation among stars in Pleiades, Hyades, and Praesepe \citep[see figure 5 in][hereafter Paper II]{fang2018}. In this work, as the third paper of this series, we continue to investigate the activity in these young open clusters by focusing on the temporal variation patterns. We have examined the dependency of activity variation on stellar properties like effective temperature, age, and rotation. We also investigated the correlation between chromospheric activity variation patterns and the strength of TiO molecular bands near 7050~\AA, which holds key clues of the coverage of active regions like cool starspots on photosphere \citep[e.g.][hereafter Paper I]{neff1995,onea1998,fang2016}.

\section{Data and sample}
\subsection{LAMOST spectra}
The LAMOST (also called the Guo Shou Jing Telescope), characterized by both wide field of view of 20 $\text{deg}^{2}$ in the sky and large effective aperture of $\sim$4 m, is a reflecting Schmidt telescope located at the Xinglong Observatory, China. A total of 4000 fibers are mounted on its focal plane, which makes it a very high efficient spectrum collecting instrument. By summer 2019, LAMOST obtained over 9 million stellar spectra with spectral resolving power of $R=\lambda/\Delta\lambda \approx1800$ (e.g., $\sim$3.6~\AA~around 6500~\AA), covering the wavelength range of 3700$-$9100~\AA~\citep[see][for more details]{zhao2006,zhao2012,luo+2015}. For AFGK-type stars, if spectra meet the signal-to-noise ratio criterion, LAMOST provides the stellar atmospheric parameters (effective temperature, surface gravity, metallicity) determined from LAMOST stellar parameter pipeline, and heliocentric radial velocity obtained by the ULYSS \citep{wu++2011,luo+2015}. Over 6 million low-resolution stellar spectra collected by summer 2019 have the above stellar parameters.
\subsection{Sample cluster members} 
Our sample stars are from three nearby young open clusters having age range from 100 to 700 Myr. The clusters are Pleiades (age$\sim$125 Myr, \citet{stau1998}; [Fe/H]$\sim+0.03$, \citet{sode2009}), Praesepe and Hyades (both have ages around 700 Myr \citet{bran2015} and their [Fe/H] values are $+0.1$ to $+0.2$ \citealt{carr2011}). Following previous series (Paper I \& II), based on the member catalogues (for Pleiades, \citet{stau2007, bouy2015}, and for Praesepe and Hyades, \citet{doug2014}), we retrieved the available multi-epoch (at least 3 epochs) LAMOST low resolution spectra collected until summer 2019 having signal-to noise ratio in $r$-band (SNR$r$) above 15. 

\begin{figure}
\centering
\includegraphics[width=\columnwidth]{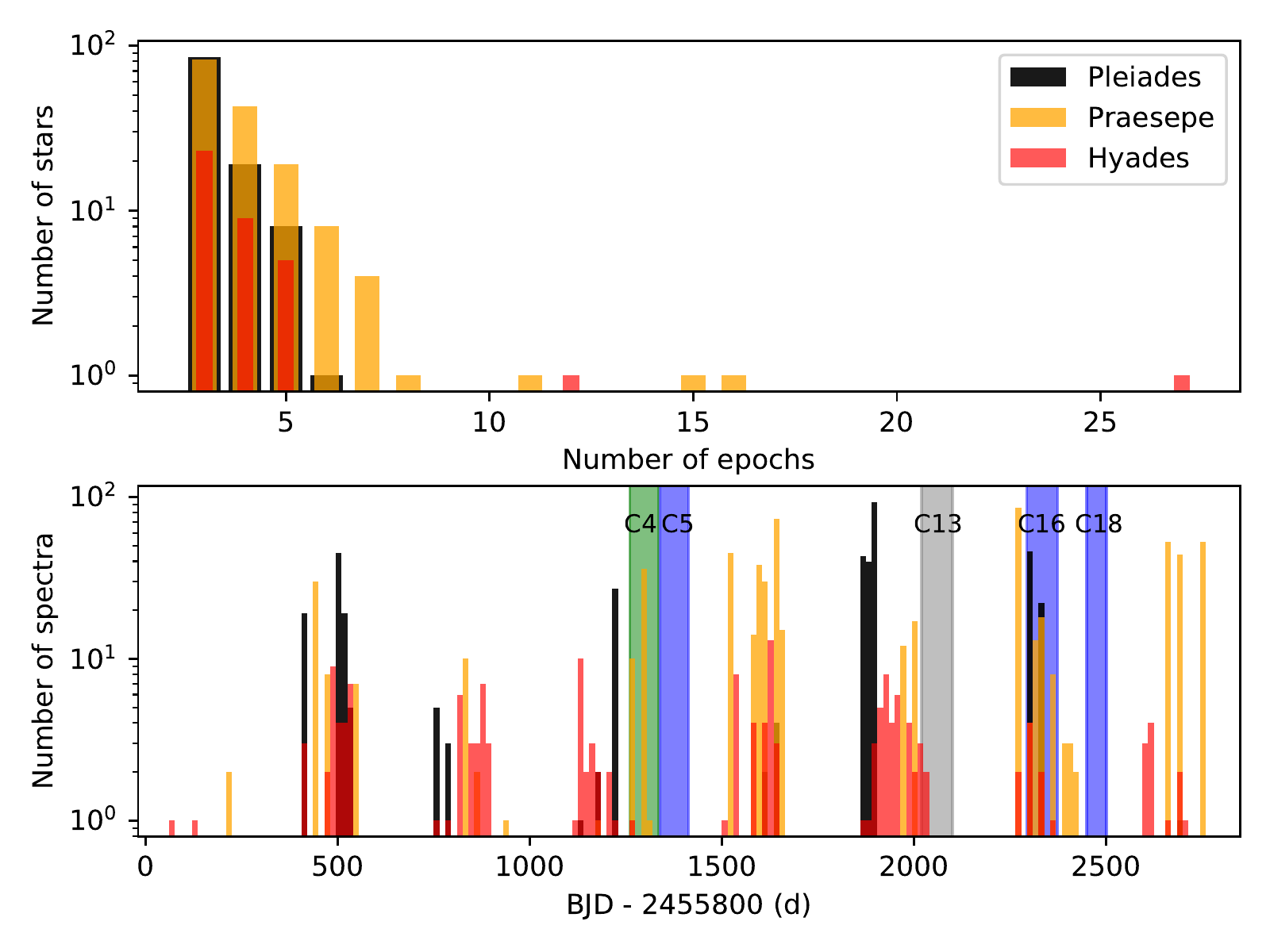}
\caption{Top Panel: Number of cluster members as a function of LAMOST observational epochs. Bottom panel: Distribution of available good quality spectra during observing time. The K2 campaigns (C4, C5, C13, C16, C18, see section 4.3.2 for more details) pointing towards the Pleiades, Praesepe and Hyades fields are marked.}
\label{fig:sample_hist}
\end{figure}

\begin{figure}
\centering
\includegraphics[width=\columnwidth]{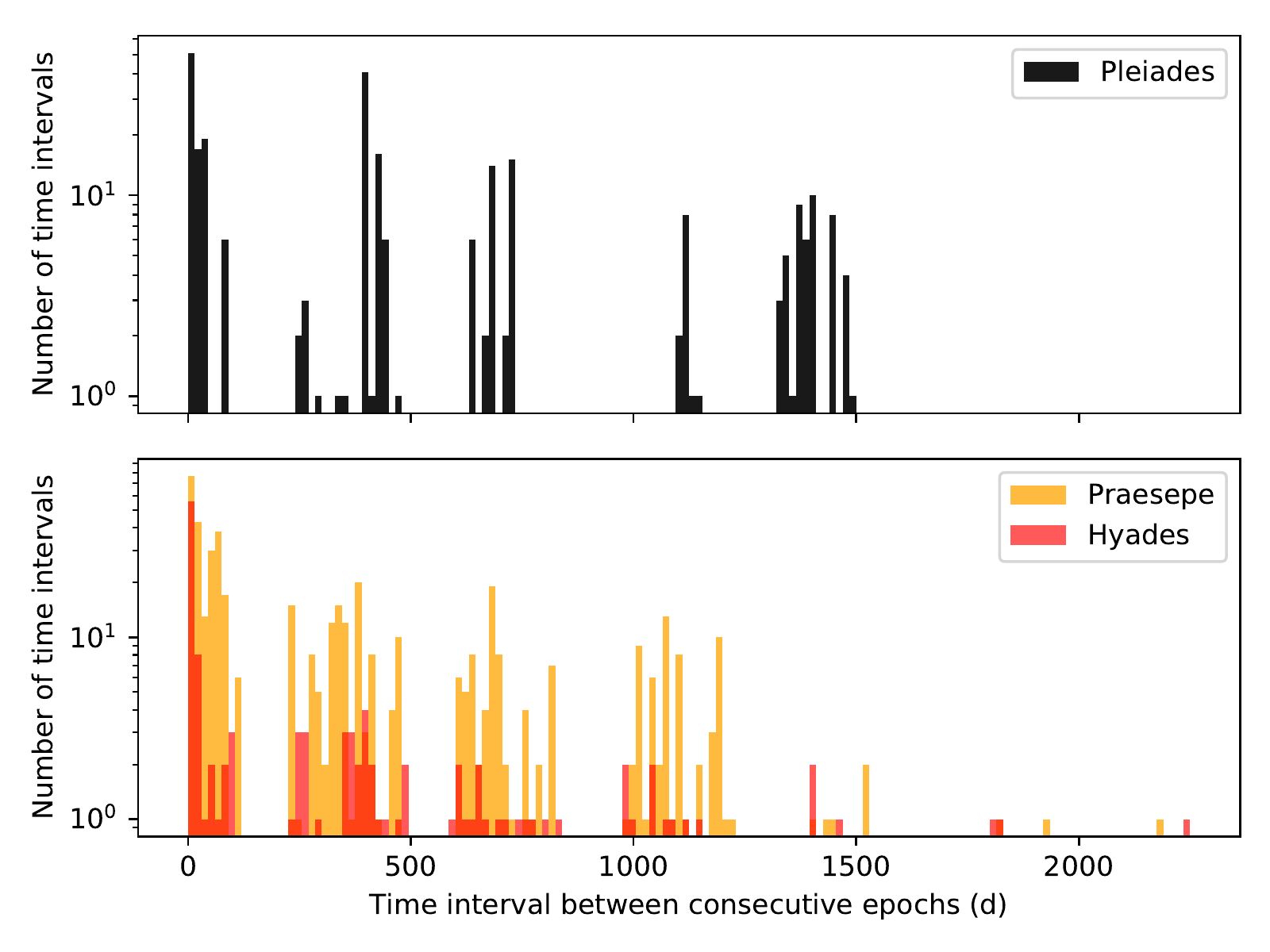}
\caption{Distribution of time intervals between consecutive epochs for sample members stars. Note y-axes are in log-scale.}
\label{fig:time_delay}
\end{figure}

We used available Gaia DR2 parallaxes and proper motions \citep{gaia2018,lind2018} to remove the contamination from field stars in our sample. For Pleiades, we adopted the distance of $136\pm12~\text{pc}$ ($\equiv \mu \pm 3\sigma$, the same below), proper motions are $\mu_{\alpha \ast}=+19.9\pm3.9~\text{mas~yr}^{-1}$ and $\mu_{\delta}=-45.5\pm4.5~\text{mas~yr}^{-1}$. For Praesepe, the adopted distance is $186\pm18~\text{pc}$, proper motions are $\mu_{\alpha \ast}=-36.1\pm4.2~\text{mas~yr}^{-1}$ and $\mu_{\delta}=-12.9\pm3.6~\text{mas~yr}^{-1}$. The adopted mean value ($\mu$) and standard deviation ($\sigma$) of each parameter were estimated by a Gaussian fitting to the distribution of Gaia DR2 measurements of candidate members from Pleiades \citep{stau2007}  and Praesepe \citep{doug2014}. Wherever Gaia measurements are unavailable,  LAMOST radial velocities (RVs) are used to remove potential field stars in our sample. The adopted RVs of Pleiades and Praesepe are $RV=+5\pm12~\text{km~s}^{-1}$, and $RV=+31\pm12~\text{km~s}^{-1}$, respectively, which were estimated by measuring the mean value and scatter of LAMOST RVs of candidates members. For Hyades, we simply removed the stars closer than 28 pc or far above 66 pc, not using their proper motions and RVs to remove potential field stars because of the proximity (thus having larger scatters). In addition, we removed several stars hotter than 6500 K from above sample as our main focus is on cool stars that have convective envelopes (e.g., from late-F to M-type). Our final sample consist of 312 late-type stars (113 Pleiades members, 160 Praesepe members, and 39 Hyades members), and each star has spectra of at least 3 epochs having SNR$r$ $>$ 15 (see Fig.~\ref{fig:sample_hist}). We note that few stars are having spectra over 10 epochs, which could be due to frequent pointing of the field by LAMOST. The lower panel of Fig.~\ref{fig:sample_hist} shows the temporal distribution of the LAMOST observations. It is clear that LAMOST spectra of these stars were collected in eight observing seasons from 2011 to 2019 with each season from autumn to summer. Fig.~\ref{fig:time_delay} shows the distribution of time interval (in Barycentric Julian Days) between two consecutive epochs, wherein the time interval has wide range (from less than a day to over five years). 

It is important to note that about 12 percent of our sample stars are binary candidates, identified from available colour--magnitude diagrams in a way described in Paper I \& II. We checked and found that these binary stars have similar variation behaviours compare to single star candidates, and removing them from our sample will not change the main conclusions. In this work, therefore, the binary candidates were not specially marked in all plots.  
\section{Quantifying activity} 
An active late-type star is often featured by the excess chromospheric emission in Hydrogen Balmer lines and Ca~{\sc ii} lines \citep[e.g.][]{stau1986,sode1993,west2004,west2015,newt2017,fang2018}, which is closely correlated with magnetic field.  For example, solar plages (the Ca~{\sc ii} K enhanced patches on the chromosphere) are found to be highly correlated with the location of the magnetic field concentrations \citep{shee2011,chat2016}, and the residual Ca~{\sc ii} HK flux is found to be approximately proportional to the square root of the surface mean magnetic field strength for stars with non-saturated chromospheres \citep{schr1989}. On the other hand, the spectrum also have footprints by photospheric temperature inhomogeneities like cool starspots. For instance, the spectrum of a most-spotted star shows excess absorption of the molecular feature like TiO bandhead at 7050~\AA, thus the relative TiO band strength can be used as an indicator of cool spot filling factor \citep[e.g.][]{vogt1979,rams1980,vogt1981,huen1987,neff1995,onea2004,fang2016}, i.e., stronger the TiO absorption  larger the cool spot coverage on its photosphere for a given star. In this work, we adopted TiO band strength near 7050~\AA~as the tracer of the coverage of cool spots, and the emission of H$\alpha$ line as the chromospheric activity indicator. 
\subsection{Measurements of activity indicators}
We used spectral index to estimate the strength of TiO molecular band near 7050~\AA, namely the ratio of mean flux within TiO absorption band ($\overline F_{\lambda,\text{Num}}$) to that in nearby continuum ($\overline F_{\lambda,\text{Den}}$), as defined by formula~\ref{equ:index}. We measured two TiO band indices, TiO2n and TiO5n, following the definitions listed in Table~\ref{tab:index_defin}. The individual measurements for each sample star were listed in Table~\ref{tab:measurements_sample}. 

\begin{equation}
\label{equ:index}
\text{Index} = \frac{\overline F_{\lambda,\text{Num}}}{\overline F_{\lambda,\text{Den}}}. 
\end{equation}

We measured the equivalent widths of H$\alpha$ and H$\beta$ lines using the formula~\ref{equ:ew}, where $F_{\lambda}$ denotes the flux in line bandpass, $F_{c}$ represents the average flux in nearby pseudo-continua. We thus obtained equivalent widths, EW$_{\text{H}\alpha}$ and EW$_{\text{H}\beta}$, following the definitions in Table~\ref{tab:index_defin}. The equivalent widths of sample stars are listed in Table.~\ref{tab:measurements_sample}.

\begin{equation}
\label{equ:ew}
\text{EW} = \int \frac{F_{\lambda}-F_{c}}{F_{c}}d\lambda.
\end{equation}

Note that these measurements are in relative manner. The measurements for stars of nearly the same spectral types are comparable. This is not true for stars of widely differing spectral type because the measurements depend not only upon the intrinsic strengths but also the nearby continuum flux. The mean values of TiO2n and EW$_{\text{H}\alpha}$ of each sample star were displayed as a function of effective temperature in upper panels of Fig.~\ref{fig:amp_error}, and also shown was their corresponding variation peak-to-peak amplitudes. The effective temperatures were estimated based on their broadband colors (for FGK-type stars in Pleiades) or spectral features like CaH and TiO strengths (for M-type stars in Pleiades, and all member stars in Praesepe and Hyades) following the procedures described in Paper II. These quantities were listed in Table~\ref{tab:information_sample}. In addition, as a comparison, the respective reference values of TiO2n and EW$_{\text{H}\alpha}$ were also shown by grey solid lines in Fig.~\ref{fig:amp_error}, which were derived based on a large sample of inactive dwarf stars following the procedure described in Paper I.
\begin{table}
\centering
\caption{Definition of spectral indices and equivalent widths}
\label{tab:index_defin}
\begin{tabular}{lcccccc}
   \hline
Index       & Numerator (\AA) & Denominator (\AA) \\
   \hline
TiO2n       & $7057-7064$     & $7042-7048$       \\
TiO5n       & $7126-7135$     & $7042-7048$       \\
   \hline
   \hline
EW        & Line bandpass (\AA) &  Pseudo-continua (\AA)  \\
   \hline
EW$_{\text{H}\alpha}$   &  $6557-6569$        &  $6547-6557,~6570-6580$ \\
EW$_{\text{H}\beta}$    &  $4855-4867$        &  $4842-4852,~4873-4883$ \\
   \hline   
\end{tabular}
\end{table}
\subsection{Measurement errors}
\begin{figure*}
\centering
\includegraphics[width=\columnwidth]{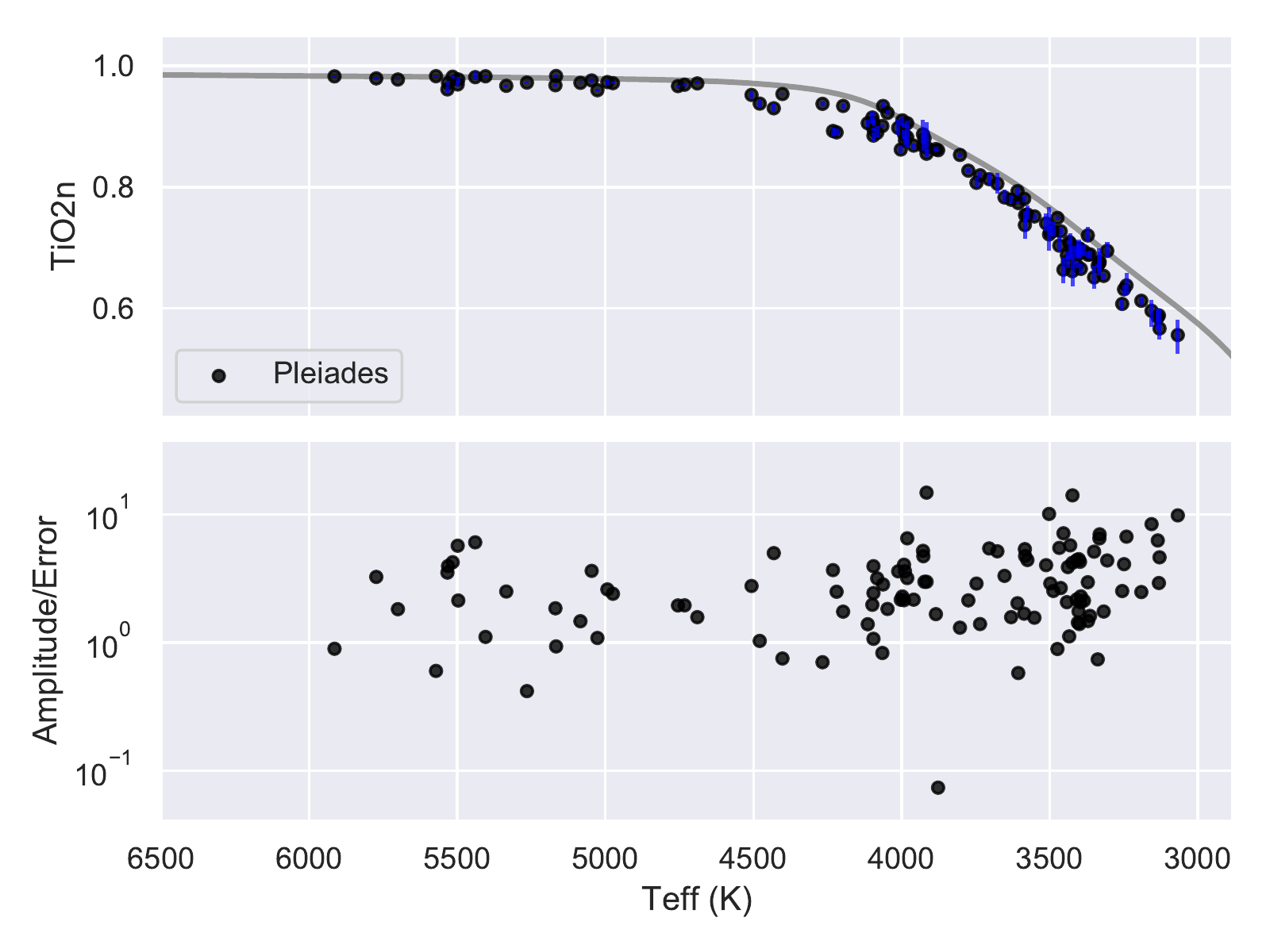}
\includegraphics[width=\columnwidth]{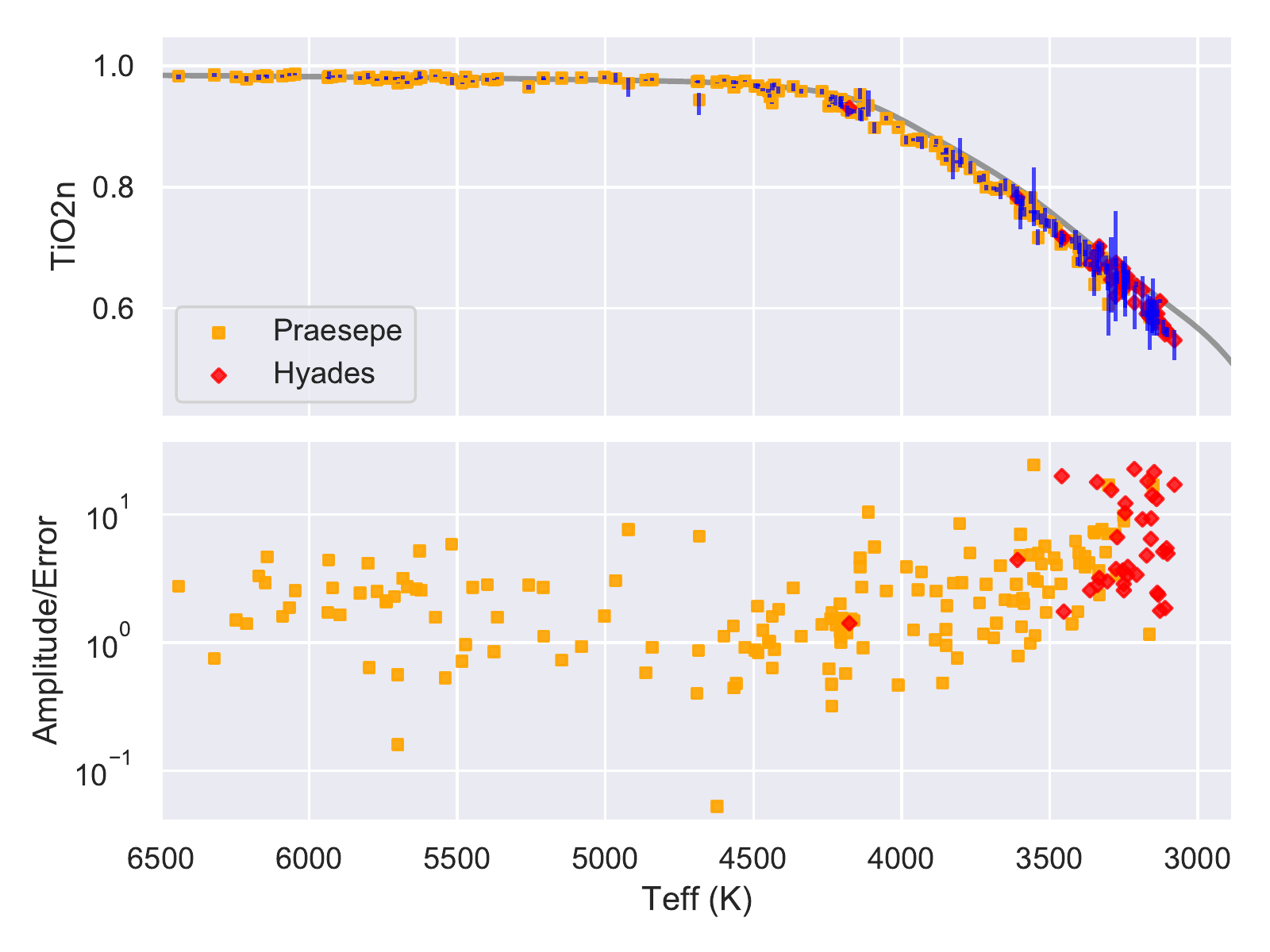}
\includegraphics[width=\columnwidth]{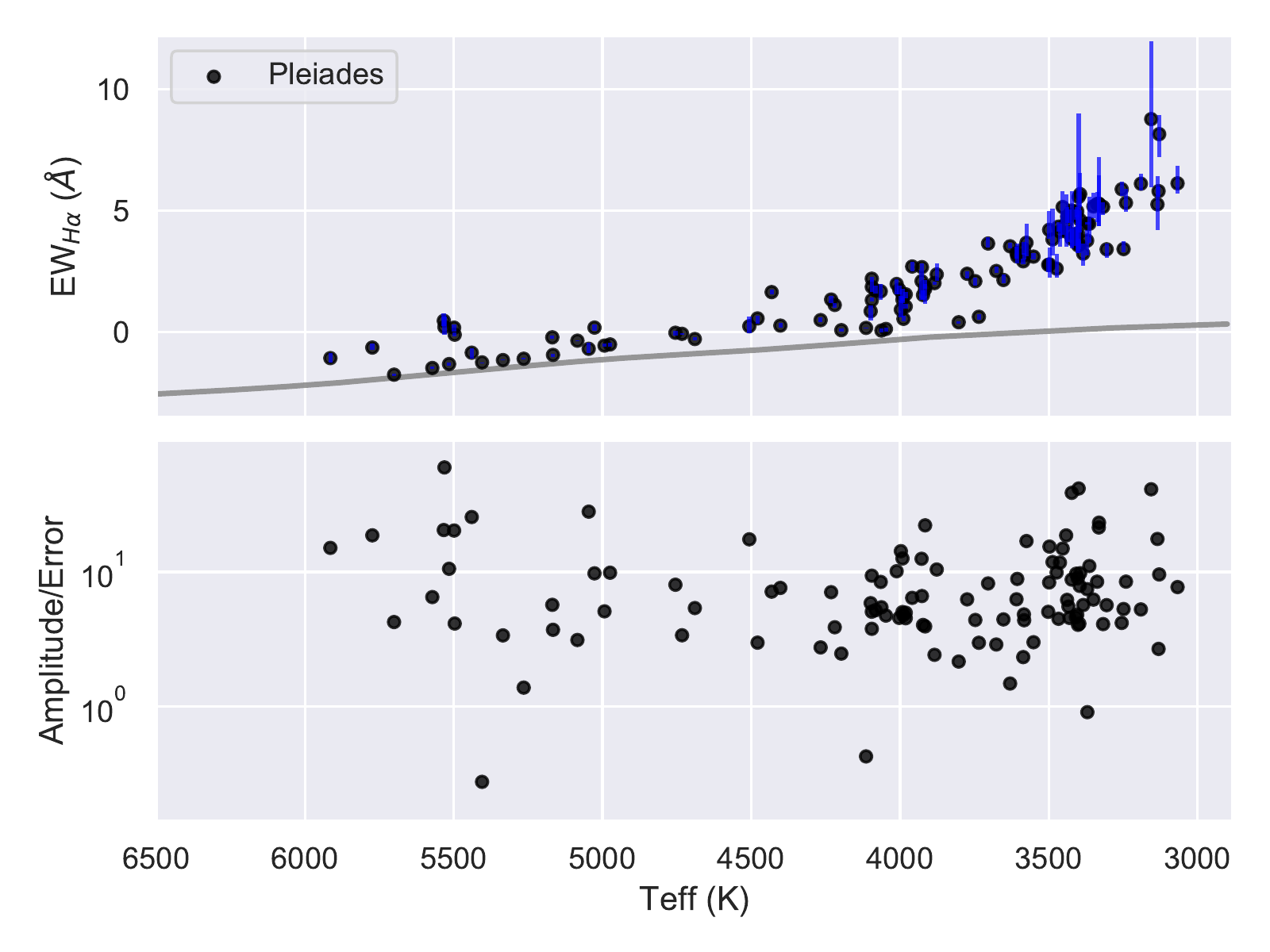}
\includegraphics[width=\columnwidth]{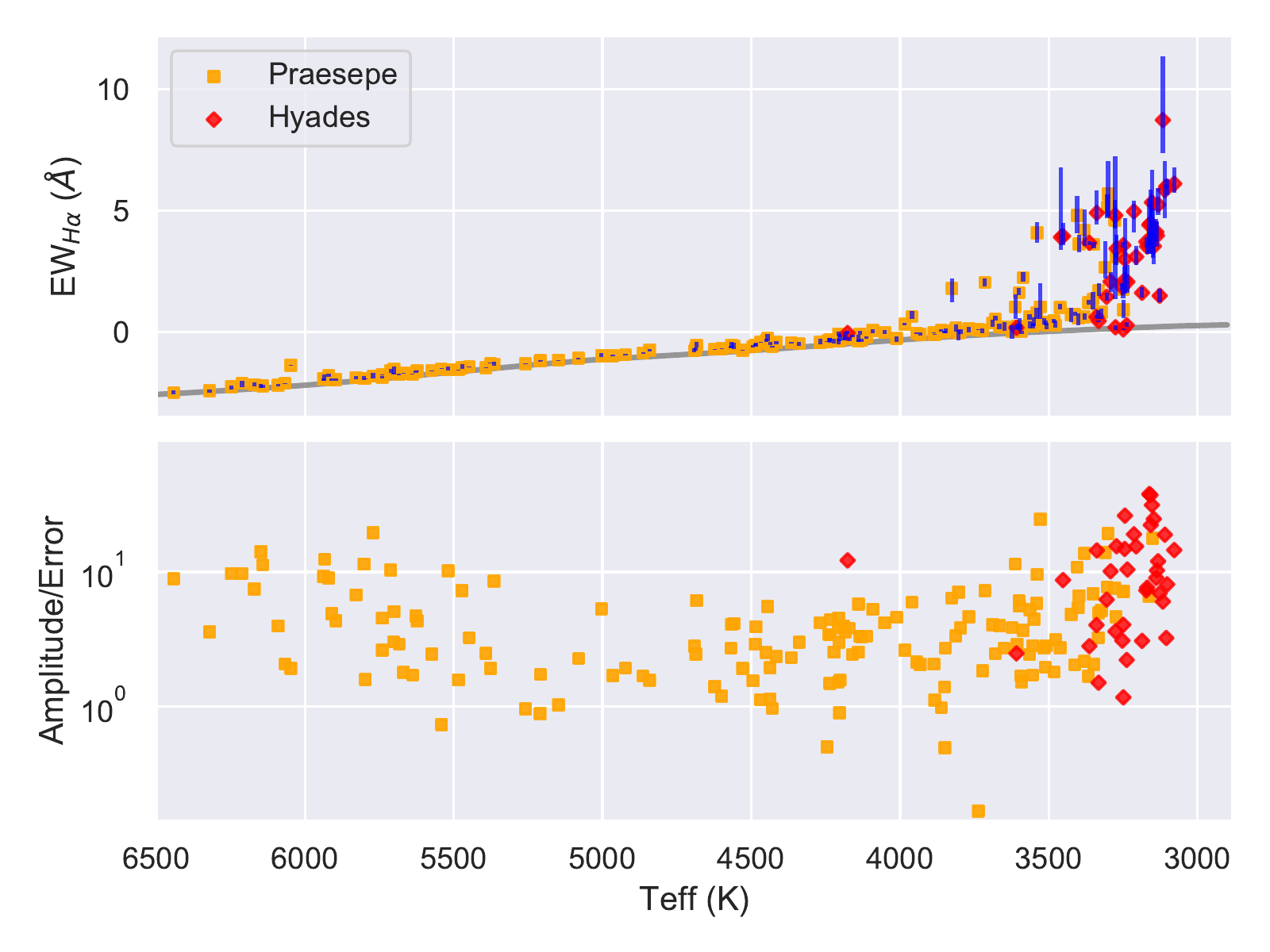}
\caption{Measured TiO2n and EW$_{\text{H}\alpha}$ against stellar effective temperature (Left: Pleiades; Right: Praesepe and Hyades), where the symbols with blue bar represents the mean value for each star, while the blue vertical bar is observed variation range (peak-to-peak amplitude) rather than the measurement error. The solid grey lines are respective standard values of inactive reference stars. The lower panel in each plot shows the ratio of the peak-to-peak amplitude to the mean value of measurement error for each star. }
\label{fig:amp_error}
\end{figure*}

Estimating the measurement errors is important particularly while investigating the temporal variation, since we should identify whether the observed variation is intrinsic or an artifact of noise. In this work, the measurement errors were estimated using the Monte Carlo simulation, i.e., we first simulated 10000 spectra resembling the observed one for each star combining with Gaussian random noise of standard deviation equal to the uncertainty in the LAMOST spectra at each wavelength (e.g. Poisson noise), and then adopted the standard deviation of the measurements based on these spectra as the measurement error. The estimated errors were listed in Table~\ref{tab:measurements_sample}. The typical errors of TiO2n are $\sim0.002$, $\sim0.005$, $\sim0.007$, for Pleiades and Praesepe members with temperatures around 5500, 4500, 3500 K, respectively. The corresponding typical errors for EW$_{\text{H}\alpha}$ are $\sim0.02$, $\sim0.05$, $\sim0.1$~\AA. On average, Hyades stars have relatively smaller errors because of the proximity. The ratio of observed peak-to-peak amplitude to the mean value of measurement errors (Amplitude/Error, hereafter $R_{\text{amp}}$) for each star is displayed in Fig.~\ref{fig:amp_error}. A simple statistics for time series simulated from Gaussian distribution shows the ratio of peak-to-peak amplitude to the standard deviation for a pure-noise time series (with 10 points) is around 3.17 (a scatter with $\sigma \sim0.31$). This ratio strongly depends on the number of points in the time series, i.e., the average is from 2.48 ($\sigma \sim0.18$) to 3.54 ($\sigma \sim0.37$) for 5 to 15 points, respectively. Considering that almost our sample stars having epochs no more than 6 (see Fig.\ref{fig:sample_hist}), in this work, we simply used the ratio $R_{\text{amp}}>6$ to separate the stars with believable variation from those stars whose variation is dominated by noise. It is clear that the observed variation amplitudes of TiO2n for many stars are dominated by noise, e.g., about 10 percent Pleiades stars and 21 percent Praesepe stars have $R_{\text{amp}}<1$, only 10 percent stars in Pleiades and Praesepe have ratios $R_{\text{amp}}>6$. Compared to TiO2n, the observed variations in EW$_{\text{H}\alpha}$ for more stars are evident, e.g., about 50 percent Pleiades stars and 24 percent Praesepe stars have ratios $R_{\text{amp}}>6$.
\begin{table*}
\caption{Measurements from the multi-epoch LAMOST spectra of representative sample. Full table is available online.}
\label{tab:measurements_sample}
\begin{tabular}{lccccccccccccccccccc}
   \hline
ID$^{a}$ & obsid$^{b}$&    BJD        &$\text{EW}_{\text{H}\alpha}$ & e$\text{EW}_{\text{H}\alpha}$ 
                                 &$\text{EW}_{\text{H}\beta}$  & e$\text{EW}_{\text{H}\beta}$ &TiO2n & eTiO2n   & TiO5n      & eTiO5n  & Cluster    \\
    &            & (+2450000)    &  (\AA)      &  (\AA)    &   (\AA)     &  (\AA)     &            &            &            &         &           \\
   \hline
  1  &    66302196  &  6213.25598  &   2.4359  &    0.0713 &    2.6312  &   0.2000   &  0.8038  &   0.0051  &   0.6627 &    0.0042 &  Pleiades  \\ 
  1  &   100902196  &  6295.07148  &   2.4249  &    0.1580 &    2.0914  &   0.4255   &  0.8237  &   0.0117  &   0.6718 &    0.0095 &  Pleiades  \\ 
  1  &   470603212  &  7662.29203  &   2.6923  &    0.0485 &    2.0960  &   0.1633   &  0.7892  &   0.0033  &   0.6509 &    0.0027 &  Pleiades  \\ 
  2  &   286111063  &  7018.07203  &  -1.0656  &    0.0704 &   -1.5015  &   0.1107   &  0.9734  &   0.0079  &   0.9177 &    0.0072 &  Pleiades  \\ 
  2  &   480714246  &  7703.24185  &  -1.1052  &    0.0190 &   -1.4841  &   0.0250   &  0.9725  &   0.0023  &   0.9315 &    0.0021 &  Pleiades  \\ 
  2  &   631316115  &  8135.02749  &  -1.1114  &    0.0100 &   -1.4272  &   0.0135   &  0.9718  &   0.0013  &   0.9248 &    0.0011 &  Pleiades  \\ 
  3  &   100905178  &  6295.07847  &   0.0330  &    0.0475 &   -0.1102  &   0.1361   &  0.9301  &   0.0047  &   0.8752 &    0.0042 &  Pleiades  \\ 
  3  &   286109165  &  7018.07203  &   0.0745  &    0.0389 &    0.1325  &   0.1111   &  0.9342  &   0.0041  &   0.8770 &    0.0037 &  Pleiades  \\ 
  3  &   480710094  &  7703.24186  &   0.1383  &    0.0414 &    0.1627  &   0.1041   &  0.9377  &   0.0043  &   0.8824 &    0.0039 &  Pleiades  \\ 
 ... &   ...        &    ...       &  ...      & ...       &  ...       &  ...       & ...      &  ...      &  ...     & ...       &  ...      \\
\hline
\multicolumn{12}{l}{a: Arbitrarily located star ID in this work (identical to the ID in Table~\ref{tab:information_sample}).}\\
\multicolumn{12}{l}{b: LAMOST spectrum ID.}\\
\end{tabular}
\end{table*}
\begin{table*}
\caption{Basic information of sample stars and their statistics of measurements. Full table is available online.}
\label{tab:information_sample}
\begin{tabular}{lccccccccccccccccccc}
   \hline
 ID &Gaia ID & RA         & Dec          &Teff      &  Period     &$\mu_{\text{EW}_{\text{H}\alpha}}$ &$A_{\text{EW}_{\text{H}\alpha}}$ 
                                                                                  &$\mu_{\text{TiO2n}}$ &$A_{\text{TiO2n}}$ &$C_{x,y}~^{a}$  \\
    &   & (deg)    & (deg)      & (K)      &  (d)        & (\AA)     &   (\AA)   &         &         &         &             \\
   \hline 
  1  &  67732257029618048  & 54.597412 & +23.108112 & 3676.31  &    3.747900  &   2.5177  &   0.2674  &   0.8056  &   0.0345  &   --    \\
  2  &  68334235349446528  & 55.128021 & +24.487307 & 5264.90  &    6.813200  &  -1.0941  &   0.0458  &   0.9726  &   0.0016  &   --   \\
  3  &  65107860213588096  & 55.307491 & +23.384953 & 4197.20  &   10.023400  &   0.0819  &   0.1053  &   0.9340  &   0.0076  &   --   \\
  4  &  68167835432980480  & 55.362019 & +24.017363 & 3155.58  &    0.359100  &   8.7672  &   5.9990  &   0.5958  &   0.0456  &   --   \\
  5  &  65151531440914560  & 55.414906 & +23.763084 & 3988.35  &   11.006300  &   0.5482  &   0.2267  &   0.8770  &   0.0151  & $-0.83$    \\
  6  &  68265588889169664  & 55.511932 & +24.210073 & 3395.80  &    0.699500  &   4.6335  &   0.9370  &   0.6922  &   0.0130  & $-0.73$    \\
  7  &  65161250950264192  & 55.512192 & +23.931648 & 3240.32  &    0.888900  &   5.3204  &   0.9405  &   0.6376  &   0.0364  &   --    \\
  8  &  68291977168100480  & 55.513741 & +24.537067 & 3399.36  &    2.257600  &   5.5613  &   5.7729  &   0.6987  &   0.0101  & $+0.17$    \\
  9  &  64450283540250240  & 55.516399 & +22.426901 & 4974.59  &    6.092648  &  -0.5050  &   0.1819  &   0.9716  &   0.0052  &   --    \\
 ... & ...                 &   ...     &    ...     &  ...     & ...          &  ...      &  ...      & ...       &  ...      &  ...     \\
\hline
\multicolumn{11}{l}{a: $C_{x,y}=C_{\text{TiO2n},\text{EW}_{\text{H}\alpha}}$ for stars having at least four LAMOST epochs.}\\
\end{tabular}
\end{table*}
\section{Results and Discussion} 
\subsection{Variability on different time-scales}\label{sec:var}
\begin{figure*}
\centering
\includegraphics[width=\columnwidth]{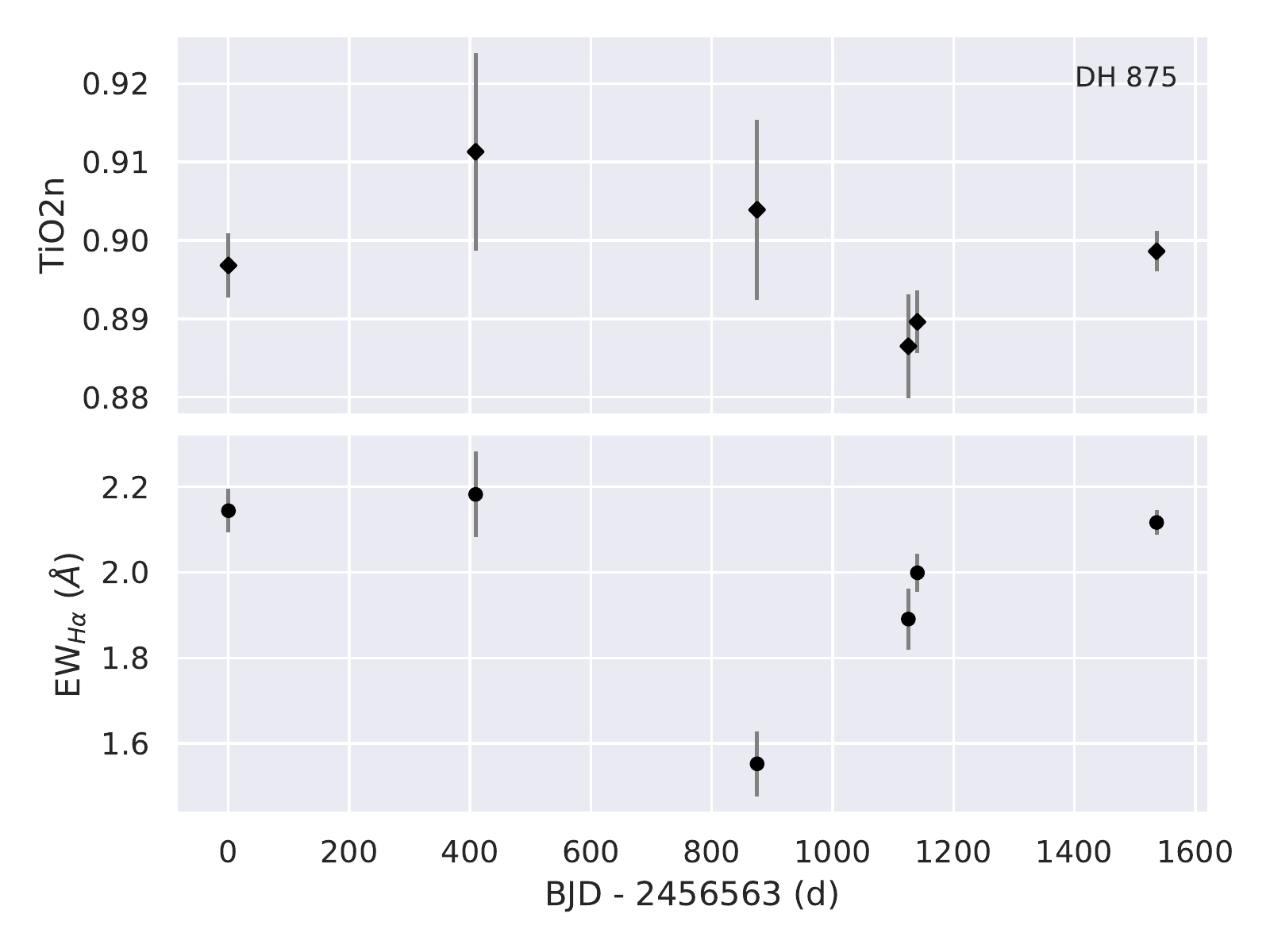}
\includegraphics[width=\columnwidth]{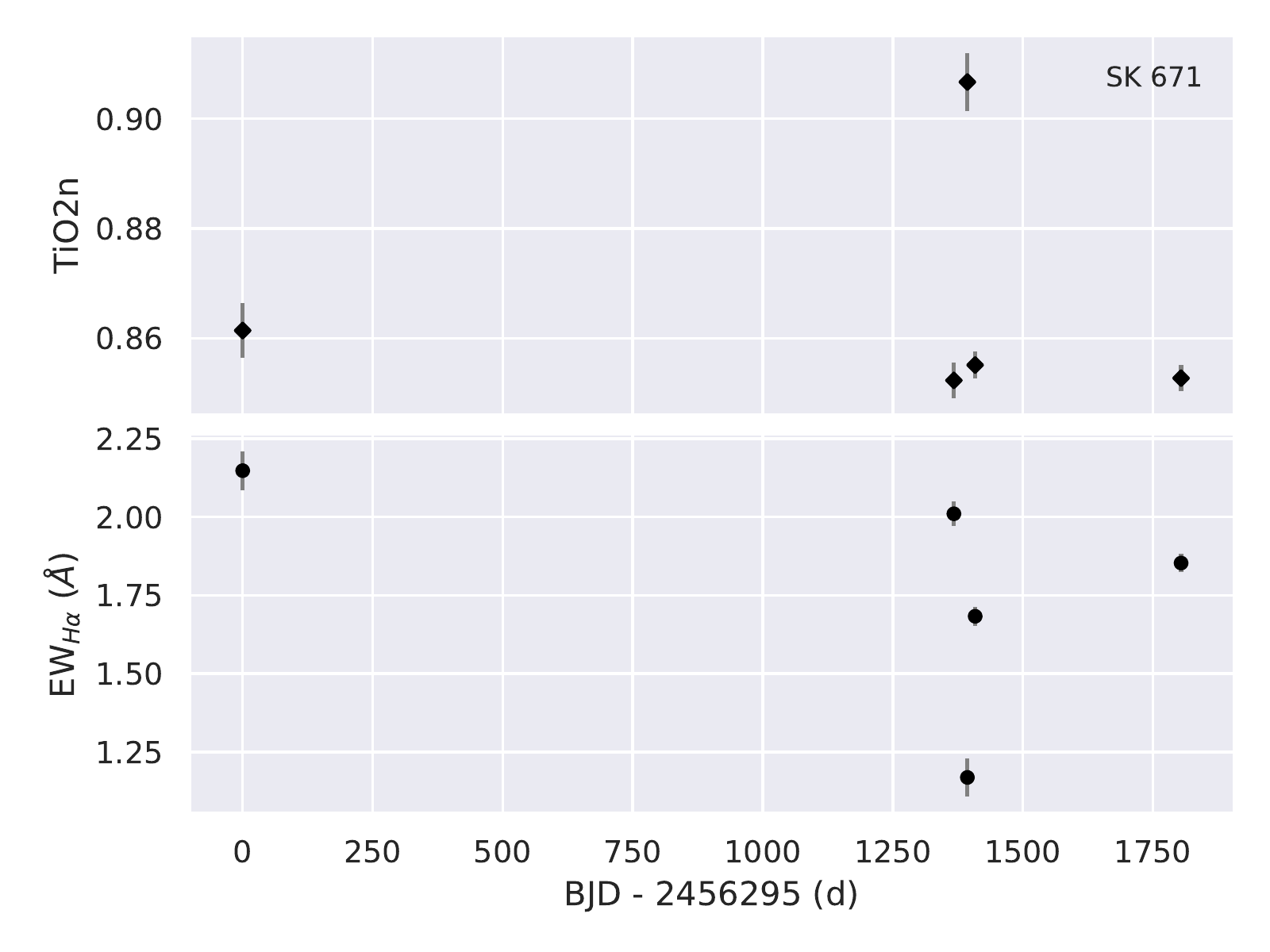}
\caption{Temporal variability of TiO2n and EW$_{\text{H}\alpha}$ for two late K-type Pleiades stars.}
\label{fig:time_ewha_tio2n_m45}
\end{figure*}

 Fig.~\ref{fig:amp_error} clearly shows that many stars (e.g. M-type stars) have detectable temporal variation in activity. Appreciable short time-scale (a few days, weeks or months) and long-term (years) variations in chromospheric emission (H$\alpha$ and Ca~{\sc ii} infrared triplet lines) for a few Pleiades stars were previously reported by \citet{sode1993}. Indeed, our results show that Pleiades stars often have time-variation behaviours in TiO2n and EW$_{\text{H}\alpha}$ in both short and long time-scales (days to years), as listed in Table~\ref{tab:measurements_sample}. Fig.~\ref{fig:time_ewha_tio2n_m45} shows typical variation patterns for two late K-type Pleiades stars, DH 875 and SK 671. 
 
\begin{figure*}
\centering
\includegraphics[width=\columnwidth]{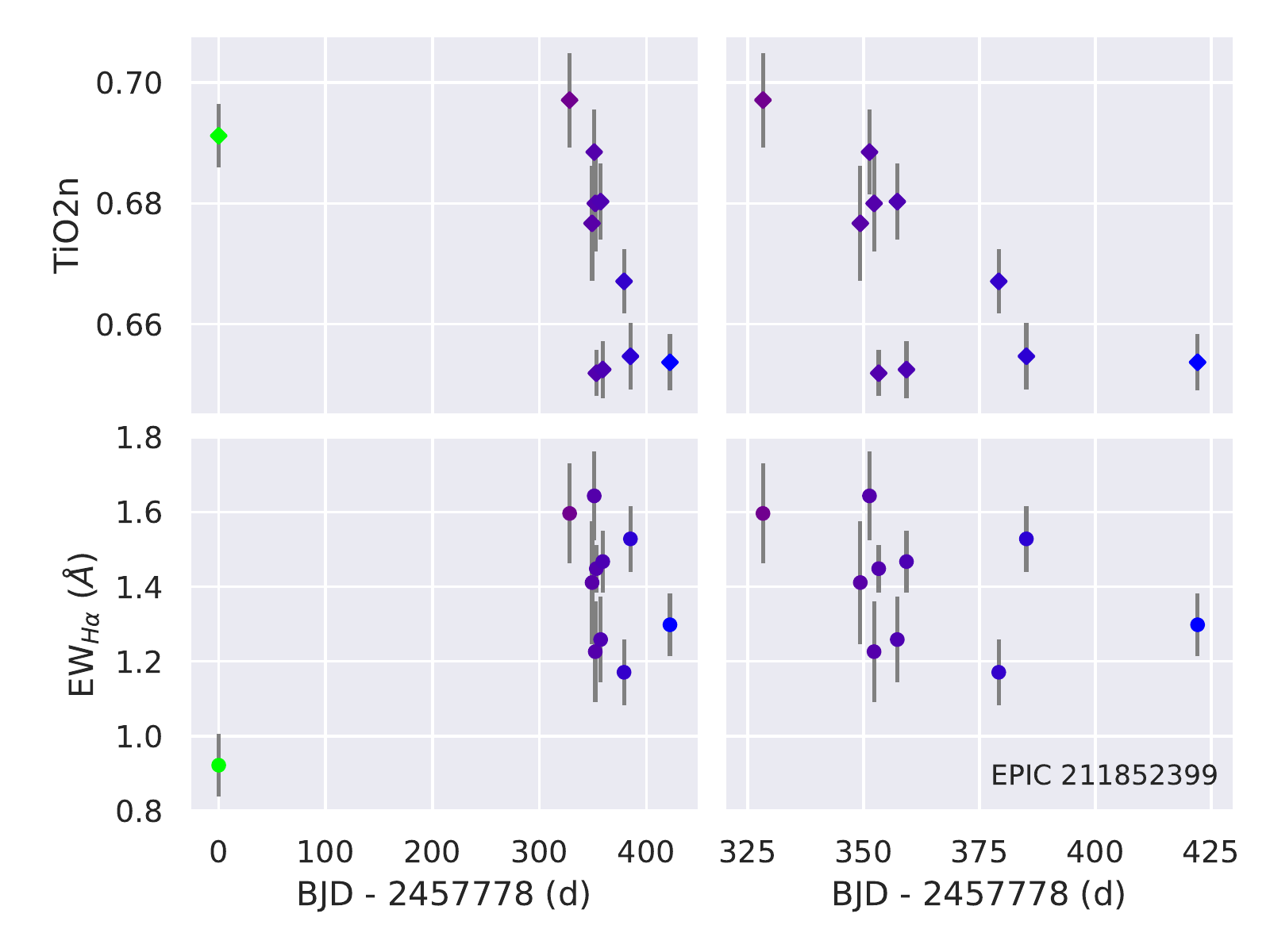}
\includegraphics[width=\columnwidth]{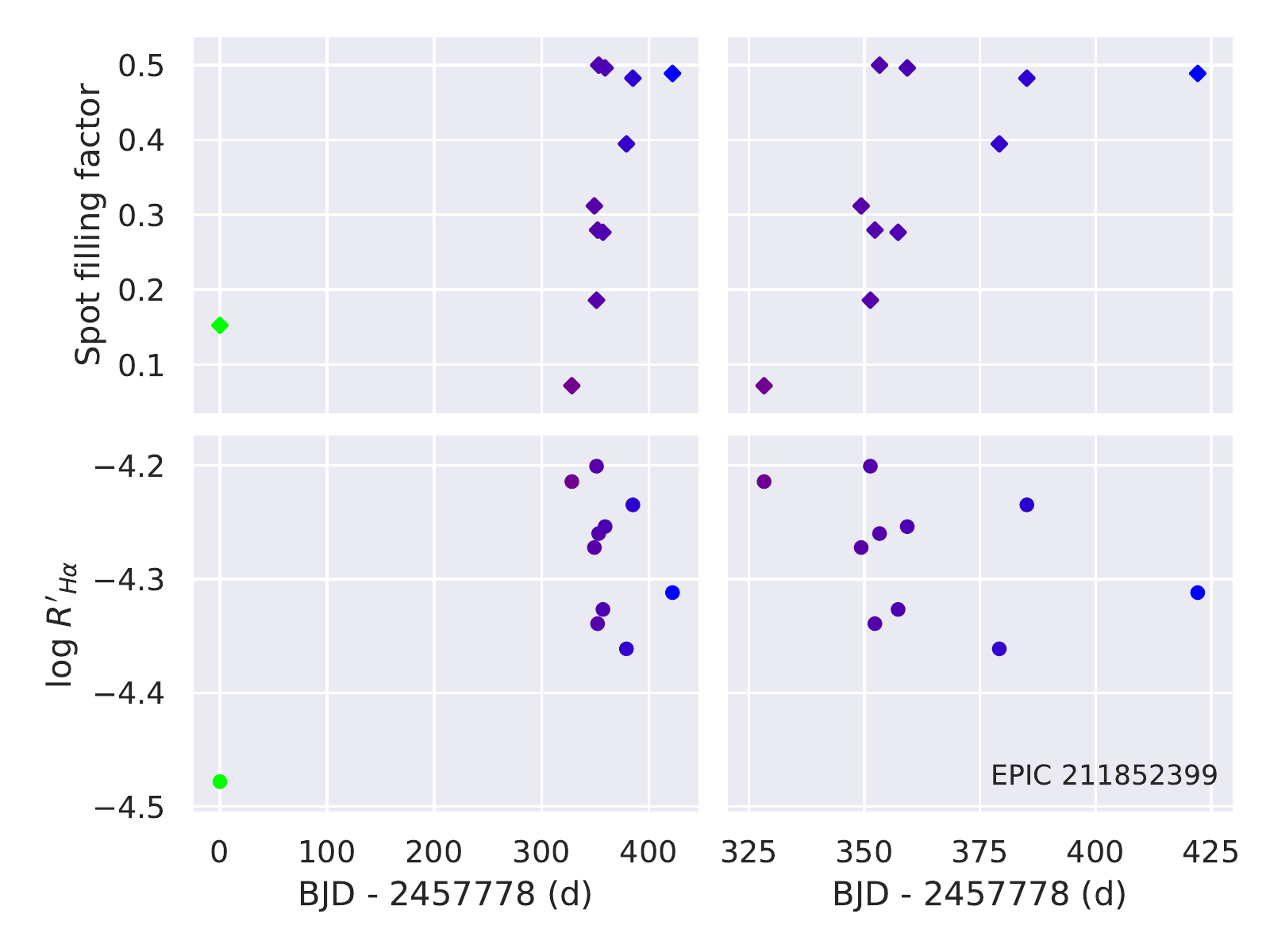}
\includegraphics[width=\columnwidth]{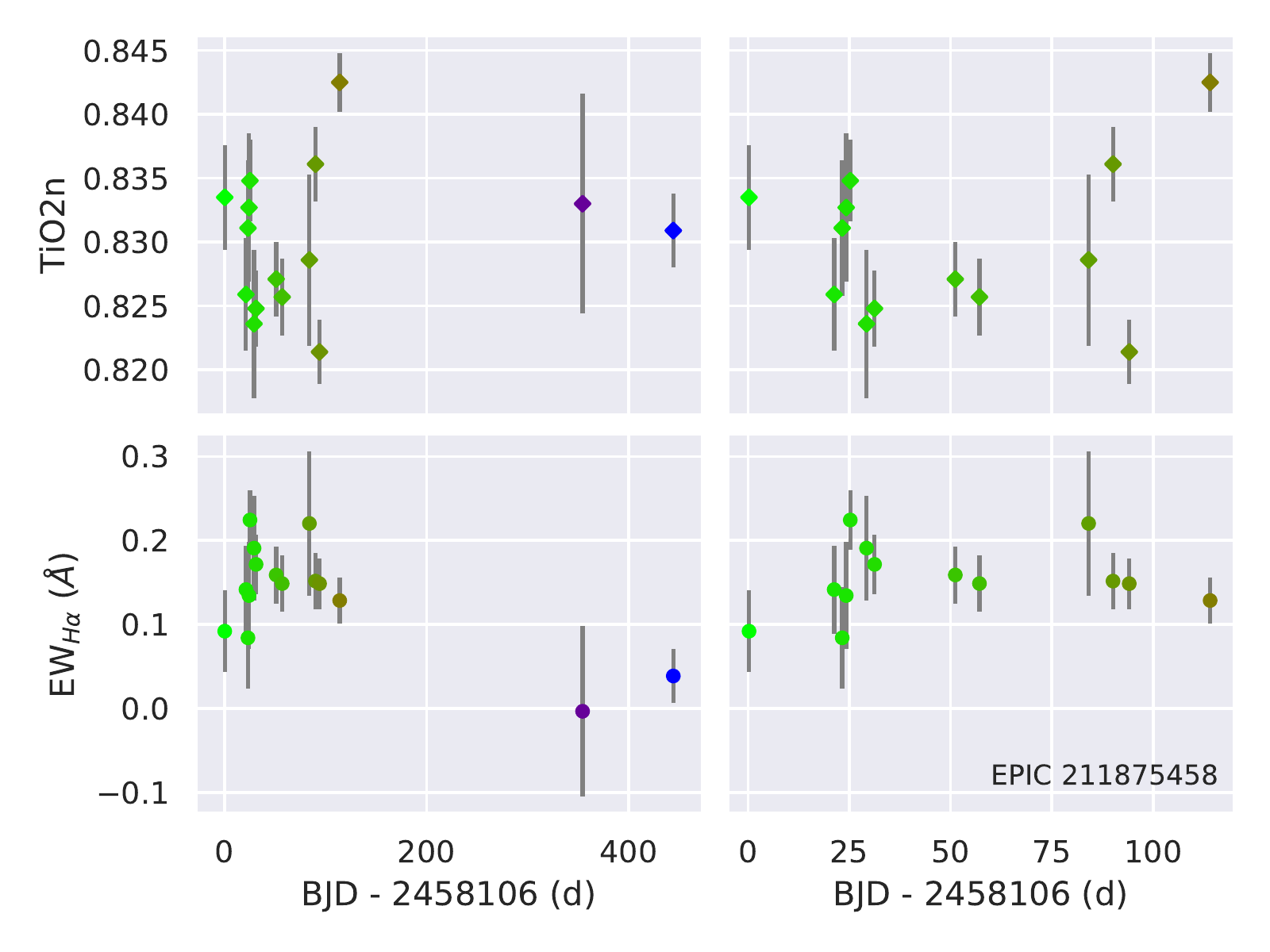}
\includegraphics[width=\columnwidth]{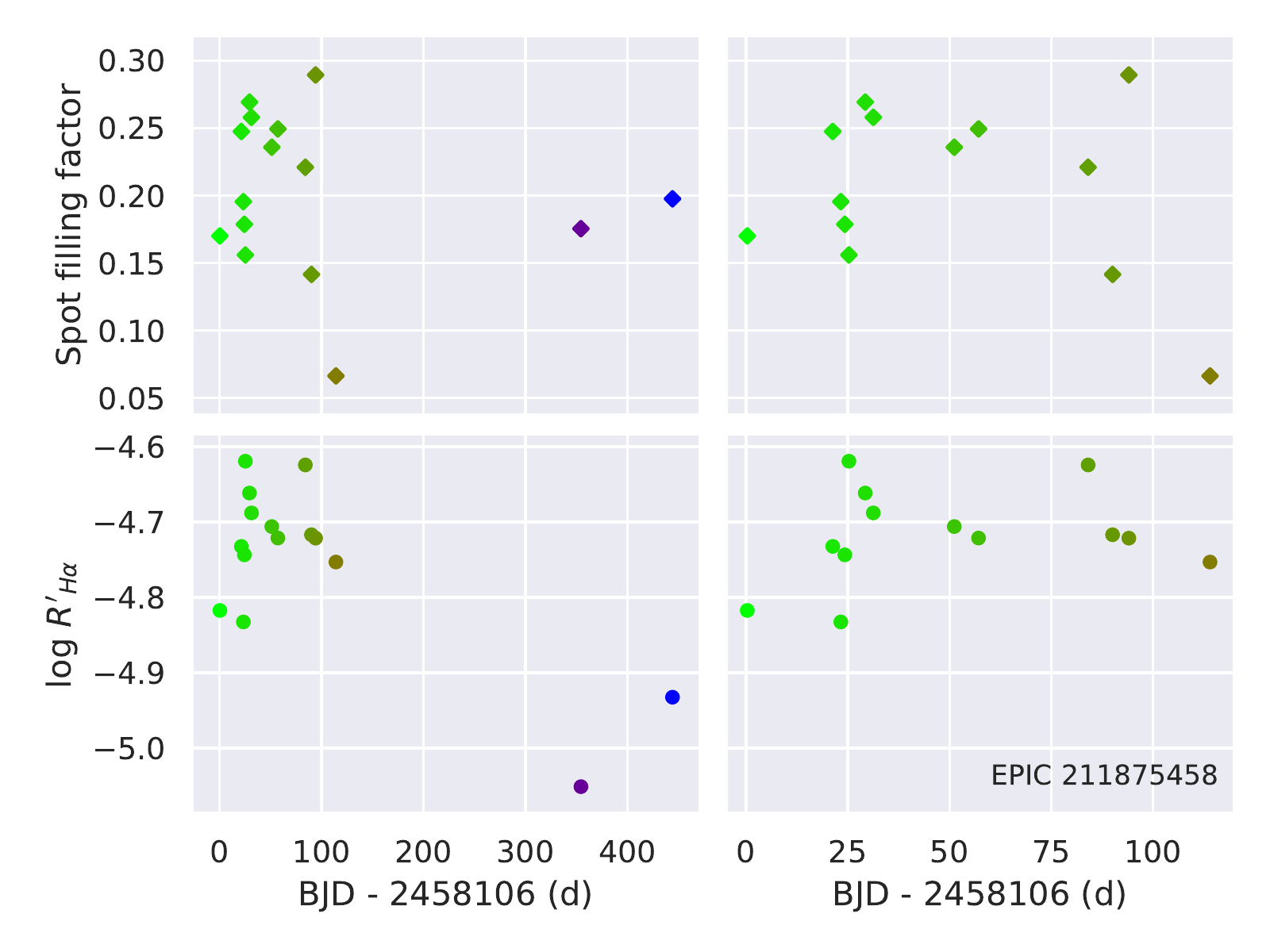}
\caption{Temporal variation of TiO2n and EW$_{\text{H}\alpha}$ (left), and TiO2n-based spot filling factors and H$\alpha$ fractional luminosities ($R'_{\text{H}\alpha}$) (right). Note the symbols are colour-coded as per observational time. The right panels in each plot are zoom-in panels of left ones.}
\label{fig:time_ewha_tio2n}
\end{figure*}

 Fig.~\ref{fig:time_ewha_tio2n} and Fig.~\ref{fig:time_ewha2} display the temporal variations in TiO2n and EW$_{\text{H}\alpha}$ for five most frequently observed stars in the Praesepe and Hyades, also shown are corresponding spot filling factors and H$\alpha$ excess fractional luminosities ($R'_{\text{H}\alpha}$). The spot filling factors were estimated by modelling their TiO2n strengths with respect to the standard TiO2n values obtained from inactive reference stars, based on the idea that the presence of cool spots could result in extra TiO band absorption (See Paper I for details). The H$\alpha$ excess fractional luminosities were obtained from excess equivalent widths using the $\chi$-method, where the H$\alpha$ excess equivalent widths were excess values compared to standard EW$_{\text{H}\alpha}$ values obtained from inactive reference stars (See Paper II for details). All of these five stars are M dwarfs (M0 to M4), which show variation in time-scales ranging from days to years. For instance, as shown in Fig.~\ref{fig:time_ewha_tio2n}, the M3 dwarf star EPIC 211852399 exhibits noticeable dispersion in both TiO2n and EW$_{\text{H}\alpha}$ during autumn 2017 to summer 2018, which implies activity variation is in short-term time-scales (days to weeks). In addition, it exhibits seasonal groupings that differ substantially from adjacent ones, which implies long-term variation of time-scales 1-2 years. In contrast, EPIC 211875458, an M0 dwarf star, exhibits little variation in TiO absorption and chromospheric emission on both short- and long-term time-scale, as shown in Fig.~\ref{fig:time_ewha_tio2n}. EPIC 202059188 shows yearly variation between consecutive seasons, e.g., both EW$_{\text{H}\alpha}$ and TiO2n during 2016-11-23 to 2017-03-27 (green-like colours) are on average different from those during the period from 2017-11-18 to 2018-02-12 (blue-like colors), showing much larger scatter in second season (see Fig.~\ref{fig:time_ewha2}). Additionally, both TiO2n and EW$_{\text{H}\alpha}$ of EPIC 202059188 are modulated by a rotation period around 0.7 day, which is responsible for the short-term variations seen within one season (see Appendix~\ref{sec:comments}). 

\begin{figure}
\centering
\includegraphics[width=\columnwidth]{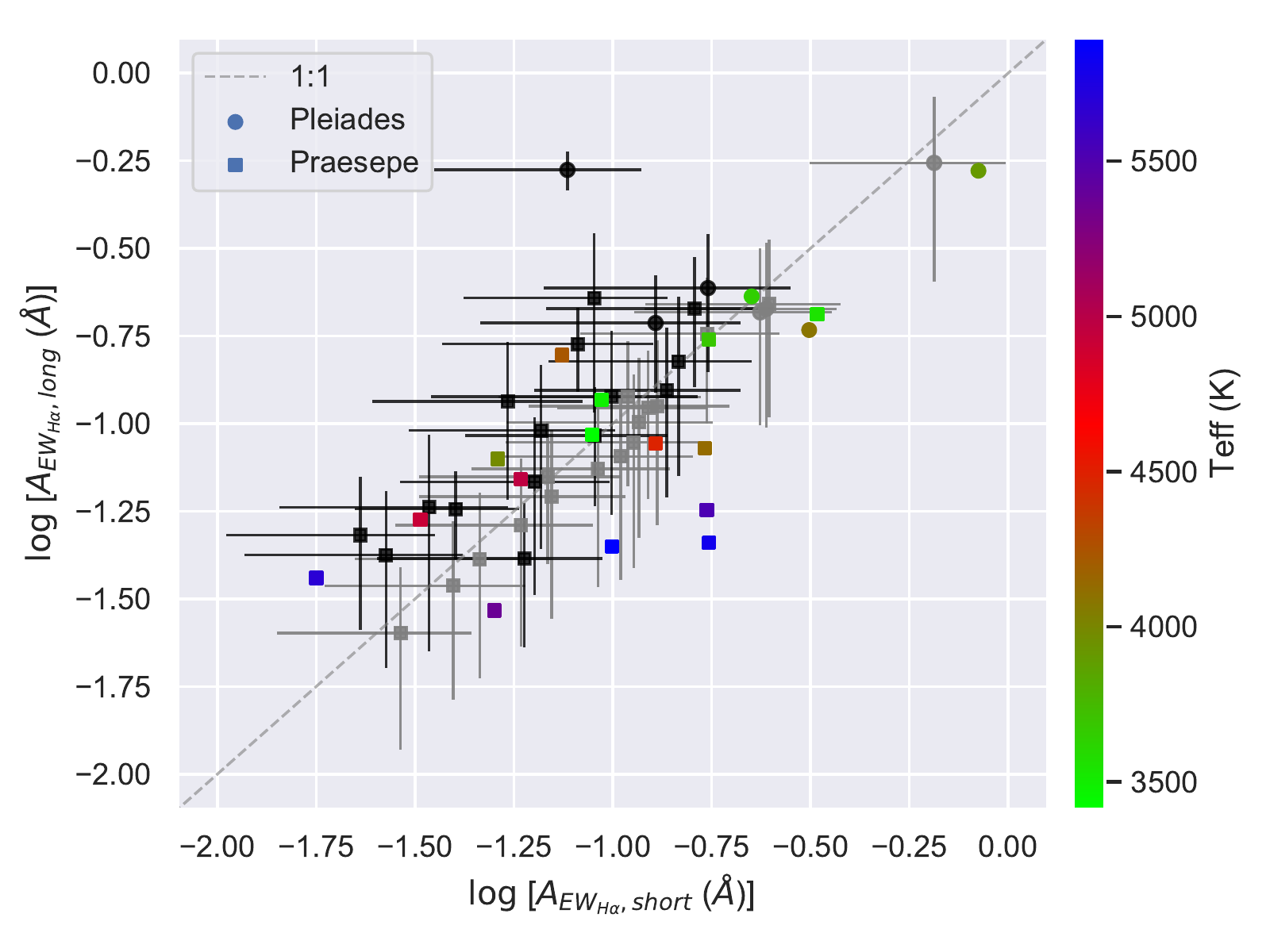}
\caption{Amplitudes of $\text{EW}_{\text{H}\alpha}$ are compared between short (within one season) and long time-scale (yearly, after averaging data within one season) for stars having at least three epochs. Note the colour contrast denotes stellar effective temperature, and all axes are in logarithmic scale. The black and grey symbols with errorbars are computed values for these stars to show the variation patterns for given observational noise, see section~\ref{sec:var} for more details. }
\label{fig:ampewha_short_long}
\end{figure}

From Fig.~\ref{fig:time_ewha_tio2n_m45}, Fig.~\ref{fig:time_ewha_tio2n}, and Fig.~\ref{fig:time_ewha2} we noticed that there is no appreciable difference in variation amplitude between different time-scales among these stars. To understand this issue further, Fig.~\ref{fig:ampewha_short_long} shows the $\text{EW}_{\text{H}\alpha}$ variation amplitudes in short time-scales (for data within one season) and long time-scales (yearly variation, after averaging data within one season) for stars having at least three epochs on both time-scales. We found no clear difference between them in this figure (along the 1:1 dashed line), but it is hard to give clear conclusion considering the sparse of our datasets and the observational uncertainties, thus we simulated two cases for each star to further understand this figure. Case-A: we simulated a case that is dominated by pure noise by sampling Gaussian random noise of standard deviation (equal to the standard deviation of observed $\text{EW}_{\text{H}\alpha}$ series) at the time of observations for each star, then calculated amplitude on the long and short time-scales for each simulation, as shown by grey symbols in Fig.~\ref{fig:ampewha_short_long}, where the error bars correspond to a scatter of 2000 simulations for each star. Case-B: we simulated a case that has long term signal (here simply a linear trend with a amplitude equal to observed long-term amplitude over the observing runs) with each epoch having random noise equal to the measurement error, as shown by black symbols in Fig.~\ref{fig:ampewha_short_long}, where the error bars correspond to a scatter of 2000 simulations for each star. As shown in Fig.~\ref{fig:ampewha_short_long}, Case-A shows very similar variation patterns on short- and long-term time-scales, as expected. Surprisingly, Case-B is very similar to Case-A for most stars (in fact, the long-term variation amplitude is on an average larger than that on short-term time-scale), indicating that the variation pattern of most stars are dominated by observation noise. Only one Pleiades star (SK 671, the star at the top-right corner of this figure) show obvious difference with Case-B. Therefore, considering the noise and sparse sampling of LAMOST observations, it is hard to discard the possibility of presence of long-term variation among these stars. Possibly, the observed patterns of long-term and short-term variation (at lest for those stars having low $R_{\text{amp}}$ ratios) are just false signals due to observational noise. But some stars still show evidence of short-term, stochastic activity variation, being lack of smooth variation cycles seen in slowly rotating old stars like the Sun \citep[e.g.][]{vaug1980}. 

All our sample stars are definitely younger than 1 Gyr, thus more intense and complex variation patterns are expected. Our results show evidence of such intense/complex variation patterns as discussed above. These observed rapid, irregular variations in activity among fast rotating young stars probably indicates the presence of complex small-scale magnetic fields,  which vary rapidly and chaotically with time that are generated by the large dynamo number \citep{park1971}. \citet{durn1981} argued that the Vaughan-Preston gap \citep{vaug+1980} represents a rapid change in dynamo number from large values to small values, corresponding abrupt shift from the complex field morphology associated with the multiple-mode dynamo to the simpler morphology of the single-mode dynamo. Alternatively, it may owe to a more profound switch in dynamos \citep{bohm2007}, i.e., a different dynamo operates in such young fast rotating stars. Moreover, active FGK-type stars younger than about 2.3 Gyr (i.e., stars with age of $0.6-2.3$ Gyr) often show both short and long period cycles \citep[][and references therein]{saar1999,bran2017}. If this scenario is still real for our sample stars, then the co-existing of short cyclic variation with periods between 1$-$3 years and long cyclic variation with periods 7$-$10 years would be expected (e.g., short cycle $P_{cyc}\sim2$ and long cycle $P_{cyc}\sim10$ years for Praesepe/Hyades K-type members). Unfortunately, we can not detect such cyclic behaviour among our FGK-type sample stars based on the current dataset, which could be due to the sparse sampling of LAMOST observations, though the whole time-span covers these short period cycles. In addition, evidences of cyclic chromospheric activity among young low-mass stars have been detected, e.g., \citet{iban2019} detected a possible activity cycle of period $\sim$5 yr in the very young and active dM1 star AU Microscopii. However, for M-type stars (in particular fully convective late-M stars), their cyclic behaviour of activity is still unclear. As mentioned above we did not detect any evident cyclic variation among our sample stars based on current sparse sampling datasets. More data points required for definitive statements for the cyclic variation, further datasets from the ongoing LAMOST survey would alleviate this issues.

\subsection{Age and rotation effect on variability}
\begin{figure*}
\centering
\includegraphics[width=\columnwidth]{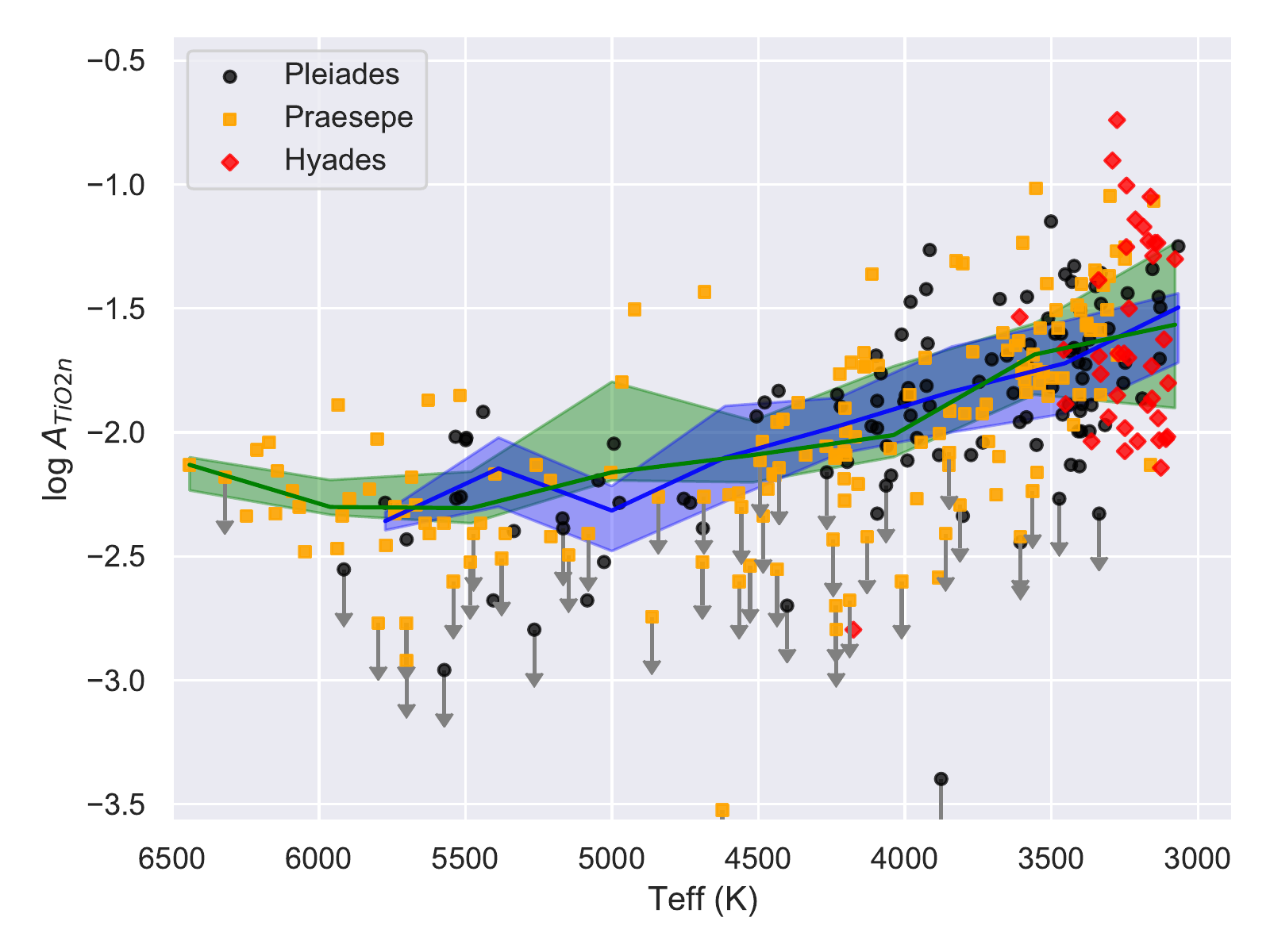}
\includegraphics[width=\columnwidth]{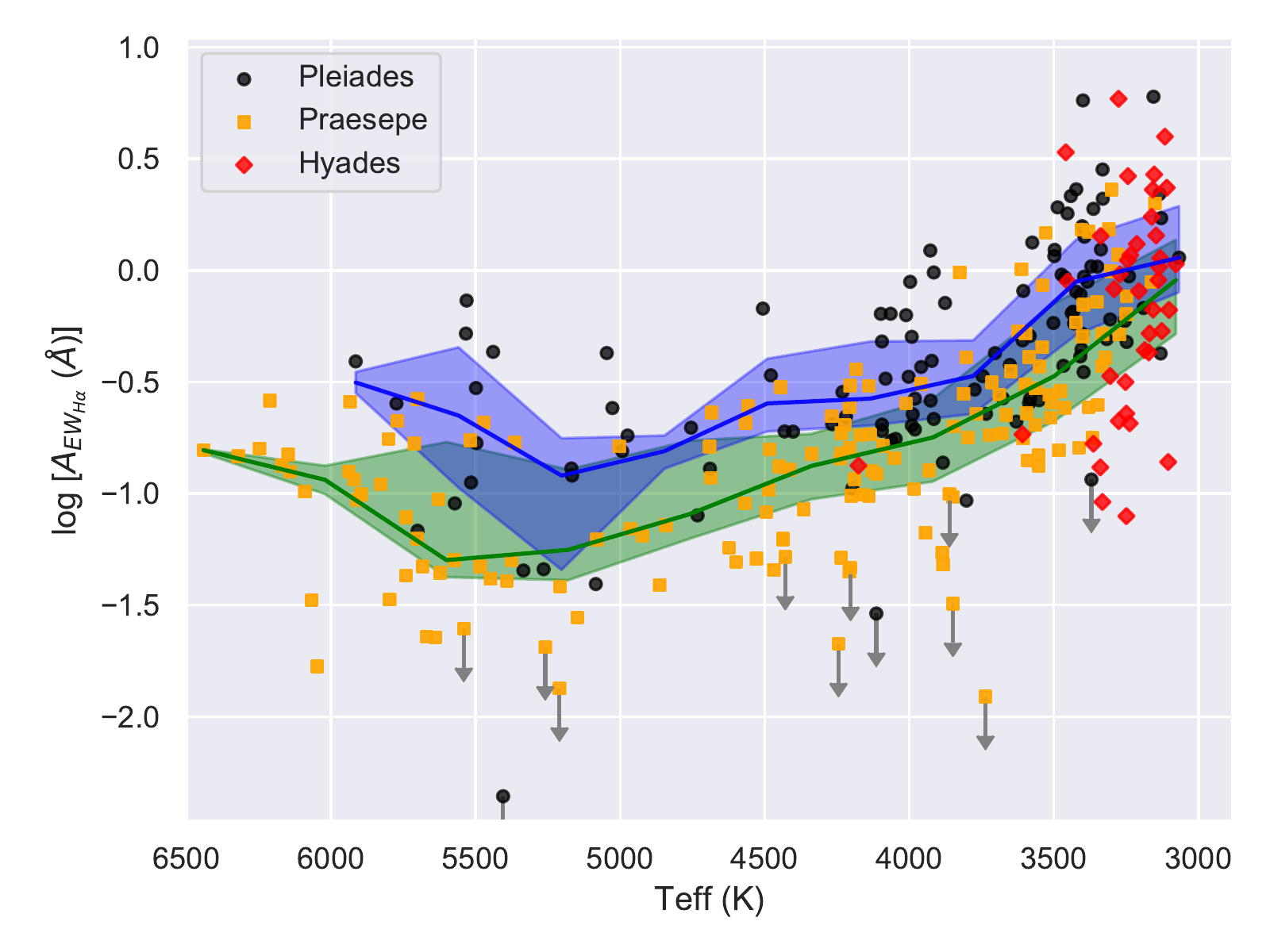}
\caption{Amplitudes of variation for TiO2n and EW$_{\text{H}\alpha}$ as a function of temperature. The solid lines (Blue: Pleiades; Green: Praesepe/Hyades) represent the temperature-binned median values, the shadows denote corresponding quartiles ($25^{\text{th}}$ and $75^{\text{th}}$ percent) for stars having $R_{\text{amp}}>1$ (the symbols with downward pointing arrows represent stars with $R_{\text{amp}}<1$, where the variation is dominated by observational noise).}
\label{fig:amp_ocs}
\end{figure*}

Fig.~\ref{fig:amp_error} shows that Pleiades stars are overall more active than stars of the same temperature in Praesepe/Hyades, i.e., Pleiades stars have stronger H$\alpha$ emissions and deeper TiO absorptions. Another remarkable feature of Fig.~\ref{fig:amp_error} is that Pleiades stars show large scatters in mean values of TiO2n and EW$_{\text{H}\alpha}$ (denote $\mu_{\text{TiO2n}}$ and $\mu_{\text{EW}_{\text{H}\alpha}}$, respectively) compared to those Praesepe/Hyades stars with similar effective temperatures, which also indicates an age-effect on stellar activity. These issues have been discussed in detail in Paper II, so, here we are interested in the behaviour of age-effect with time-variation. We presented the peak-to-peak amplitude of variations in TiO2n and EW$_{\text{H}\alpha}$ (denote as $A_{\text{TiO2n}}$ and $A_{\text{EW}_{\text{H}\alpha}}$, respectively) as a function of temperature in Fig.~\ref{fig:amp_ocs}. Note that the stars in Praesepe/Hyades are slightly metal-rich compare to Pleiades stars. We noticed such a difference in metallicity could not result in detectable effect on activity variation, thus we simply discuss the age-effect on activity variability by comparing Pleiades sample with respect to Praesepe/Hyades sample. $A_{\text{EW}_{\text{H}\alpha}}$ shows a weak statistical difference between these two sample that the younger stars have intense variability, as illustrated by lines in Fig.~\ref{fig:amp_ocs}, which denotes the temperature-binned median value for Pleiades stars (blue line) and Praesepe/Hyades stars (green line), where the corresponding shades represent the quartiles in each temperature bin. A simple two-sided Kolmogorov-Smirnov test shows that the $A_{\text{EW}_{\text{H}\alpha}}$ distribution of the two sample are probably different, e.g., for stars with $3000 <\text{Teff}< 3750~K$, $p$-value$\sim0.01$, for stars with $3750<\text{Teff}<4500~K$, $p$-value$\sim4\times10^{-06}$. In fact, similar features have been detected among older Sun-like stars, i.e., compared to those less active solar-age stars, the younger active ones show intense chromospheric Ca~{\sc ii} HK variability in both short and long time-scales \citep[e.g.][]{radi1998}. Therefore, the results suggest that an age-effect on activity variation among cool stars younger than 700 Myr is still valid, i.e., the younger stars including M-type stars tend to have greater chromospheric activity variability. However, as shown in Fig.~\ref{fig:amp_ocs}, TiO2n distribution of these two sample are almost the same, e.g., the Kolmogorov-Smirnov test indicates a $p$-value$\sim0.21$ for stars with $3000 <\text{Teff}< 3750~K$, and $p$-value$\sim0.13$ for stars with $3750<\text{Teff}<4500~K$, probably due to that TiO2n has large noise compared to its intrinsic variation. 

\begin{figure}
\centering
\includegraphics[width=\columnwidth]{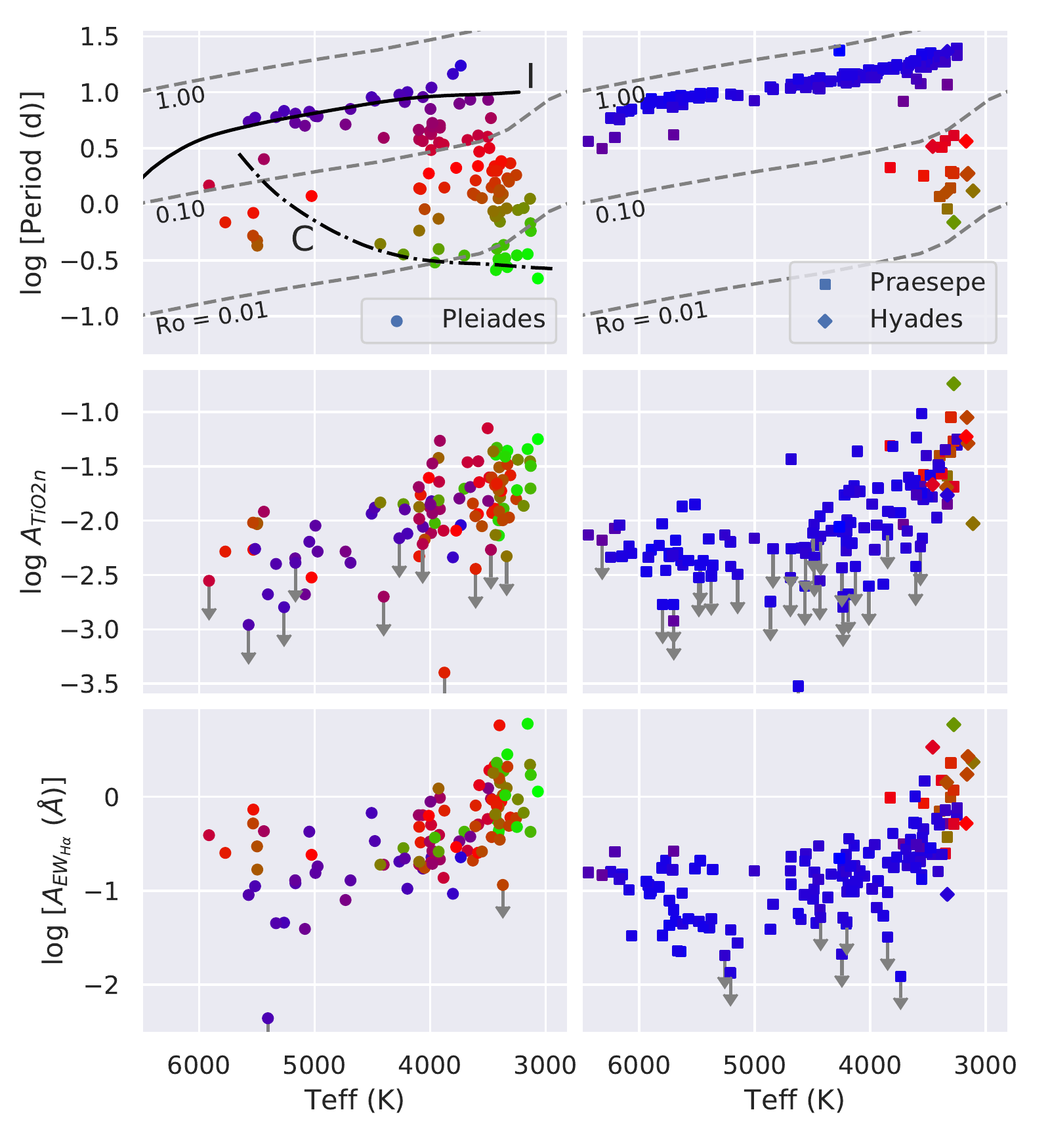}
\caption{Rotation periods and amplitudes of variation for TiO2n and EW$_{\text{H}\alpha}$ as a function of temperature are for stars with available rotation period. Each star is colour-coded by its Rossby number (Ro), as illustrated by grey dashed line in top panels. The I and C rotation sequences in Pleiades were illustrated by black solid and dash-dotted lines in top panels, respectively. The symbols with downward pointing arrows represent stars with $R_{\text{amp}}<1$. }
\label{fig:amp_rot}
\end{figure}

Stellar rotation is known to be related to the activity level, e.g., faster rotators tend to have larger starspot coverages \citep[e.g.][]{fang2016}, stronger chromospheric emissions \citep[e.g.][]{stau1997, doug2014,newt2017,fang2018} and X-ray emissions \citep[e.g.][]{pizz2003,mama2008}, thus a correlation between rotation and variation level should also be expected. In fact, Fig.~\ref{fig:amp_ocs} shows that there exists large scatter in variation amplitude even for stars in the same cluster. It is reasonable to connect the scatter to the stellar rotation rate diversity, considering the fact that there exist noticeable differences in rotation periods between stars of the same temperature in young open clusters \citep[e.g.][]{stau2016}. To understand further on this scenario, we provided variation amplitudes as a function of temperature in Fig.~\ref{fig:amp_rot}, wherein shown were stars with known rotation periods. For comparison purpose, we also displayed the corresponding rotation periods in the upper panels. The periods were collected from the literature (For Pleiades, \citet{hart2010,cove2016,rebu2016a}; For Praesepe, \citet{ague2011,delo2011,kova2014,rebu2017}; For Hyades, \citet{hart2011,arms2015,doug2016}); for a few stars in Praesepe, such as EPIC 211852399 and EPIC 211875458, the periods were obtained by us based on K2 data (see Appendix~\ref{sec:comments}). The colour contrast in Fig.~\ref{fig:amp_rot} denotes the Rossby number, Ro, the normalized rotation period by the convective turnover time (the convective turnover time was estimated by using the correlation reported by \citet{wrig2011}, see Paper I/II for details). It is known that the open cluster members mainly locate in two rotation sequences in the rotation-colour diagram \citep{barn2003}, i.e. I (Interface) and C (Convective) sequence, corresponding to slow and fast rotators, respectively, as shown by top panels in Fig.~\ref{fig:amp_rot}. The figure shows a tendency that GK-type C sequence stars in Pleiades have larger variation amplitude of EW$_{\text{H}\alpha}$ compared to I sequence stars of the same temperature. We found no evident difference between different rotators among M-type Pleiades stars, partially due to most of them are very fast rotators that reside in activity-saturation regime \citep[when Ro $\lesssim$ 0.1, e.g.,][]{fang2018}. For Praesepe/Hyades sample, almost GK-type stars rotate slow (e.g. Ro $\sim$ 0.5), located in the I sequence (those stars with blue colours), there is no clear trend among these stars. Among M-type stars some stars are fast rotators in C sequence (red or green colours), whilst others are slow rotators (in I sequence), as shown by upper-right panel of Fig.~\ref{fig:amp_rot}. Thus, there exist a tendency that faster rotators have larger variation amplitude of EW$_{\text{H}\alpha}$. Unlike EW$_{\text{H}\alpha}$, TiO2n shows no clear difference among all sample stars with different rotation periods, probably due to its small intrinsic variation amplitude which could be dominated by observational noise.

\begin{figure*}
\centering
\includegraphics[width=\columnwidth]{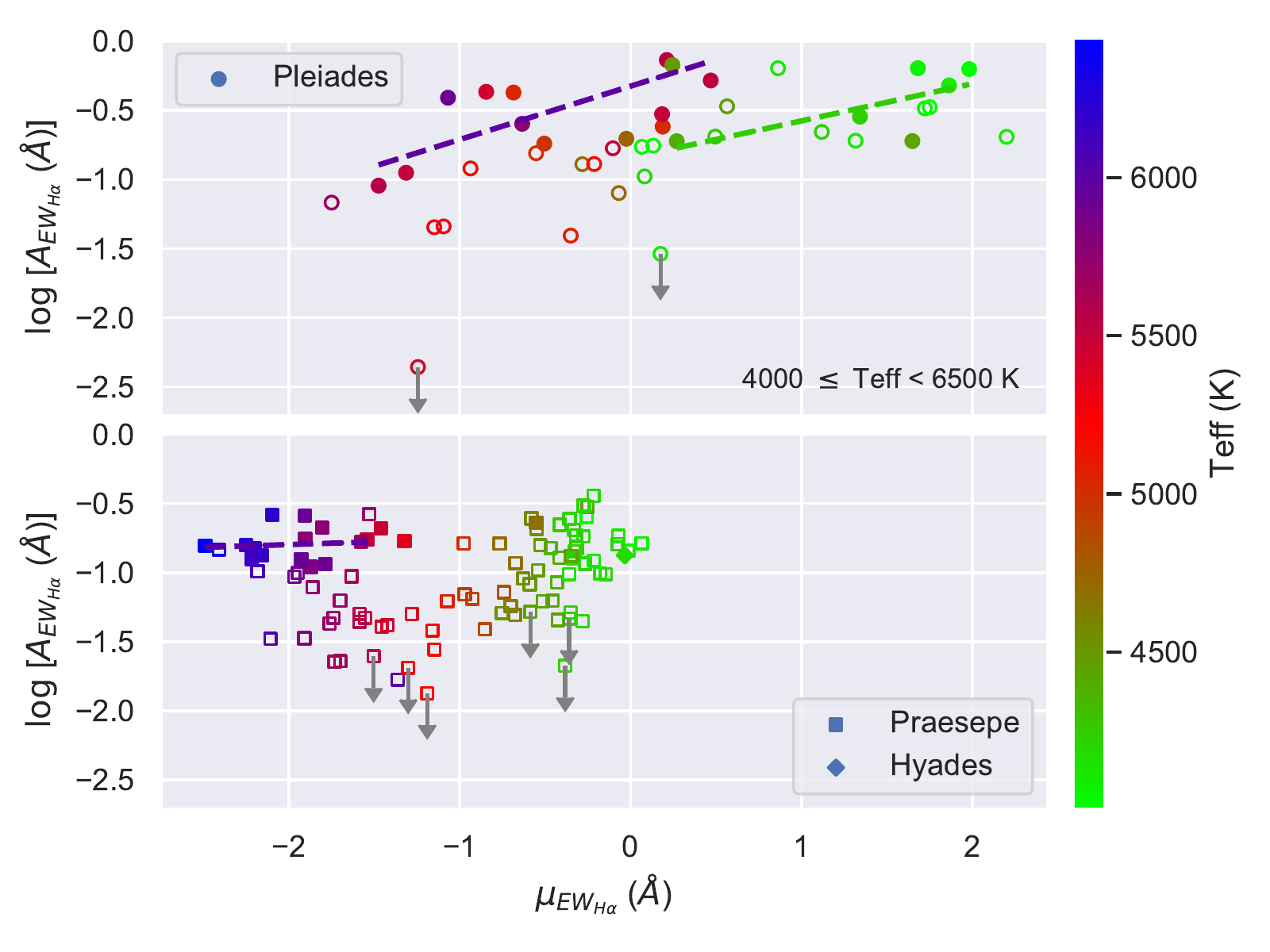}
\includegraphics[width=\columnwidth]{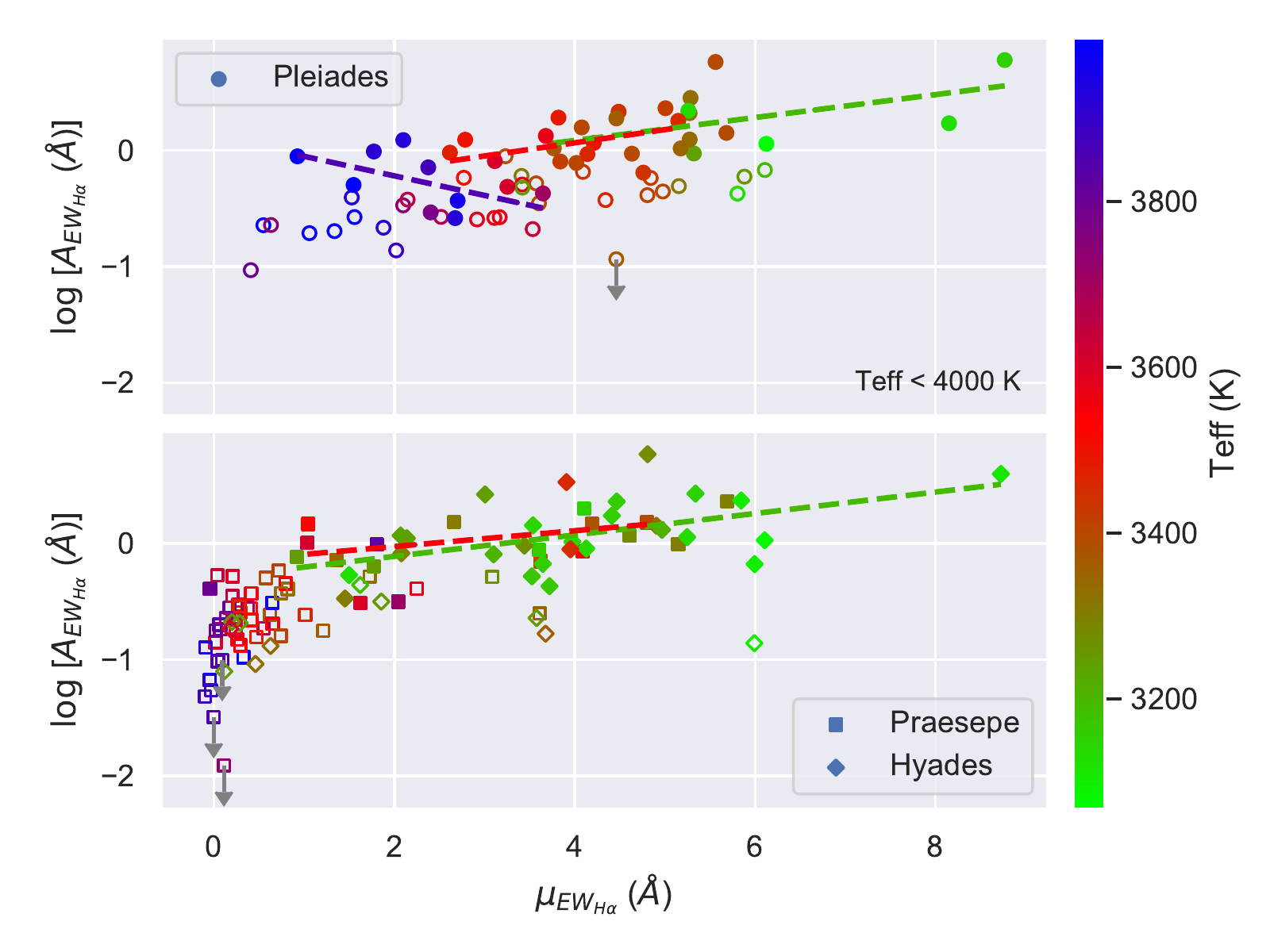}
\includegraphics[width=\columnwidth]{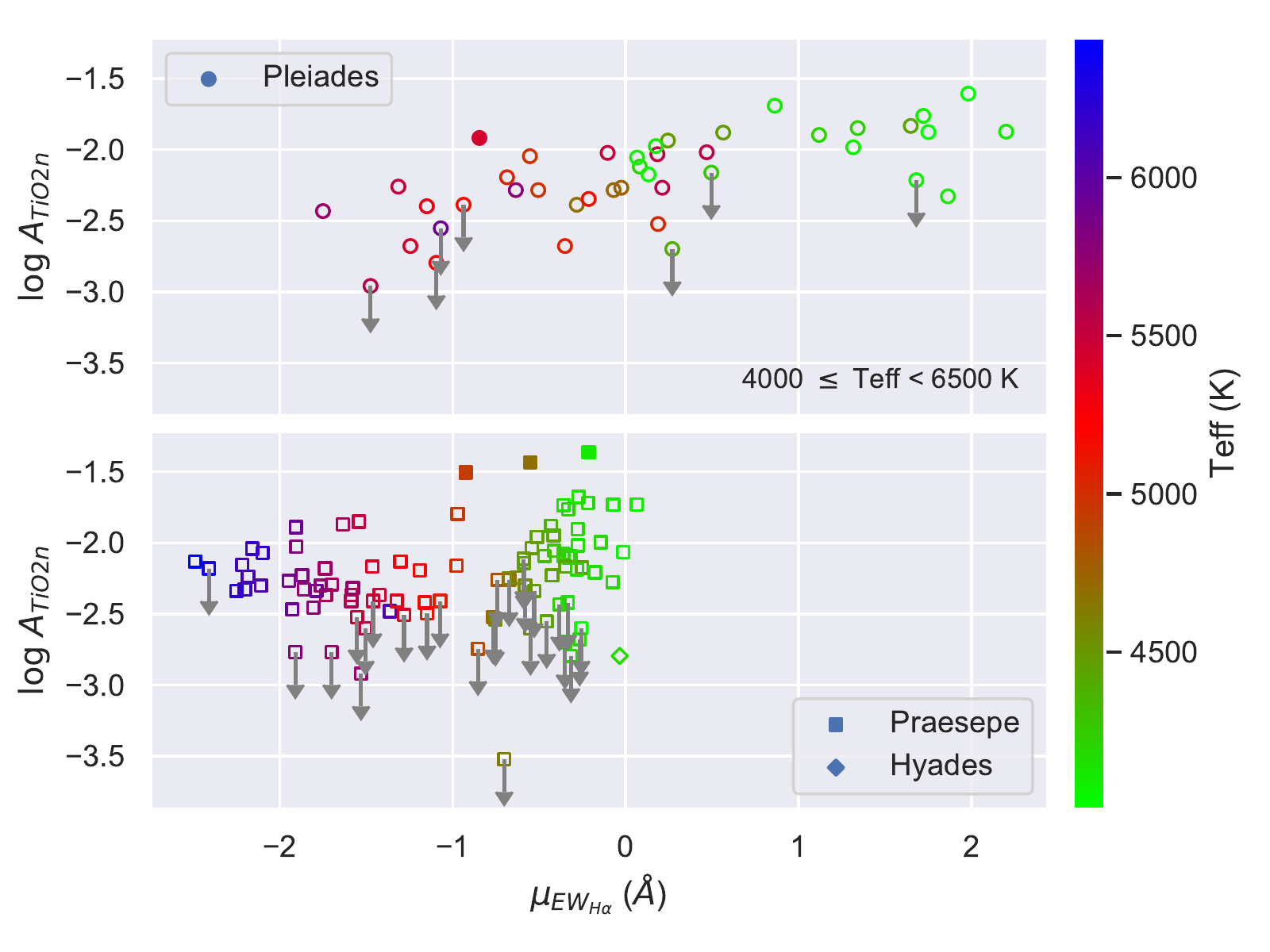}
\includegraphics[width=\columnwidth]{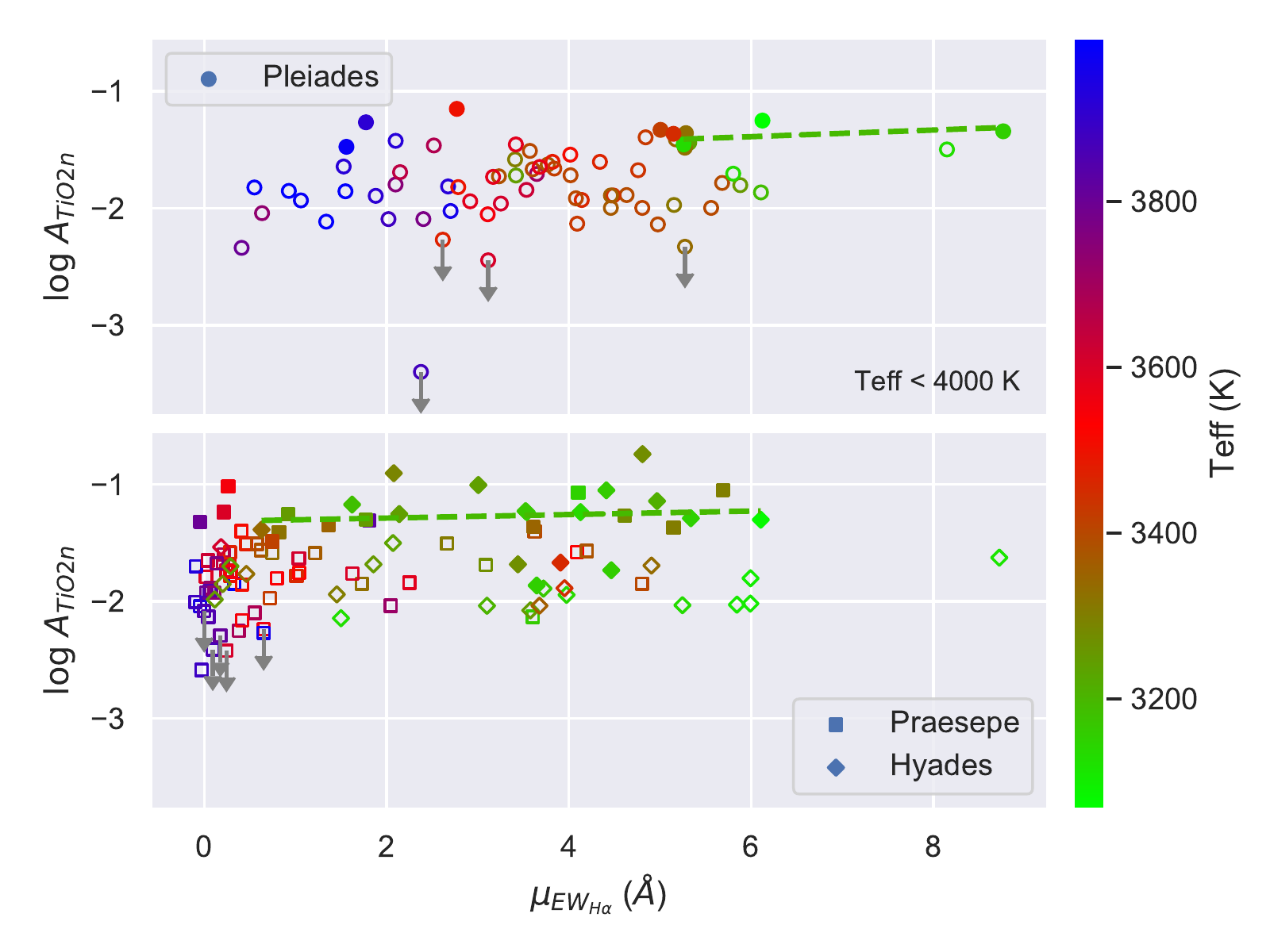}
\caption{The amplitudes of EW$_{\text{H}\alpha}$ and TiO2n as a function of mean EW$_{\text{H}\alpha}$ ($\mu_{\text{EW}_{\text{H}\alpha}}$) for each sample star. Note the colour contrast denotes the temperature (left panels: Teff $\geq 4000$ K, right panels: Teff $< 4000$ K). Open symbols represent stars with $R_{\text{amp}} \leq 6$ (downward pointing arrows with grey colours indicate the star has a ratio $R_{\text{amp}}<1$), and filled symbols are stars with $R_{\text{amp}}>6$. The colour-coded, dashed lines represent linear fitting to corresponding temperature-binned samples for stars with $R_{\text{amp}}>6$. Note the fit is not shown if enough stars are not available in a temperature bin. }
\label{fig:mean_amp}
\end{figure*}

In fact, the age/rotation effects on variation level discussed above equivalently means that more active stars have intense variations. Indeed, previous studies have shown that both chromospheric Ca~{\sc ii} HK and photometric variability among Sun-like stars are related to average chromospheric activity by power laws in both short-term \citep{radi1998} and long-term (year-to-year) time-scales \citep[e.g.][]{lock2007,radi2018}. Fig.~\ref{fig:mean_amp} shows the variation of EW$_{\text{H}\alpha}$ as a function of the mean value for each sample star, where the colour represents the temperature. We noticed a trend that stronger H$\alpha$ emission corresponds to larger variation amplitude for M-type sample stars. We use a Pearson's analysis to test the statistical significance of the linear correlation between $\log A_{\text{EW}_{\text{H}\alpha}}$ and $\mu_{\text{EW}_{\text{H}\alpha}}$ for stars having $R_{\text{amp}}>6$. Table~\ref{tab:r_tio2_ewha} show the results of Pearson's $r$ test, which were conducted within respective temperature group to remove the dependency of the equivalent width on the nearby continuum flux. Note the result is not shown if enough stars having $R_{\text{amp}}>6$ are not available in a temperature bin. Only mid/late M-type stars seem to have clear correlation, as illustrated by dashed lines in upper panels of Fig.~\ref{fig:mean_amp} and the $r$ values in Table~\ref{tab:r_tio2_ewha}. FGK-type Pleiades stars show some weak evidences of correlation between $\log A_{\text{EW}_{\text{H}\alpha}}$ and $\mu_{\text{EW}_{\text{H}\alpha}}$. The lower plots of Fig.~\ref{fig:mean_amp} show the variation amplitude of TiO2n against the mean value of EW$_{\text{H}\alpha}$. The observed variation in TiO2n is dominated by observational noise for most stars, thus it is hard to say whether there exists a relation between them. 

We noticed that most M-type stars in these three open clusters have very similar chromospheric emission levels, locating in a rotation-activity saturation region \citep[e.g.][]{fang2018}. This will partially reduce the differences in activity variation level between these stars in Pleiades and Praesepe/Hyades sample. In other words, in these active rapidly rotating stars, chromospheric emission reaches a maximum saturated level, and thus such indicators more likely ceases to be an effective diagnostic of magnetic activity variability.  
\begin{table*}
\centering
\caption{Pearson's $r$ tests of $\log A_{\text{EW}_{\text{H}\alpha}}$ against $\mu_{\text{EW}_{\text{H}\alpha}}$ and $\log A_{\text{TiO2n}}$ against $\mu_{\text{EW}_{\text{H}\alpha}}$ for stars with $R_{\text{amp}}>6$}
\label{tab:r_tio2_ewha}
\begin{tabular}{lcccccccccccccc}
   \hline
             & \multicolumn{3}{c}{$\log A_{\text{EW}_{\text{H}\alpha}}$ against $\mu_{\text{EW}_{\text{H}\alpha}}$} 
             & \multicolumn{3}{c}{$\log A_{\text{TiO2n}}$ against $\mu_{\text{EW}_{\text{H}\alpha}}$}  \\
Teff (K)     & $r$ ($p$-value)     & $r$ ($p$-value)     & $r$ ($p$-value)       & $r$ ($p$-value)     & $r$ ($p$-value)     & $r$ ($p$-value)      \\ 
             & Pleiades            & Praesepe/Hyades     & All               & Pleiades            & Praesepe/Hyades     & All             \\
   \hline
$3000-3400$  &  $+0.49~(7.4 \times 10^{-2})$   &  $+0.56~(3.7 \times 10^{-4})$   &  $+0.60~(4.7 \times 10^{-6})$   &  $+0.45~(3.7 \times 10^{-1})$  &  $+0.10~(6.4 \times 10^{-1})$  &  $-0.02~(9.0 \times 10^{-1})$   \\
$3400-3700$  &  $+0.43~(1.1 \times 10^{-1})$   &  $+0.35~(4.4 \times 10^{-1})$   &  $+0.36~(1.0 \times 10^{-1})$   &  --  &   -- &  $-0.41~(3.6 \times 10^{-1})$   \\
$3700-4000$  &  $-0.55~(1.3 \times 10^{-1})$   &  --                              &  $-0.26~(4.2 \times 10^{-1})$   &  --  &  --  & --   \\
$4000-4500$  &  $+0.68~(1.3 \times 10^{-1})$   &  --                              &  $+0.80~(2.9 \times 10^{-2})$   &  --  &  --  &  --  \\
$4500-5000$  &  --                              &  --                              &  $+0.75~(2.5 \times 10^{-1})$   &  --  &  --  &  --   \\
$5000-5500$  &  --                              &  --                              &  $+0.30~(5.7 \times 10^{-1})$   &  --  &  --  &  --   \\
$5500-6500$  &  $+0.86~(2.9 \times 10^{-2})$   &  $+0.09~(7.6 \times 10^{-1})$   &  $+0.73~(2.6 \times 10^{-4})$   &  --  &  --  &  --   \\   
   \hline   
\end{tabular}
\end{table*}

\subsection{Variation correlation between indicators} \label{sec:corr}
\subsubsection{Correlation between spectral features} \label{sec:corr_amp}
\begin{figure*}
\centering
\includegraphics[width=\columnwidth]{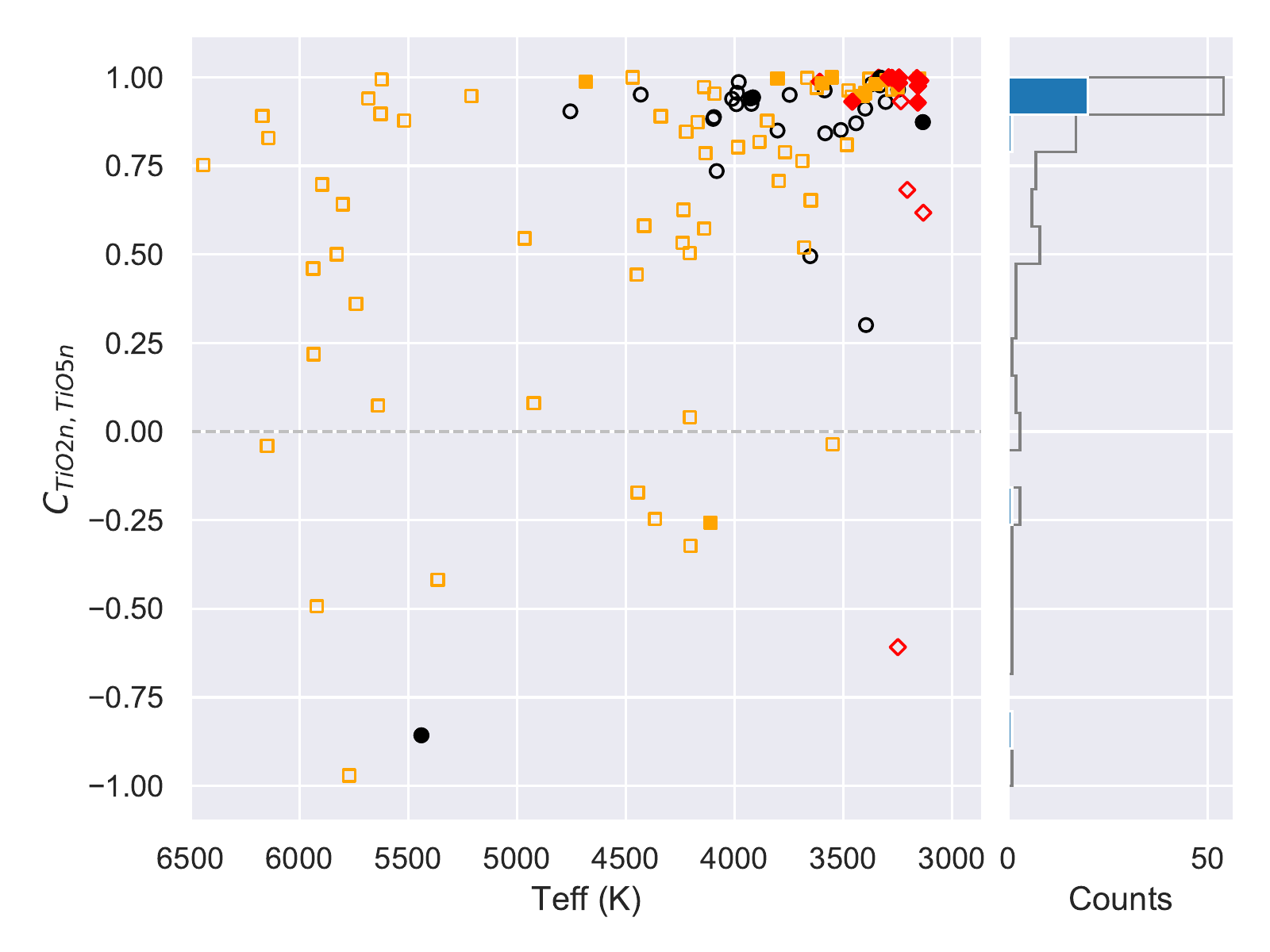}
\includegraphics[width=\columnwidth]{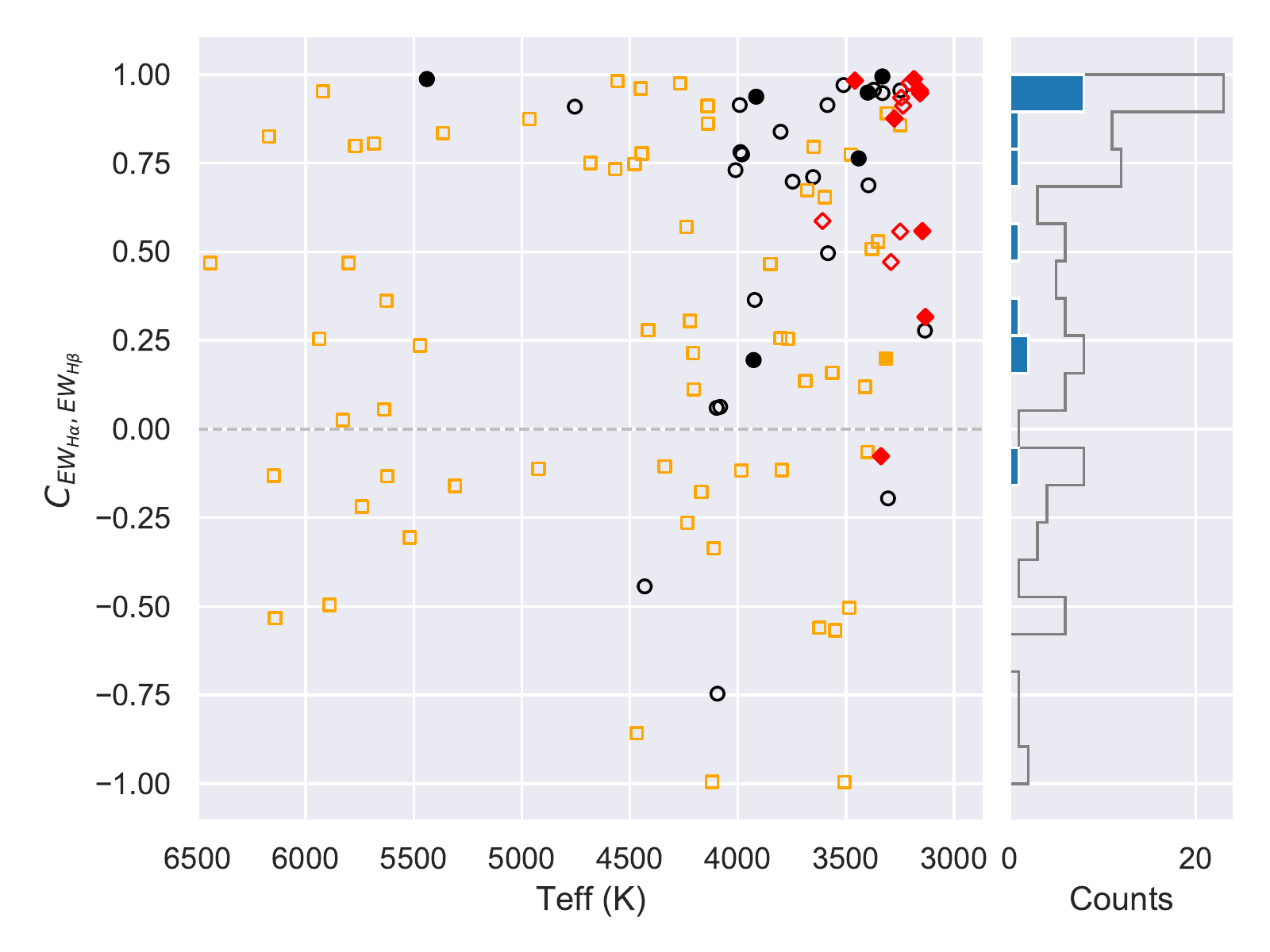}
\caption{Measured zero-normalized cross-correlations between TiO2n and TiO5n ($C_{\text{TiO2n,TiO5n}}$), and between Hydrogen Balmer emissions ($C_{\text{EW}_{\text{H}\alpha},\text{EW}_{\text{H}\beta}}$) as a function of temperature for stars with at least four epochs. Open symbols (circle: Pleiades; square: Praesepe; diamond: Hyades) represent stars with $1 < R_{\text{amp}} \leq 6$, and filled symbols are stars with $R_{\text{amp}}>6$. The right panel of each plot shows corresponding $C_{x,y}$ distribution among entire sample stars (grey-colour histogram for all stars having ratios $R_{\text{amp}}>1$, blue-colour histogram for stars having values of $R_{\text{amp}}>6$).}
\label{fig:cross0}
\end{figure*}

\begin{figure*}
\centering
\includegraphics[width=\columnwidth]{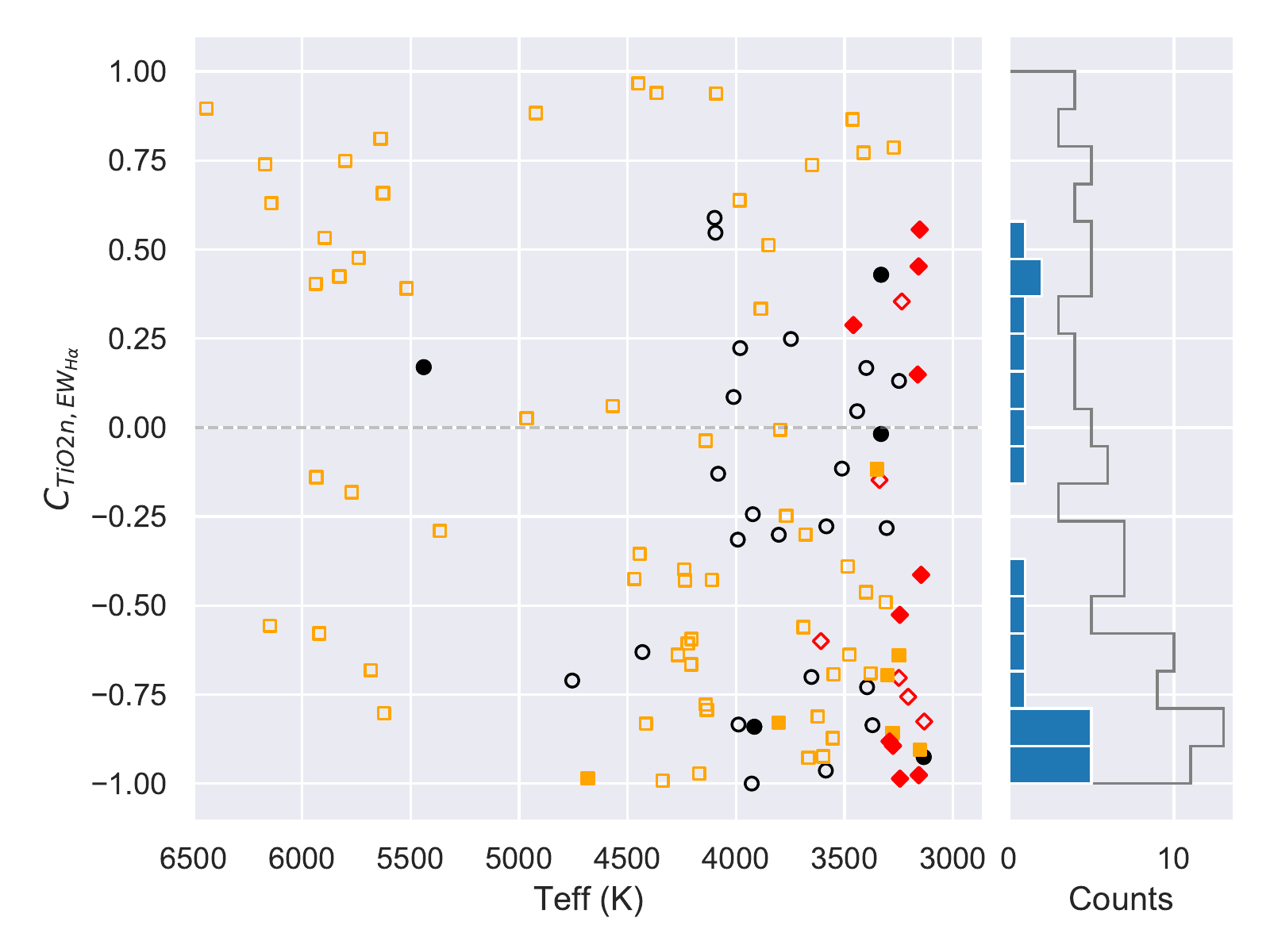}\\
\includegraphics[width=\columnwidth]{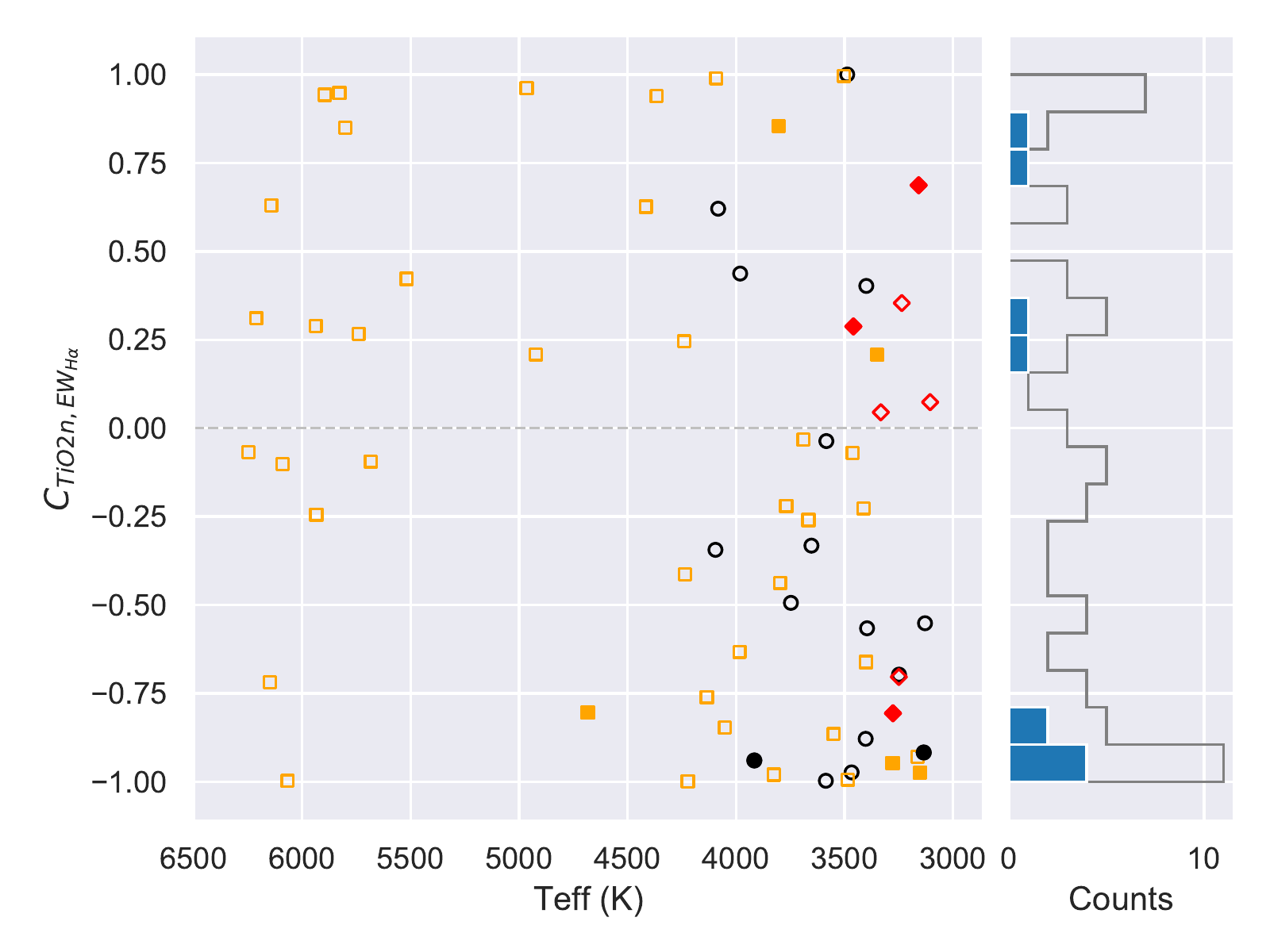}
\includegraphics[width=\columnwidth]{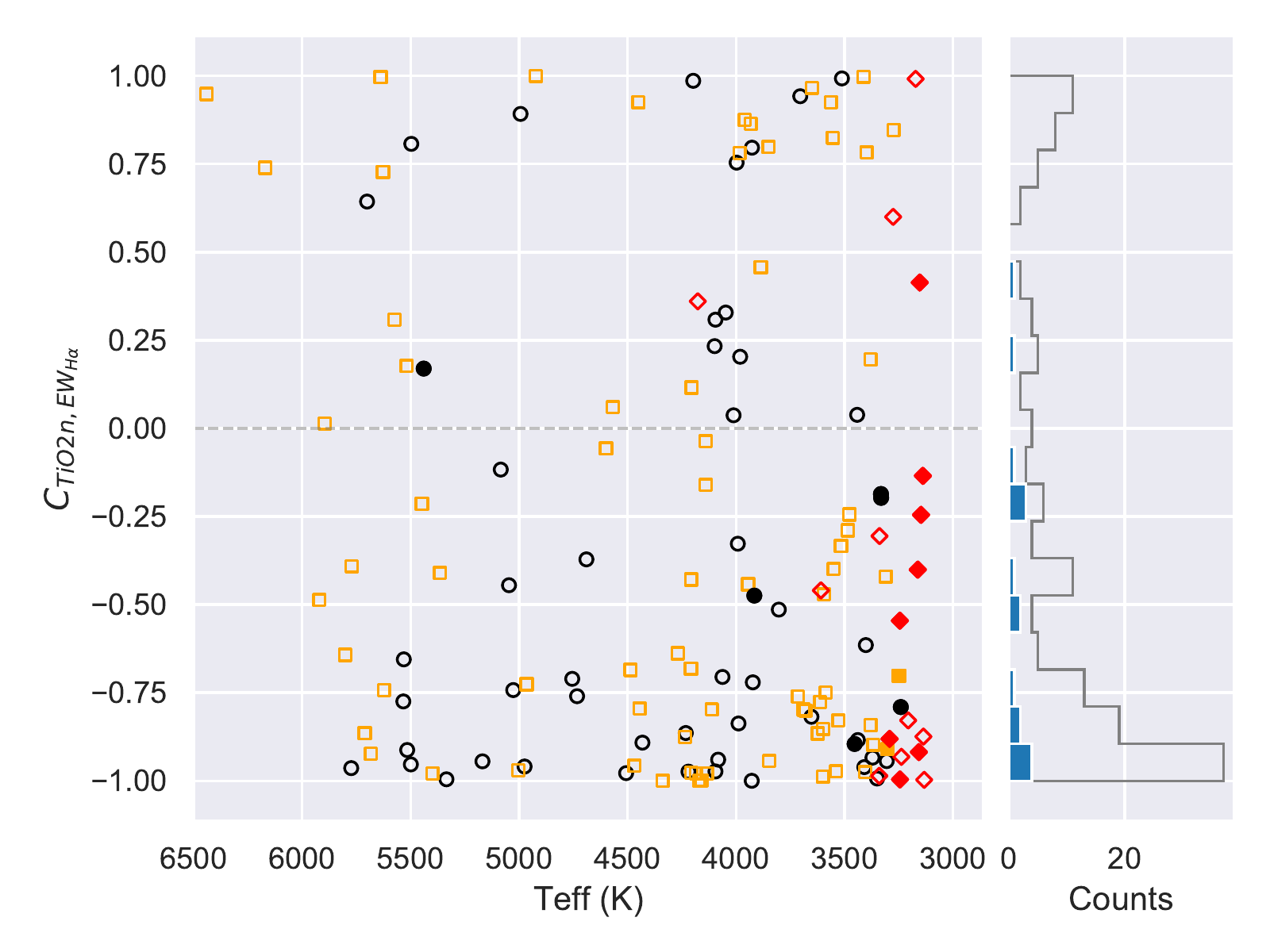}
\caption{Same as Fig.~\ref{fig:cross0} but for zero-normalized cross-correlations between TiO2n and EW$_{\text{H}\alpha}$ ($C_{\text{TiO2n},\text{EW}_{\text{H}\alpha}}$). Top plot: correlation for stars totally having at least four epochs. Lower left: correlation on short-term time-scales only based on observations within one season (for stars with at least three epochs within one season). Lower right: correlations on long-term time-scales for stars with at least three yearly epochs after averaging every season. In each plot, right panel shows the $C_{x,y}$ distribution among entire sample stars grey-colour histogram for all stars having ratios $R_{\text{amp}}>1$, blue-colour histogram for stars having values of $R_{\text{amp}}>6$. }
\label{fig:cross1}
\end{figure*}

To understand the connection among activity tracers, we focused on the variation pattern of one activity indicator with respect to another one, i.e., are temporal variations of two activity indicators in phase? To detect a level of correlation between temporal changes in different indicators, we measured the zero-normalized cross-correlation between two time-series (e.g., $x$, $y$ time-series with each having $N$ epochs), as follows, 
\begin{equation}
C_{x,y} =\frac{1}{N}\sum\limits_{i=1}^{N} \frac{(x_{i}-\mu_{x})}{\sigma_{x}}\frac{(y_{i}-\mu_{y})}{\sigma_{y}} ,
\label{equ:cross_define}
\end{equation}
where $\mu$ and $\sigma$ is the mean value and the population standard deviation of the time-series, respectively. Note that in the case of this definition, the resulted cross-correlation between two indicators for any star has a value of $-1\leq C_{x,y} \leq 1$, and it is in fact equivalent to the Pearson correlation coefficient $r$. Fig.~\ref{fig:cross0} shows the measured correlations between TiO2n and TiO5n, and between EW$_{\text{H}\alpha}$ and EW$_{\text{H}\beta}$, where only those with at least four epochs having $R_{\text{amp}}>1$ were displayed. It is clear that the TiO2n temporal variation tends to have positive correlation with respect to TiO5n (75 percent stars with $R_{\text{amp}}>1$ have $C_{\text{TiO2n,TiO5n}} > 0.6$, and 91 percent stars with $R_{\text{amp}}>6$ have $C_{\text{TiO2n,TiO5n}} > 0.6$). For EW$_{\text{H}\alpha}$ and EW$_{\text{H}\beta}$, it shows similar feature, e.g., about 67 percent of stars with $R_{\text{amp}}>6$ have $C_{\text{EW}_{\text{H}\alpha},\text{EW}_{\text{H}\beta}} > 0.7$. Such results confirm their intrinsic correlation between these indicators, just as expected, since they share similar physical mechanisms. The correlation between TiO2n and $\text{EW}_{\text{H}\alpha}$ however shows a different case, as shown by top plot of Fig.~\ref{fig:cross1}. It shows no correlation between this two temporal variations as the $C_{x,y}$ value randomly resides in the figure. However, over a third of them (about 40 percent) have $C_{x,y}<-0.6$ showing a negative correlation, as shown by histogram in the right side of the plot. The two-sided Kolmogorov-Smirnov statistical test shows that such a distribution is significantly different ($p$-value $\sim5.7\times10^{-4}$) from the null distribution of uniform random distribution within $-1\leq C_{x,y} \leq 1$, which means that such an aggregation toward $C_{x,y}=-1$ is real and believable. To further investigate the variation correlation between them on different time-scales, we displayed the short-term (within one season) and long-term (year-to-year) variability correlations, as shown by lower left and lower right plots in this figure, respectively. Both of them shows similar features seen in upper plot, showing that stars tend to have anti-correlations on both short-term and long-term variation. The short-term anti-correlation might be explained by the co-location of cool starspots and plages like seen on the Sun \citep{mand2017}, while the long-term anti-correlation is the reflection of spot-dominated activity because of the youth \citep[e.g.][]{radi2018}.   

\subsubsection{Correlation with K2 brightness variation}
\label{sec:corr_k2}
Due to the loss of two reaction wheels on the $Kepler$ spacecraft, the K2 mission started a project since 2014 that pointing near the ecliptic, sequentially observing fields as it orbits the Sun \citep{howe2014}. Several open clusters were monitored by K2 mission, e.g., the field including Pleiades and Hyades was observed during Campaign 4 (2015-02-07 to 2015-04-23), and Hyades field was covered again during Campaign 13 (2017-03-08 to 2017-05-27), Praesepe field was monitored during Campaign 5 (2015-04-27 to 2015-07-10), 16 (2017-12-07 to 2018-02-25), and 18 (2018-05-12 to 2018-07-02). The K2 data have been used to investigate the variable stars in the Pleiades \citep[e.g.][]{rebu2016a,rebu2016b,stau2016}, Praesepe \citep{rebu2017,doug2017} and Hyades \citep{doug2016}. 

\begin{figure*}
\centering
\includegraphics[width=\columnwidth]{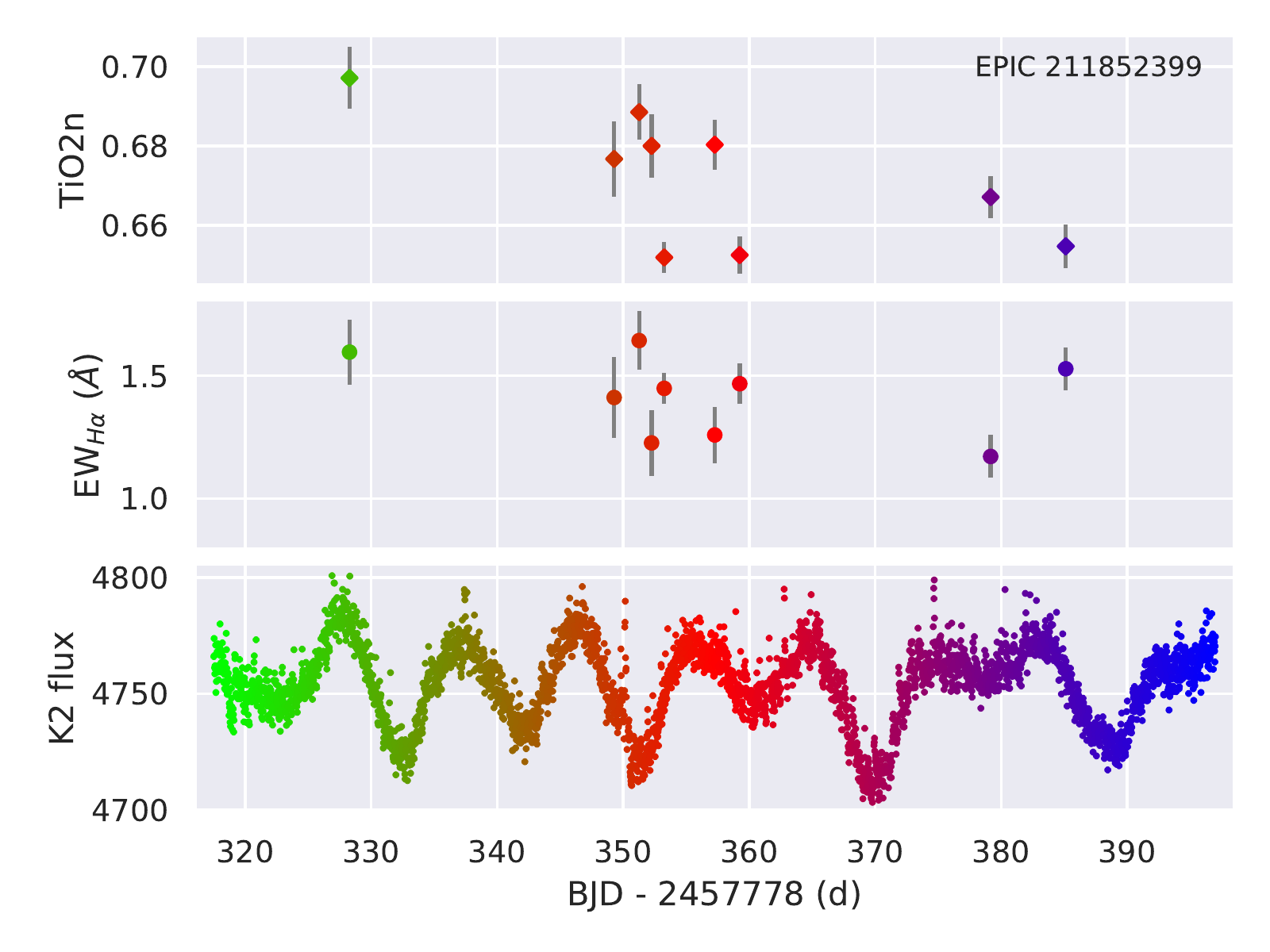}
\includegraphics[width=\columnwidth]{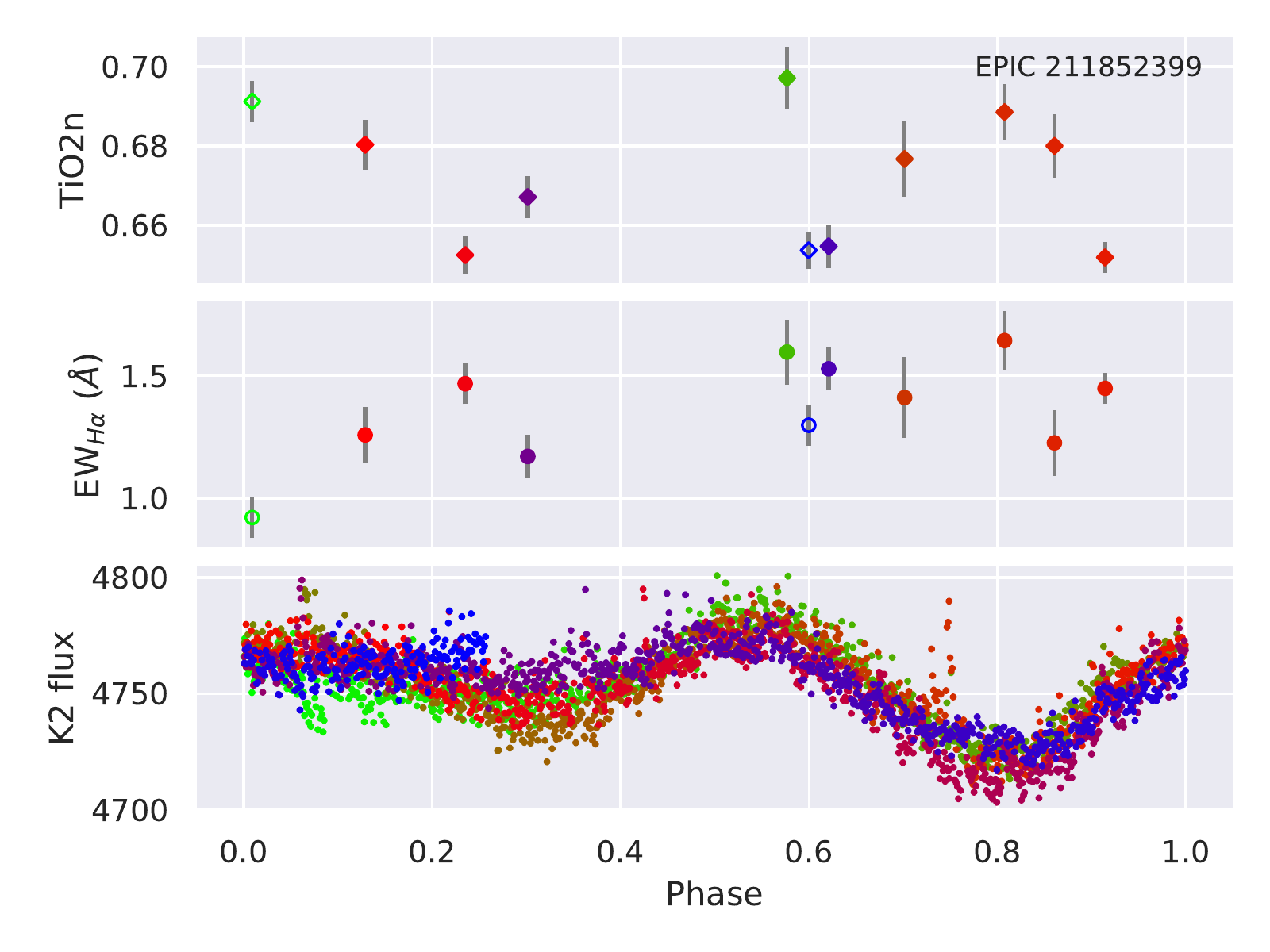}
\includegraphics[width=\columnwidth]{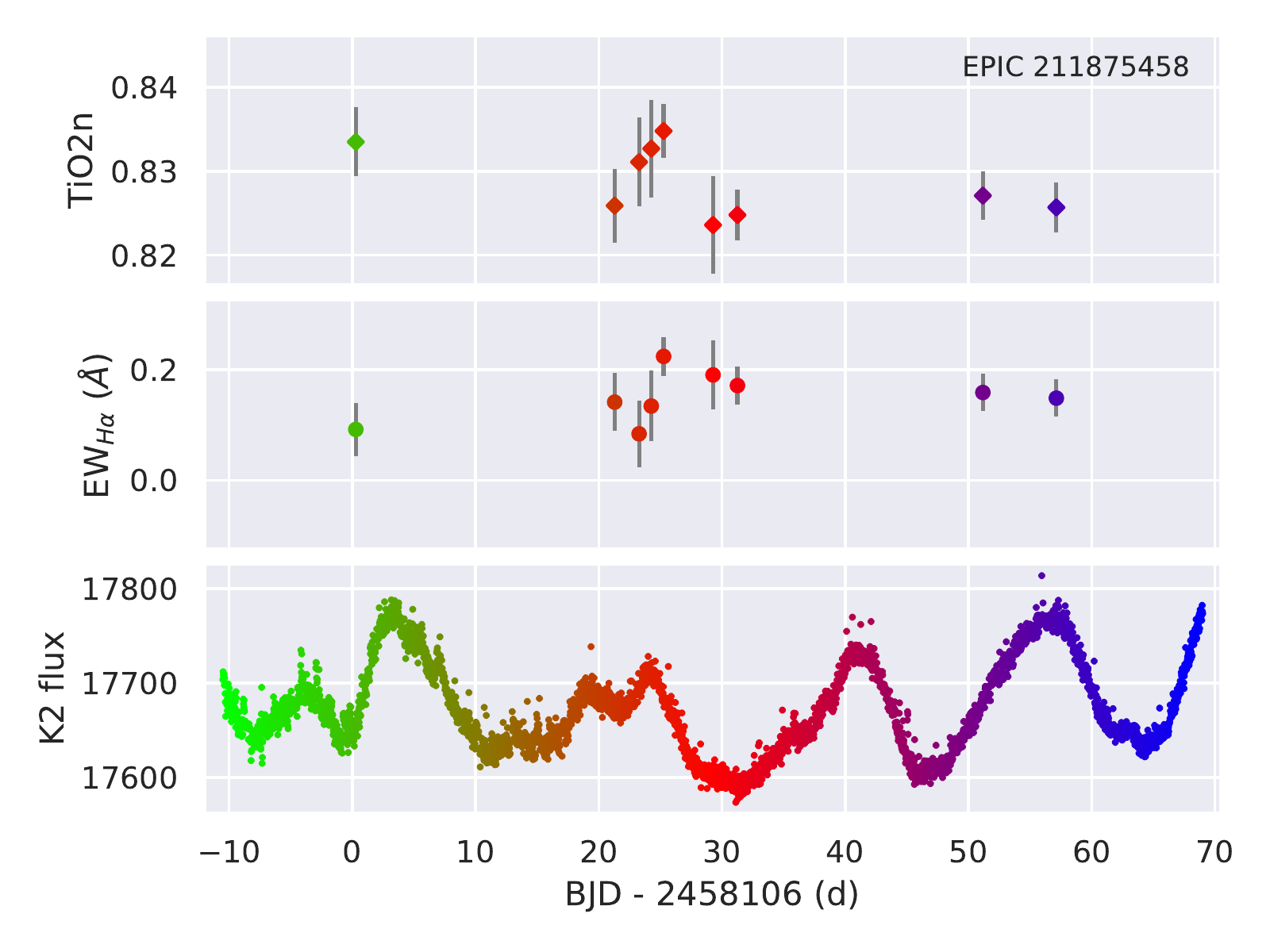}
\includegraphics[width=\columnwidth]{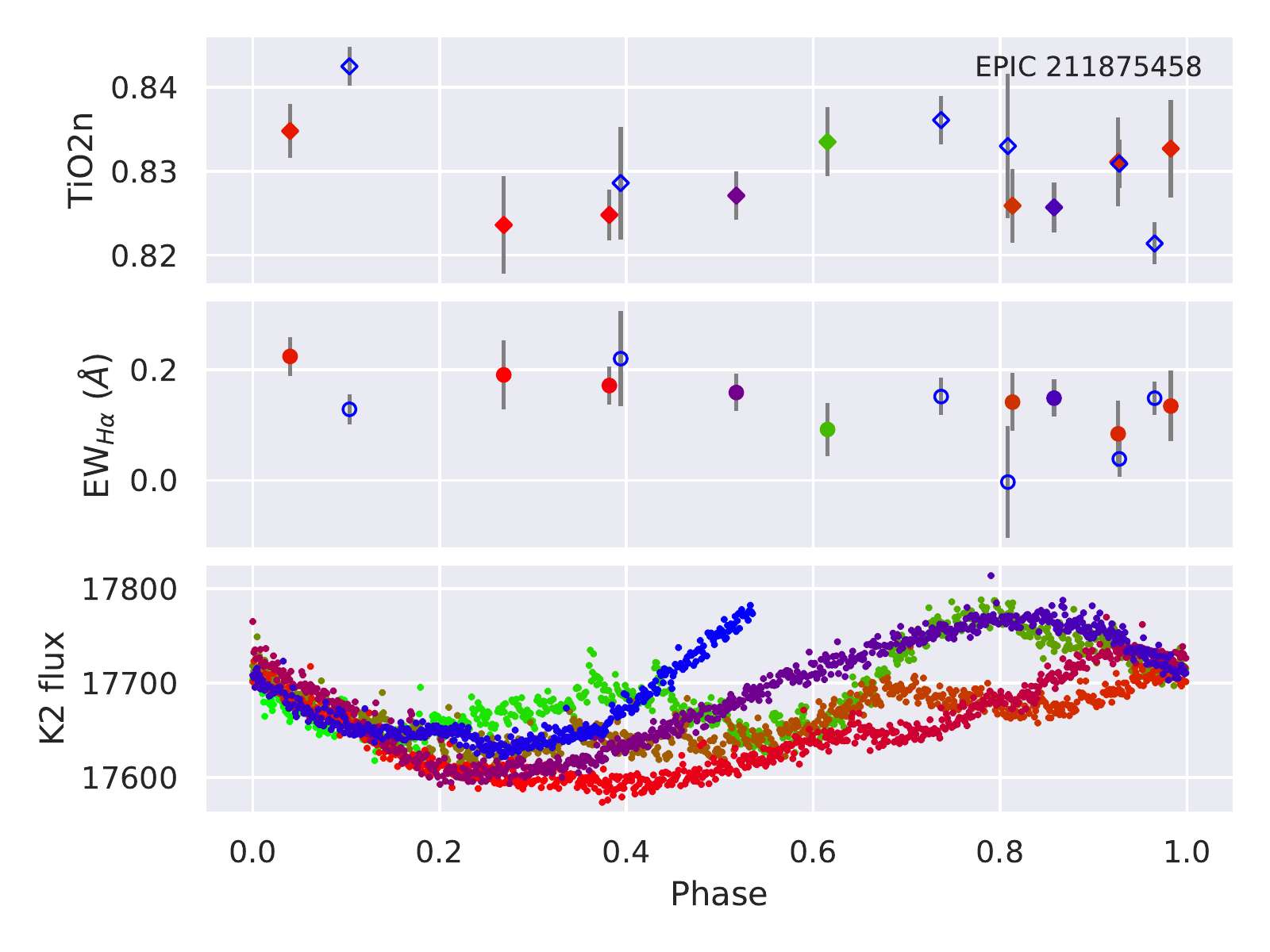}
\caption{Two examples (EPIC 211852399 and EPIC 211875458) showing temporal variabilities in LAMOST spectral features and K2 photometry (K2 PDCSAP flux in electrons per second). Left: as a function of observation time (grey bars denote errorbars); Right: as a function of rotational phase, which were folded with a period of $P=18.6750$, $P=17.5336$ days for EPIC 211852399 and EPIC 211875458, respectively. Note that they are colour-coded with observation time; open symbols denote measurements collected by LAMOST during runs outside of K2 campaigns, e.g., before (green) or after (blue) K2 campaigns.}
\label{fig:k2lc_1}
\end{figure*}

Many of our sample stars were monitored by K2 mission, however, only a small fraction of them have simultaneous LAMOST observations (see Fig.~\ref{fig:sample_hist}). In Fig.~\ref{fig:k2lc_1} and Fig.~\ref{fig:k2lc_2} shown were K2 PDCSAP LCs of five M-type stars that were frequently observed by LAMOST, wherein superposed with their LAMOST measurements. Inspection of the plots gives the impression that their LC shapes vary on time-scales of days and weeks (comparable to their rotation periods), such as EPIC 211852399 and EPIC 211875458, indicating a rapid evolution of spot configuration on the surface. However, there is no evident correlation in phase between K2 photometry and LAMOST spectroscopic features (TiO2n and $\text{EW}_{\text{H}\alpha}$) on them, as shown by Fig.~\ref{fig:k2lc_1}. In contrast, the other three stars show somewhat evident correlation in phase between LAMOST spectral features and K2 flux (though the observations are non-contemporaneous), e.g., both EPIC 211892240 and EPIC 202059188 show anti-correlation between $\text{EW}_{\text{H}\alpha}$ and K2 flux, and direct correlation between TiO2n and K2 flux (see Appendix~\ref{sec:comments} for more details), as shown by Fig.~\ref{fig:k2lc_2}.

\subsubsection{Co-location of cool starspots and bright plages?}
It is well known that plages show a latitude dependence similar to dark sunspots, and furthermore large solar plages are mostly spatially associated with dark sunspots \citep[e.g.][]{mand2017}. In this regard, we expected to detect some degree of anti-correlation variation patterns between the H$\alpha$ emission and photospheric activity tracers (such as TiO2n and K2 photometry). As discussed above, we indeed detected an aggregation toward anti-correlated variation in phase between chromospheric emission and spot coverage ($\text{EW}_{\text{H}\alpha}$ and TiO2n). Additionally, we found an anti-correlation between $\text{EW}_{\text{H}\alpha}$ and K2 flux in EPIC 211892240 and EPIC 202059188, perhaps indicative of large plages that accompany its most-spotted hemisphere. Similar association between plages and starspots has been detected in other active stars, e.g., 12 Oph and 61 Cyg A \citep{dorr1982}, EK Dra \citep{jarv2007}, AP 149 \citep{fang2010}, and the RS CVn star II Peg \citep{berd1999}. 

Evidences also show that the plages do not always spatially co-site with starspots \citep[e.g.][]{morr2018}. As seen in Fig.~\ref{fig:cross1}, many stars have measured zero-normalized cross-correlations between TiO2n and H$\alpha$ emission with $|C_{\text{TiO2n},\text{EW}_{\text{H}\alpha}}|<0.5$, which suggests the spot coverage does not change in phase with chromospheric emission strength. Furthermore, we found no correlation between H$\alpha$ emission and the K2 flux on EPIC 211852399 and EPIC 211875458. Such results may indicate that the plages are distributed axisymmetrically on the stellar surface. Alternatively, there may exist many small size plages and starspots, considering that small solar plages are found to be less spatially correlated to the sunspots, unlike the large ones \citep{mand2017}. Furthermore, as shown in Fig.~\ref{fig:k2lc_1} and Fig.~\ref{fig:k2lc_2}, these five frequently observed stars tend to always be heavy spotted during whole period of light variation (that is modulated by asymmetric starspots), and H$\alpha$ emission show similar behaviour that the evident emissions were detected in all phases, both of which suggest that there exist basal portions of spot coverage and chromospheric emission. For instance, the detected spots consist of two parts, the asymmetric part resulting in light variation, and a basal, symmetric part having no contribution to light variation (e.g. polar-like spots, and/or numerous, axis-symmetrically distributed spots). Therefore, the analysis of simultaneous photometric and spectroscopic observations and further investigation on spot configuration would be important. However, such analyses is beyond the scope of this paper, perhaps a dedicated study is needed. 
\section{Conclusion}
The multi-epoch good quality LAMOST spectra of Pleiades, Praesepe and Hyades candidate members are collected, and the temporal variations in TiO2n and $\text{EW}_{\text{H}\alpha}$ are investigated. We found the young stars show rapid variability in both spot coverage and chromospheric emission in the order of time-scales from days to months. We also found long-term variation in the time-scale of years. However, we did not detect any cyclic behaviour based on current datasets. 

The activity variability shows an age-dependence, i.e., younger stars tend to show intense variation than older stars of similar temperatures. Furthermore, we found the variation is correlated with rotation, e.g., among GK-type stars in Pleiades and M-type stars in Praesepe/Hyades, fast rotators have larger variation amplitude. In other words, more active stars tend to have more temporal variations, which is supported by our results in the framework of variation amplitude versus mean chromospheric emission. To sum up in a word, we got a picture among stars younger than 700 Myr: younger and faster rotators have higher activity levels, and also show intense variations.  

There is a tendency of anti-correlation between temporal variation of TiO2 and EW$_{\text{H}\alpha}$ in both short-term and long-term variation, suggesting a connection between photospheric and chromospheric activity. Additionally, appreciable anti-correlation in the rotational phase between H$\alpha$ emission and K2 brightness is detected in some M dwarfs, indicating spatial co-location of the plages with cool starspots. However, we noticed that cool stars do not always show such co-location features, i.e., among several M-type stars both TiO2n and $\text{EW}_{\text{H}\alpha}$ are non-correlated with K2 light variation in phase. Furthermore, spot coverage and H$\alpha$ emission were evident in all rotational phases of several M dwarfs, suggesting the presence of a basal exponent of activity that perhaps due to multiple small active regions on their outer atmospheres, e.g., numerous, axis-symmetrically distributed small-scale spots on photosphere, and multiple small flares such as nanoflares on chromosphere. 

Finally, we emphasize that this work is a preliminary attempt to understand the time variation pattern of stellar activity in young late-type stars. Considering the sparse sampling of LAMOST datasets and the measurement uncertainty, conclusions based on it must be drawn with great caution, the features shown in this work are therefore instructive more than conclusive. More observational work is required to further understand the variation pattern among these young cool stars. Fortunately, the related issues would continually benefit from more available data provided by the ongoing LAMOST spectral survey.  
\section*{Acknowledgements} 
We thank the referee for the observant comments and constructive suggestions that helped to improve the manuscript. This study is supported by the National Natural Science Foundation of China (NSFC) under grant No. 11988101, 11890694 and National Key R\&D Program of China No. 2019YFA0405502. This work is sponsored (in part) by the Chinese Academy of Sciences (CAS), through a grant to the CAS South America Center for Astronomy (CASSACA) in Santiago, Chile. Y.B.K thanks the support from NSFC grant no. 11850410437. This work has made use of LAMOST data. The Guo Shou Jing Telescope (the Large sky Area Multi-Object fiber Spectroscopic Telescope, LAMOST) is a National Major Scientific Project built by the Chinese Academy of Sciences. Funding for the project has been provided by the National Development and Reform Commission. LAMOST is operated and managed by National Astronomical Observatories, Chinese Academy of Sciences. Some of the data presented in this paper were obtained from the Mikulski Archive for Space Telescopes (MAST). This paper includes data collected by the Kepler (K2) mission. Funding for the Kepler mission is provided by the NASA Science Mission directorate. This research has made use of NASA's Astrophysics Data System (ADS) Abstract Service, and of the VizieR catalogue access tool and the cross-match service provided by CDS, Strasbourg, France. This research has also made use of TOPCAT \citep{tayl2005}, Astropy \citep{astr2013,astr2018}, Matplotlib \citep{hunt2007}, as well as NumPy \citep{walt2011,olip2015} and SciPy \citep{jone2001}.



\bibliographystyle{mnras}

\begin{thebibliography}{99}
\bibitem[\protect\citeauthoryear{Ag{\"u}eros et al.}{2011}]{ague2011} 
Ag{\"u}eros M. A., Covey K. R., Lemonias J. J., et al., 2011, \apj, 740, 110

\bibitem[\protect\citeauthoryear{Alekseev \& Kozhevnikova}{2018}]{alek2018} 
Alekseev I. Y. \& Kozhevnikova A. V., 2018, Astronomy Reports, 62, 396

\bibitem[\protect\citeauthoryear{Armstrong et al.}{2015}]{arms2015} 
Armstrong D. J., Kirk J., Lam K. W. F., et al., 2015, \aap, 579, 19

\bibitem[\protect\citeauthoryear{Astropy Collaboration et al.}{2013}]{astr2013}
Astropy Collaboration, Robitaille T. P., Tollerud E. J., Greenfield P., et al., 2013, \aap, 558, A33

\bibitem[\protect\citeauthoryear{Astropy Collaboration et al.}{2018}]{astr2018} 
Astropy Collaboration, Price-Whelan A. M., SipH ocz B. M., Günther H. M., et al., 2018, \aj, 156, 123

\bibitem[\protect\citeauthoryear{Baliunas et al.}{1995}]{bali1995} 
Baliunas S. L., Donahue R. A., Soon W. H., et al., 1995, \apj, 438, 269

\bibitem[\protect\citeauthoryear{Baliunas et al.}{1998}]{bali1998} 
Baliunas S. L., Donahue R. A., Soon W. \& Henry G. W., 1998, Cool Stars, Stellar Systems, and the Sun, 154, 153

\bibitem[\protect\citeauthoryear{Barnes}{2003}]{barn2003} 
Barnes S. A., 2003, \apj, 586, 464

\bibitem[\protect\citeauthoryear{Berdyugina et al.}{1999}]{berd1999} 
Berdyugina S. V., Ilyin I. \& Tuominen, I., 1999, \aap, 349, 863

\bibitem[\protect\citeauthoryear{B\"{o}hm-Vitense}{2007}]{bohm2007} 
B\"{o}hm-Vitense, E., 2007, \apj, 657, 486

\bibitem[\protect\citeauthoryear{Bouy et al.}{2015}]{bouy2015} 
Bouy H. et al., 2015, \aap, 577, A148

\bibitem[\protect\citeauthoryear{Brandenburg et al.}{2017}]{bran2017} 
Brandenburg A., Mathur S. \& Metcalfe T. S., 2017, \apj, 845, 79

\bibitem[\protect\citeauthoryear{Brandt \& Huang}{2015}]{bran2015}
Brandt T. D. \& Huang C. X., 2015, \apj, 807, 24

\bibitem[\protect\citeauthoryear{Carrera \& Pancino}{2011}]{carr2011}
Carrera R. \& Pancino E., 2011, \aap, 535, A30

\bibitem[\protect\citeauthoryear{Chatterjee et al.}{2016}]{chat2016}
Chatterjee S., Banerjee D. \& Ravindra B., 2016, \apj, 827, 87

\bibitem[\protect\citeauthoryear{Covey et al.}{2016}]{cove2016} 
Covey K. R., Agüeros M. A., Law N. M., et al., 2016, \apj, 822, 81

\bibitem[\protect\citeauthoryear{Crespo-Chac{\'o}n et al.}{2007}]{cres2007}
Crespo-Chac{\'o}n I., Micela G., Reale F., Caramazza M., López-Santiago J., \& Pillitteri I., 2007, \aap, 471, 929

\bibitem[\protect\citeauthoryear{Cui et al.}{2012}]{cui+2012} 
Cui X.-Q et al., 2012, Res. Astron. Astrophys., 12, 1197

\bibitem[\protect\citeauthoryear{Davenport et al.}{2014}]{dave2014} 
Davenport J. R. A., Hawley S. L., Hebb L., et al., 2014, \apj, 797, 122

\bibitem[\protect\citeauthoryear{Delorme et al.}{2011}]{delo2011}
Delorme P., Collier Cameron A., Hebb L., Rostron J., Lister T. A., Norton A. J., Pollacco D. \& West R. G., 2011, \mnras, 413, 2218

\bibitem[\protect\citeauthoryear{Dorren \& Guinan}{1982}]{dorr1982}
Dorren J. D. \& Guinan E. F., 1982, \aj, 87, 1546

\bibitem[\protect\citeauthoryear{Douglas et al.}{2014}]{doug2014}
Douglas S. T., Ag{\"u}eros M. A., Covey K. R., 2014, \apj, 795, 161

\bibitem[\protect\citeauthoryear{Douglas et al.}{2016}]{doug2016}
Douglas S. T., Ag{\"u}eros M. A., Covey K. R., Cargile P. A., Barclay T., Cody A., Howell S. B. \& Kopytova T., 2016, \apj, 822, 47

\bibitem[\protect\citeauthoryear{Douglas et al.}{2017}]{doug2017}
Douglas S. T., Ag{\"u}eros M. A., Covey K. R., \& Kraus A., 2017, \apj, 842, 83

\bibitem[\protect\citeauthoryear{Durney et al.}{1981}]{durn1981}
Durney B. R., Mihalas D. \& Robinson R. D., 1981, \pasp, 1981, 93, 537

\bibitem[\protect\citeauthoryear{Egeland et al.}{2015}]{egel2015}
Egeland R., Metcalfe T. S., Hall J. C., \& Henry G. W., 2015, \apj, 812, 12

\bibitem[\protect\citeauthoryear{Fang et al.}{2010}]{fang2010}
Fang X.-S., Gu S.-H., Cheung S.-L., Hui H.-K., Kwok C.-T., Leung K.-C., 2010, Res. Astron. Astrophys., 10, 253

\bibitem[\protect\citeauthoryear{Fang et al.}{2016}]{fang2016}
Fang X.-S., Zhao G., Zhao J.-K., Chen Y.-Q., Bharat Kumar Y., 2016, \mnras, 463, 2494

\bibitem[\protect\citeauthoryear{Fang et al.}{2018}]{fang2018}
Fang X.-S., Zhao G., Zhao, J.-K., Bharat Kumar, Y., 2018, \mnras, 476, 908

\bibitem[\protect\citeauthoryear{Gaia Collaboration et al.}{2018}]{gaia2018}
Gaia Collaboration, Brown A. G. A., Vallenari A., Prusti T., et al., 2018, \aap, 616, A1


\bibitem[\protect\citeauthoryear{Giles et al.}{2017}]{gile2017} 
Giles H. A. C., Collier Cameron A. \& Haywood, R. D., 2017, \mnras, 472, 1618

\bibitem[\protect\citeauthoryear{Gray et al.}{2015}]{gray2015} 
Gray R. O., Saken, J. M., Corbally, C. J., et al., 2015, \aj, 150, 203

\bibitem[\protect\citeauthoryear{Hall et al.}{2007}]{hall2007} 
Hall J. C., Lockwood G. W. \& Skiff B. A., 2007, \aj, 133, 862

\bibitem[\protect\citeauthoryear{Hartman et al.}{2010}]{hart2010} 
Hartman J. D., Bakos G. {\'A}., Kov{\'a}cs G., Noyes R. W., 2010, \mnras, 408, 475

\bibitem[\protect\citeauthoryear{Hartman et al.}{2011}]{hart2011}
Hartman J. D., Bakos G. {\'A}., Noyes R. W., SipH ocz B., Kov{\'a}cs G., Mazeh T., Shporer A. \& P{\'a}l A., 2011, \aj, 141, 166

\bibitem[\protect\citeauthoryear{Hawley et al.}{2014}]{hawl2014}
Hawley S. L., Davenport J. R. A., Kowalski A. F., Wisniewski J. P., Hebb L., Deitrick R., \& Hilton E. J., 2014, \apj, 797, 121

\bibitem[\protect\citeauthoryear{Howell et al.}{2014}]{howe2014}
Howell S. B., Sobeck C., Haas, M., et al., 2014, \pasp, 126, 398

\bibitem[\protect\citeauthoryear{Huenemoerder \& Ramsey}{1987}]{huen1987}
Huenemoerder D. P., \& Ramsey L. W., 1987, ApJ, 319, 392

\bibitem[\protect\citeauthoryear{Hunter}{2007}]{hunt2007}
Hunter J. D., 2007, Computing in Science and Engineering, 9, 90

\bibitem[\protect\citeauthoryear{Iba{\~n}ez Bustos et al.}{2019}]{iban2019}
Iba{\~n}ez Bustos R. V., Buccino A. P., Flores M., Martinez C. I., Maizel D., Messina S., \& Mauas P. J. D., 2019, \mnras, 483, 1159

\bibitem[\protect\citeauthoryear{Janson et al.}{2014}]{jans2014}
Janson M., Bergfors C., Brandner W., Kudryavtseva N., Hormuth F., Hippler S. \& Henning T., 2014, \apj, 789, 102

\bibitem[\protect\citeauthoryear{J{\"a}rvinen et al.}{2007}]{jarv2007}
J{\"a}rvinen S. P., Berdyugina S. V., Korhonen H., Ilyin I., Tuominen I., 2007, \aap, 472, 887

\bibitem[\protect\citeauthoryear{J{\"a}rvinen et al.}{2008}]{jarv2008}
J{\"a}rvinen S. P., Korhonen H., Berdyugina S. V., Ilyin I., Strassmeier K. G., Weber M., Savanov I., \& Tuominen I., 2008, \aap, 488, 1047

\bibitem[\protect\citeauthoryear{Jones et al.}{2001}]{jone2001}
Jones E., Oliphant T., Peterson P., et al., 2001, SciPy: Open source scientific tools for Python

\bibitem[\protect\citeauthoryear{Kov{\'a}cs et al.}{2014}]{kova2014}
Kov{\'a}cs G., Hartman J. D., Bakos G. {\'A}., et al., 2014, \mnras, 442, 2081

\bibitem[\protect\citeauthoryear{Lindegren et al.}{2018}]{lind2018}
Lindegren L., Hern{\'a}ndez J., Bombrun A., et al., 2018, \aap, 616, A2

\bibitem[\protect\citeauthoryear{Lockwood et al.}{2007}]{lock2007}
Lockwood G. W., Skiff B. A., Henry G. W., Henry S., Radick R. R., Baliunas S. L., Donahue R. A. \& Soon W., 2007, \apjs, 171, 260

\bibitem[\protect\citeauthoryear{Luo et al.}{2015}]{luo+2015} 
Luo A.-L. et al., 2015, Res. Astron. Astrophys., 15, 1095

\bibitem[\protect\citeauthoryear{Mamajek \& Hillenbrand}{2008}]{mama2008} 
Mamajek E. E. \& Hillenbrand L. A., 2008, \apj, 687, 1264

\bibitem[\protect\citeauthoryear{Mandal et al.}{2017}]{mand2017} 
Mandal S., Chatterjee S., Banerjee D., 2017, \apj, 835, 158

\bibitem[\protect\citeauthoryear{Metcalfe et al.}{2013}]{metc2013} 
Metcalfe T. S., Buccino A. P., Brown, B. P., et al., 2013, \apjl, 763, L26

\bibitem[\protect\citeauthoryear{Morris et al.}{2018}]{morr2018} 
Morris B. M., Curtis J. L., Douglas S. T., Hawley S. L., Ag{\"u}eros M. A., Bobra M. G. \& Agol E., 2018, \aj, 156, 203

\bibitem[\protect\citeauthoryear{Neff et al.}{1995}]{neff1995} 
Neff J. E., O'Neal D., \& Saar S. H., 1995, ApJ, 452, 879

\bibitem[\protect\citeauthoryear{Newton et al.}{2017}]{newt2017} 
Newton E. R., Irwin J., Charbonneau D., Berlind P., Calkins M. L. \& Mink J., 2017, \apj, 834, 85

\bibitem[\protect\citeauthoryear{O'Neal et al.}{1998}]{onea1998} 
O'Neal D., Neff J. E., \& Saar S. H., 1998, \apj, 507, 919

\bibitem[\protect\citeauthoryear{O'Neal et al.}{2004}]{onea2004} 
O'Neal D., Neff J. E., Saar S. H., \& Cuntz M., 2004, AJ, 128, 1802

\bibitem[\protect\citeauthoryear{Oliphant}{2015}]{olip2015} 
Oliphant T. E., 2015, Guide to NumPy (2nd ed.; New York: Continuum Press)

\bibitem[\protect\citeauthoryear{Parker}{1971}]{park1971} 
Parker E. N., 1971, \apj, 165, 139

\bibitem[\protect\citeauthoryear{Pizzolato et al.}{2003}]{pizz2003} 
Pizzolato N., Maggio A., Micela G., Sciortino S. \& Ventura P., 2003, \aap, 397, 147

\bibitem[\protect\citeauthoryear{Radick et al.}{1998}]{radi1998} 
Radick R. R., Lockwood G. W., Skiff B. A. \& Baliunas S. L., 1998, \apjs, 118, 239

\bibitem[\protect\citeauthoryear{Radick et al.}{2018}]{radi2018} 
Radick R. R., Lockwood G. W., Henry G. W., Hall J. C., \& Pevtsov A. A., 2018, \apj, 855, 75

\bibitem[\protect\citeauthoryear{Ramsey \& Nations}{1980}]{rams1980} 
Ramsey L. W., \& Nations H. L., 1980, ApJ, 239, 121

\bibitem[\protect\citeauthoryear{Rebull et al.}{2016a}]{rebu2016a} 
Rebull L. M., Stauffer J. R., Bouvier J., et al., 2016a, \aj, 152, 113

\bibitem[\protect\citeauthoryear{Rebull et al.}{2016b}]{rebu2016b} 
Rebull L. M., Stauffer J. R., Bouvier J., et al., 2016b, \aj, 152, 114

\bibitem[\protect\citeauthoryear{Rebull et al.}{2017}]{rebu2017} 
Rebull L. M., Stauffer J. R., Hillenbrand L. A., et al., 2017, \apj, 839, 92

\bibitem[\protect\citeauthoryear{Saar \& Brandenburg}{1999}]{saar1999}
Saar S. H. \& Brandenburg A., 1999, \apj, 524, 295

\bibitem[\protect\citeauthoryear{Savanov et al.}{2003}]{sava2003}
Savanov I., Strassmeier K., Romanyuk I., Kudryavtsev D., 2003, Information Bulletin on Variable Stars, 5440, 1

\bibitem[\protect\citeauthoryear{Schrijver et al.}{1989}]{schr1989}
Schrijver C. J., Cot{\'e} J., Zwaan C. \& Saar S. H., 1989, \apj, 337, 964

\bibitem[\protect\citeauthoryear{Sheeley et al.}{2011}]{shee2011}
Sheeley Jr. N. R., Cooper T. J. \& Anderson J. R. L., 2011, \apj, 730, 51

\bibitem[\protect\citeauthoryear{Soderblom et al.}{1993}]{sode1993} 
Soderblom D. R., Stauffer J. R., Hudon J. D., Jones B. F., 1993, \apjs, 85, 315

\bibitem[\protect\citeauthoryear{Soderblom et al.}{2009}]{sode2009}
Soderblom D. R., Laskar T., Valenti J. A., Stauffer J. R. \& Rebull L. M., 2009, \aj, 138, 1292

\bibitem[\protect\citeauthoryear{Stauffer \& Hartmann}{1986}]{stau1986} 
Stauffer J. R. \& Hartmann L. W., 1986, \apjs, 61, 531

\bibitem[\protect\citeauthoryear{Stauffer et al.}{1997}]{stau1997}
Stauffer J. R., Balachandran S. C., Krishnamurthi A., Pinsonneault M., Terndrup D. M. \& Stern R. A, 1997, \apj, 475, 604

\bibitem[\protect\citeauthoryear{Stauffer et al.}{1998}]{stau1998}
Stauffer J. R., Schultz G. \& Kirkpatrick J. D., 1998, \apjl, 499, L199

\bibitem[\protect\citeauthoryear{Stauffer et al.}{2007}]{stau2007}
Stauffer J. R., Hartmann L. W., Fazio, G. G., et al., 2007, \apjs, 172, 663

\bibitem[\protect\citeauthoryear{Stauffer et al.}{2016}]{stau2016}
Stauffer J., Rebull L., Bouvier J., et al., 2016, \aj, 152, 115

\bibitem[\protect\citeauthoryear{Taylor}{2005}]{tayl2005}
Taylor M. B., 2005, in Shopbell P., Britton M., Ebert R., eds, ASP Conf. Ser. Vol. 347, Astronomical Data Analysis Software and Systems XIV. Astron. Soc. Pac., San Francisco, p. 29

\bibitem[\protect\citeauthoryear{Vogt}{1979}]{vogt1979} 
Vogt S. S., 1979, \pasp, 91, 616

\bibitem[\protect\citeauthoryear{Vogt}{1981}]{vogt1981} 
Vogt S. S., 1981, \apj, 247, 975

\bibitem[\protect\citeauthoryear{Van Der Walt et al.}{2011}]{walt2011} 
Van Der Walt S., Colbert S. C., Varoquaux G., 2011, Comput. Sci. Eng., 13, 22

\bibitem[\protect\citeauthoryear{Vanderplas}{2015}]{vand2015}
Vanderplas J., 2015, gatspy: General tools for Astronomical Time Series in Python, Zenodo, doi:10.5281/zenodo.14833

\bibitem[\protect\citeauthoryear{VanderPlas \& Ivezi{\'c}}{2015}]{vand+2015}
VanderPlas J. T., Ivezić Ž., 2015, \apj, 2015, 812, 18

\bibitem[\protect\citeauthoryear{Vaughan \& Preston}{1980}]{vaug+1980}
Vaughan A. H. \& Preston G. W., 1980, \pasp, 92, 385

\bibitem[\protect\citeauthoryear{Vaughan}{1980}]{vaug1980}
Vaughan A. H., 1980, \pasp, 92, 392

\bibitem[\protect\citeauthoryear{West et al.}{2004}]{west2004}
West A. A., Hawley S. L., Walkowicz L. M., et al., 2004, \aj, 128, 426

\bibitem[\protect\citeauthoryear{West et al.}{2015}]{west2015}
West A. A., Weisenburger K. L., Irwin J., Berta-Thompson Z. K., Charbonneau D., Dittmann J. \& Pineda J. S., 2015, \apj, 812, 3

\bibitem[\protect\citeauthoryear{Willamo et al.}{2019}]{will2019}
Willamo T., Hackman T., Lehtinen J. J., et al., 2019, \aap, 622, A170

\bibitem[\protect\citeauthoryear{Wilson}{1978}]{wils1978}
Wilson O. C., 1978, \apj, 226, 379

\bibitem[\protect\citeauthoryear{Wright et al.}{2011}]{wrig2011}
Wright N. J., Drake J. J., Mamajek E. E. \& Henry G. W., 2011, \apj, 743, 48

\bibitem[\protect\citeauthoryear{Wu et al.}{2011}]{wu++2011}
Wu Y., Luo A.-L., Li H.-N., et al., 2011, Res. Astron. Astrophys., 11, 924

\bibitem[\protect\citeauthoryear{Zhao et al.}{2006}]{zhao2006}
Zhao G., Chen Y.-Q., Shi J.-R., Liang Y.-C., Hou J.-L., Chen L., Zhang H.-W. \& Li A.-G., 2006, ChJAA, 6, 265

\bibitem[\protect\citeauthoryear{Zhao et al.}{2012}]{zhao2012}
Zhao G., Zhao Y.-H., Chu Y.-C., Jing Y.-P., Deng L.-C., 2012, Res. Astron. Astrophys., 12, 723

\end{thebibliography}



\appendix
\section{Comments on most frequently observed Stars}
\label{sec:comments}
Among our sample, five stars in Praesepe and Hyades with spectral types from M0 to M4 were frequently observed by LAMOST (each has over 10 epochs). Detailed comments on these stars are provided below.
\subsection{EPIC 211852399}
EPIC 211852399 ($\alpha_{\text{J2000}}=132.^\circ751672$, $\delta_{\text{J2000}}=18.^\circ051046$) is an M3-type Praesepe member with a distance of about 185 pc \citep{gaia2018}. It shows clear variability in TiO2n and H$\alpha$ emission, and in both short- and long-term. The surface of this star is covered by cool starspots with spot filling factors ranging from less than 10 to about 50 percent, and has chromospheric emission with $\log R^{'}_{\text{H}\alpha}$ from $-4.5$ to $-4.2$ dex  (see Fig.~\ref{fig:time_ewha_tio2n}). We found no clear correlation between TiO2n and $\text{EW}_{\text{H}\alpha}$ among all epochs, e.g., $C_{\text{TiO2n},\text{EW}_{\text{H}\alpha}}\sim0.2$. It was observed by K2 during the C16 campaign (from December 2017 to February 2018), showed intense light variations, e.g., the shape varies in time-scales of weeks, as shown by Fig.~\ref{fig:k2lc_1}, which probably due to rapid change of cool spots. Its K2 LCs indicate a period of $P=18.6750$ days, which was obtained by using multiterm Lomb-Scargle periodograms \citep{vand+2015} via gatspy package \citep{vand2015}. There is no clear correlation in the phase between chromospheric emission and K2 photometry (see Fig.~\ref{fig:k2lc_1}).

\subsection{EPIC 211875458}
EPIC 211875458 ($\alpha_{\text{J2000}}=133.^\circ084377$, $\delta_{\text{J2000}}=18.^\circ371503$) is M0 dwarf in Praesepe located at a distance of around 180 pc \citep{gaia2018}. Both chromospheric emission and TiO absorption of this star did not change much over the observing period from December 2017 to March 2019, as shown in Fig.~\ref{fig:time_ewha_tio2n}, and found no clear correlation between them (for all observing epochs, $C_{\text{TiO2n},\text{EW}_{\text{H}\alpha}}=-0.25$; for epochs from December 2017 to April 2018, $C_{\text{TiO2n},\text{EW}_{\text{H}\alpha}}=-0.22$). This star is covered by spots with spot filling factors around 20 $\%$ and has chromospheric H$\alpha$ emission with $\log R^{'}_{\text{H}\alpha}\sim-4.7$ dex. This star was monitored by K2 mission during the observing campaign from December 2017 to February 2018, as shown by the bottom left plot in Fig.~\ref{fig:k2lc_1}.  We got a rotation period of $P=17.5336$ days based on its K2 LCs using the Lomb-Scargle periodograms mentioned above. It’s clear that this star has rapid variation in LC shape in shorter time-scale of one or two rotation periods, as shown by phase-folded LCs in Fig.~\ref{fig:k2lc_1}, indicating fast evolution of spot configuration (e.g., the emergence of new spots, disappearance and/or movement of old ones). In contrast, as mentioned above, its spectral features show little variation during the K2 observing campaign. There is no clear correlation between the K2 broadband flux and spectroscopic features like H$\alpha$ emission in phase.

\subsection{EPIC 211892240}
EPIC 211892240 ($\alpha_{\text{J2000}}=132.^\circ093878$, $\delta_{\text{J2000}}=18.^\circ612448$) is M2-type Praesepe member, located at a distance of $\sim$187 pc \citep{gaia2018}. It exhibits the characteristic of irregular temporal behaviour in chromospheric H$\alpha$ emission and TiO absorption on time-scales of days and weeks from February 2016 to April 2018, as shown in Fig.~\ref{fig:time_ewha2}. It has spot coverages with spot filling factors around 40 percent, and a typical H$\alpha$ emission of $\log R^{'}_{\text{H}\alpha}\lesssim-4.6$ dex. We found weak anti-correlations between TiO2n and H$\alpha$ emission over time ($C_{\text{TiO2n},\text{EW}_{\text{H}\alpha}}=-0.46$ for all epochs; $C_{\text{TiO2n},\text{EW}_{\text{H}\alpha}}=-0.66$ for epochs from December 2017 to April 2018). It has a rotation period between $P=20.7332$ days \citep{rebu2017} and $P=21.32$ days \citep{doug2017}. By using the Lomb-Scargle periodograms, we re-estimated its rotation period based on the K2 LCs collected during campaigns C5 and C18 (note that any systematic offset between LCs from these two campaigns was removed by making them have the same average flux), and got a similar period of $P=21.2513$ days, which was used to fold the observations in phase in this work (see Fig.~\ref{fig:k2lc_2}). Its K2 LCs shape varies very little within 3 years from C5 (2015-04-27 to 2015-07-10) to C18 (2018-05-12 to 2018-07-02), indicative of a spot re-configuration during that time. Both TiO2n and $\text{EW}_{\text{H}\alpha}$ are rotationally modulated. A period analysis for TiO2n and $\text{EW}_{\text{H}\alpha}$ indicates a similar period of around $21.43$ days. More interestingly, though with larger time gap, LAMOST spectral features correlate in phase with K2 flux, e.g., its TiO2n (collected during the season from December 2017 to April 2018) has a clear direct correlation in phase with K2 flux collected during the Campaign 5, expectedly that TiO2n increases (less spotted) as it becomes brighter. H$\alpha$ emission is as expected but a bit weak anti-correlation with K2 flux in phase.

 \subsection{EPIC 202059188}
The Hyades candidate member EPIC 202059188 ($\alpha_{\text{J2000}}=92.^\circ573998$, $\delta_{\text{J2000}}=22.^\circ572115$) is a very low-mass star with spectral type of M3-M4, at a closer distance about 29 pc \citep{gaia2018}. This star was much more frequently observed by LAMOST from November 2016 to February 2018, showing rapid changes in the activity (see Fig.~\ref{fig:time_ewha2}). Note that, however, its spectra may have possible contamination by a very near star whose angular distance is about 1.\arcsec9 \citep{jans2014}. It is rapid rotator with period of $P\sim0.69$ day \citep{arms2015}, derived based on K2 data, which shows very high activity, e.g., large spot coverages around 60 percent and strong chromospheric H$\alpha$ emission with $\log R^{'}_{\text{H}\alpha}$ up to $-3.6$ dex (see Fig.~\ref{fig:time_ewha2}). Such high values put it in the chromospheric activity saturation regime \citep[H$\alpha$ emission saturates at $\log R^{'}_{\text{H}\alpha}\approx-3.7$ dex, e.g.][]{fang2018}. The $C_{\text{TiO2n},\text{EW}_{\text{H}\alpha}}=-0.89$ among all available epochs means that they strongly anti-correlated with each other during the observing periods from November 2016 to February 2018. It was monitored by K2 from April to May in 2014, as shown by Fig.~\ref{fig:k2lc_2}. Interestingly, $\text{EW}_{\text{H}\alpha}$ is anti-correlated with non-simultaneous K2 flux in phase, as shown in Fig.~\ref{fig:k2lc_2} (see Fig.~\ref{fig:k2lc_3} for a much more clear display), indicating the spatial co-location of chromospherically active regions like plages with cool starspots which rotationally modulate K2 flux. This suggests the presence of large organized regions of surface activity which can persist for many rotation periods. However, given the large time-gap between K2 and LAMOST observations and considering uncertainties present in the period analysis (the correlation in phase between them is thus very sensitive to rotation period), caution should be exercised in the interpretation of such a co-location. The period analysis of $\text{EW}_{\text{H}\alpha}$ and TiO2n of this star indicates a period of $P\simeq0.73$ days, very close to the value derived based on K2 flux. Therefore, a solid conclusion could be obtained with certainty for this star, namely, both TiO2n and $\text{EW}_{\text{H}\alpha}$ are rotationally modulated by a period of around 0.7 days with anti-correlation between these two quantities, also indicating a co-location characteristic of bright plages and dark starspots.

\subsection{EPIC 202083206}
\begin{figure*}
\centering
\includegraphics[width=\columnwidth]{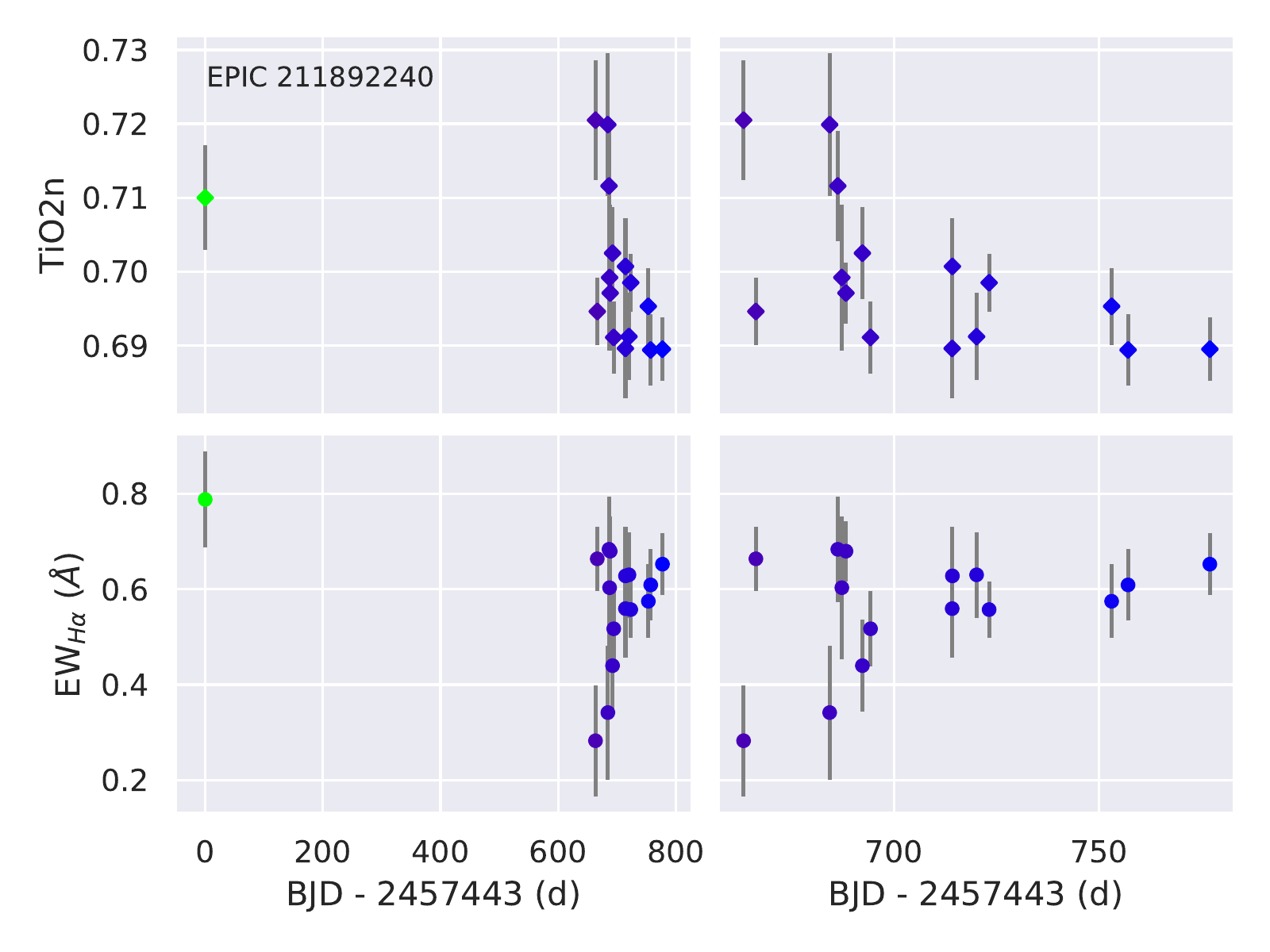}
\includegraphics[width=\columnwidth]{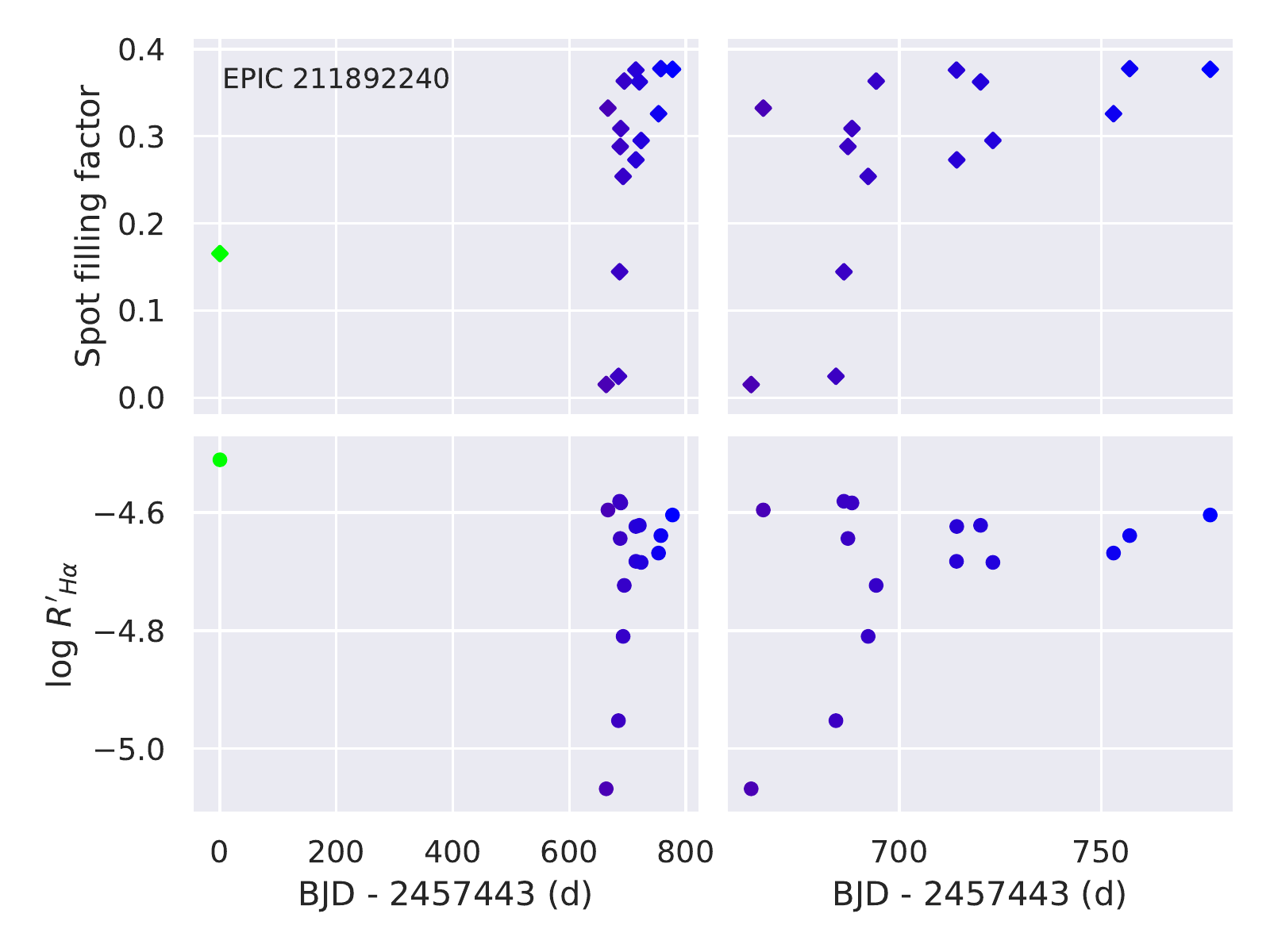}
\includegraphics[width=\columnwidth]{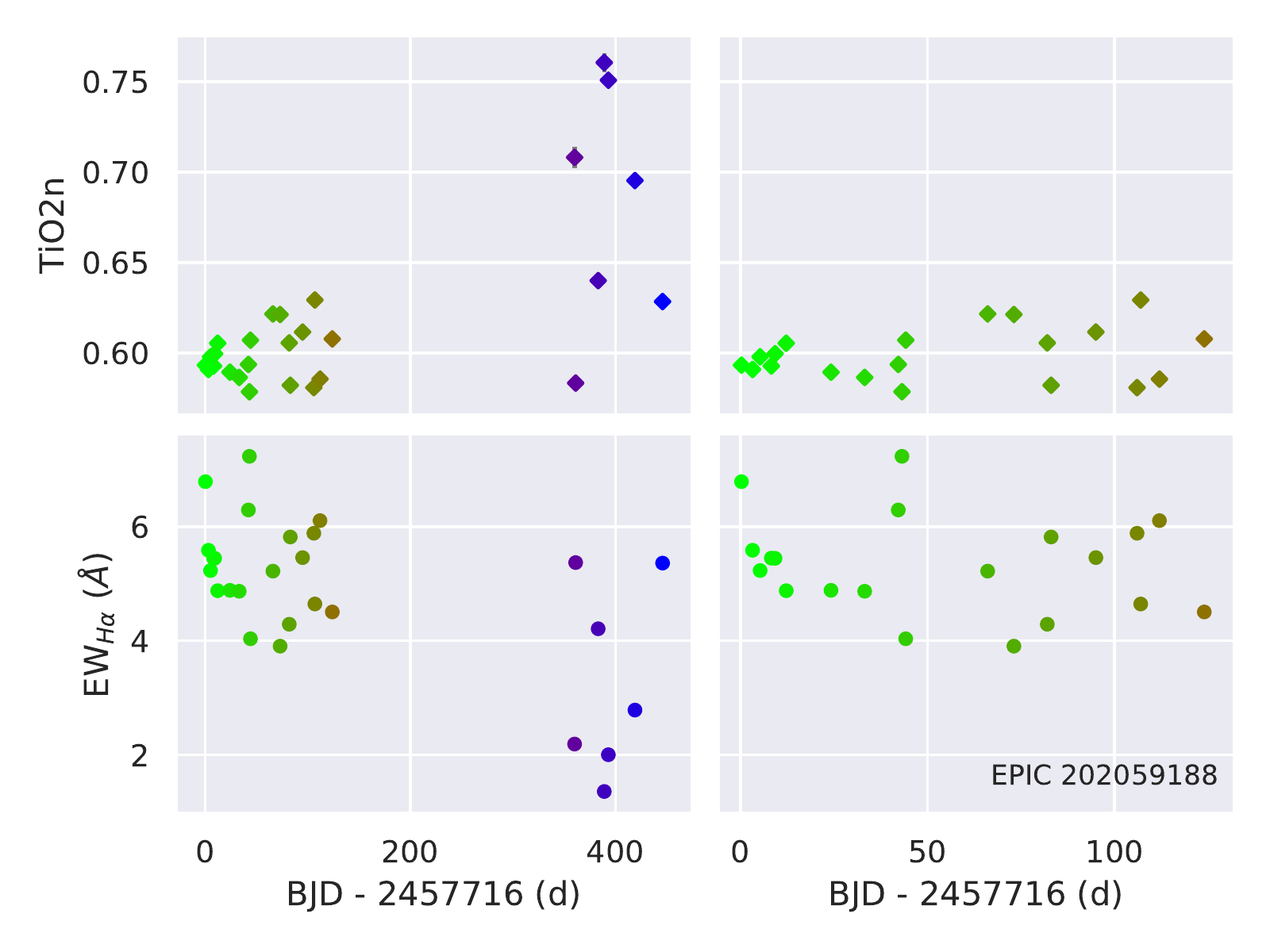}
\includegraphics[width=\columnwidth]{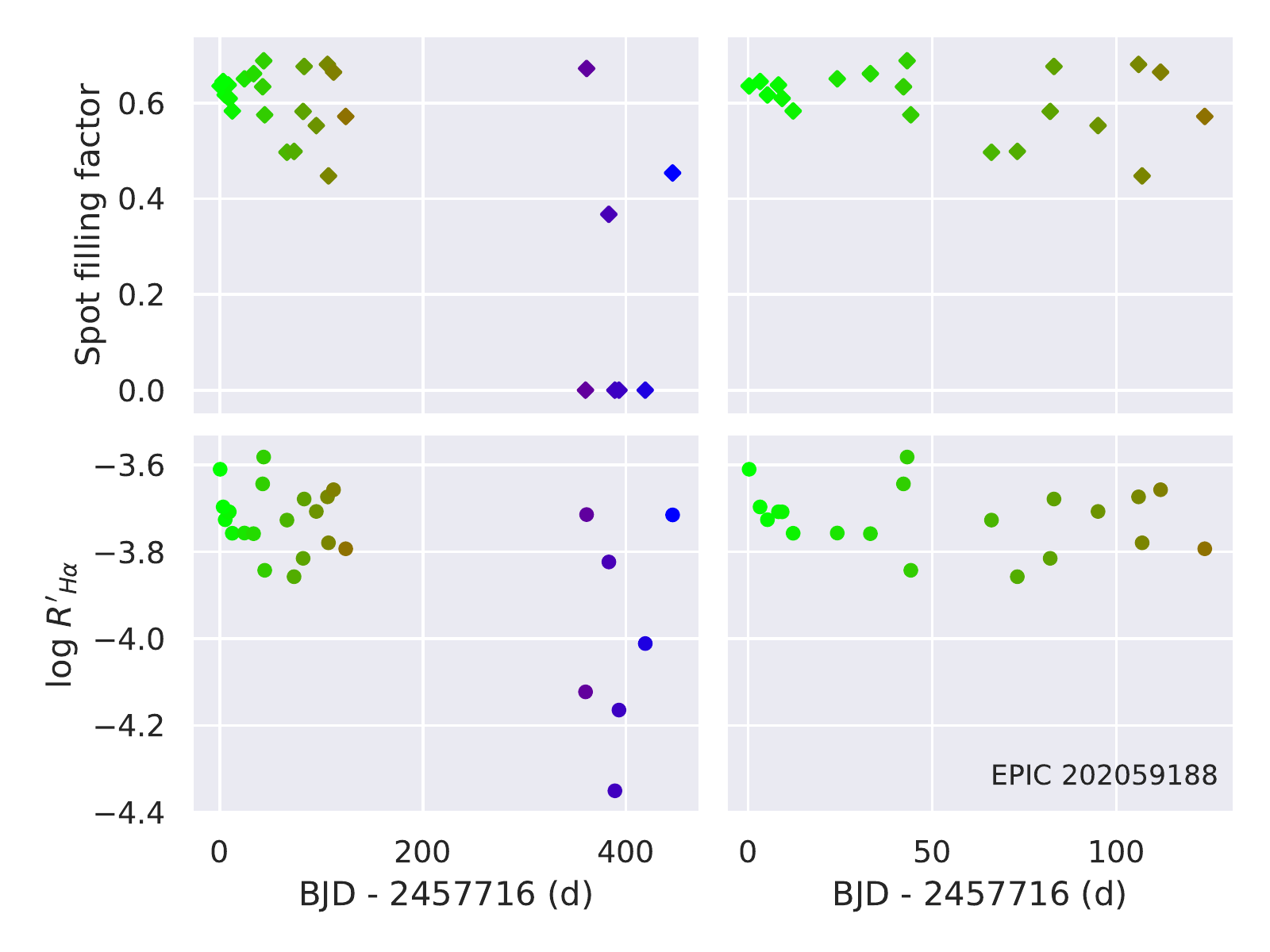}
\includegraphics[width=\columnwidth]{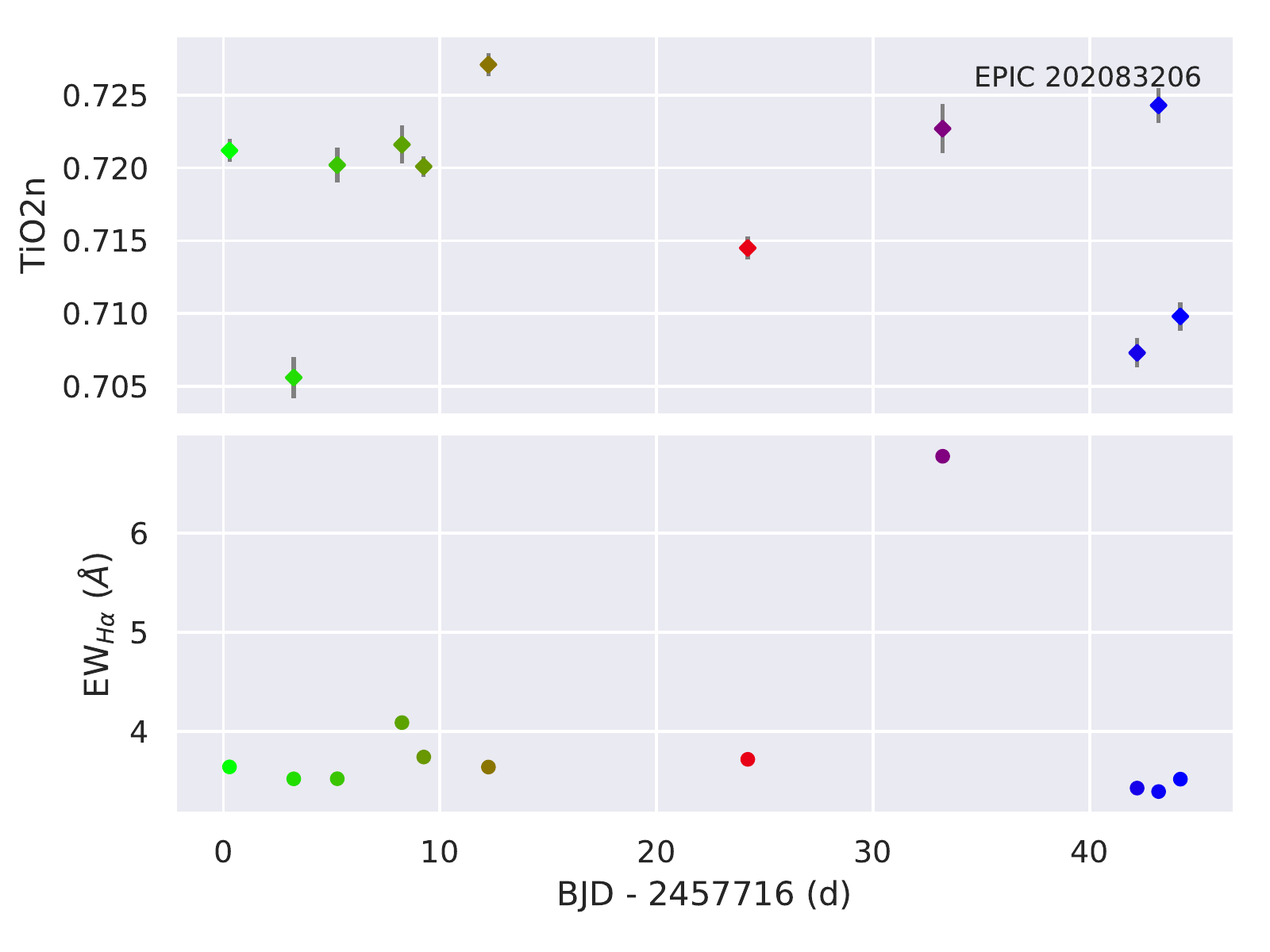}
\includegraphics[width=\columnwidth]{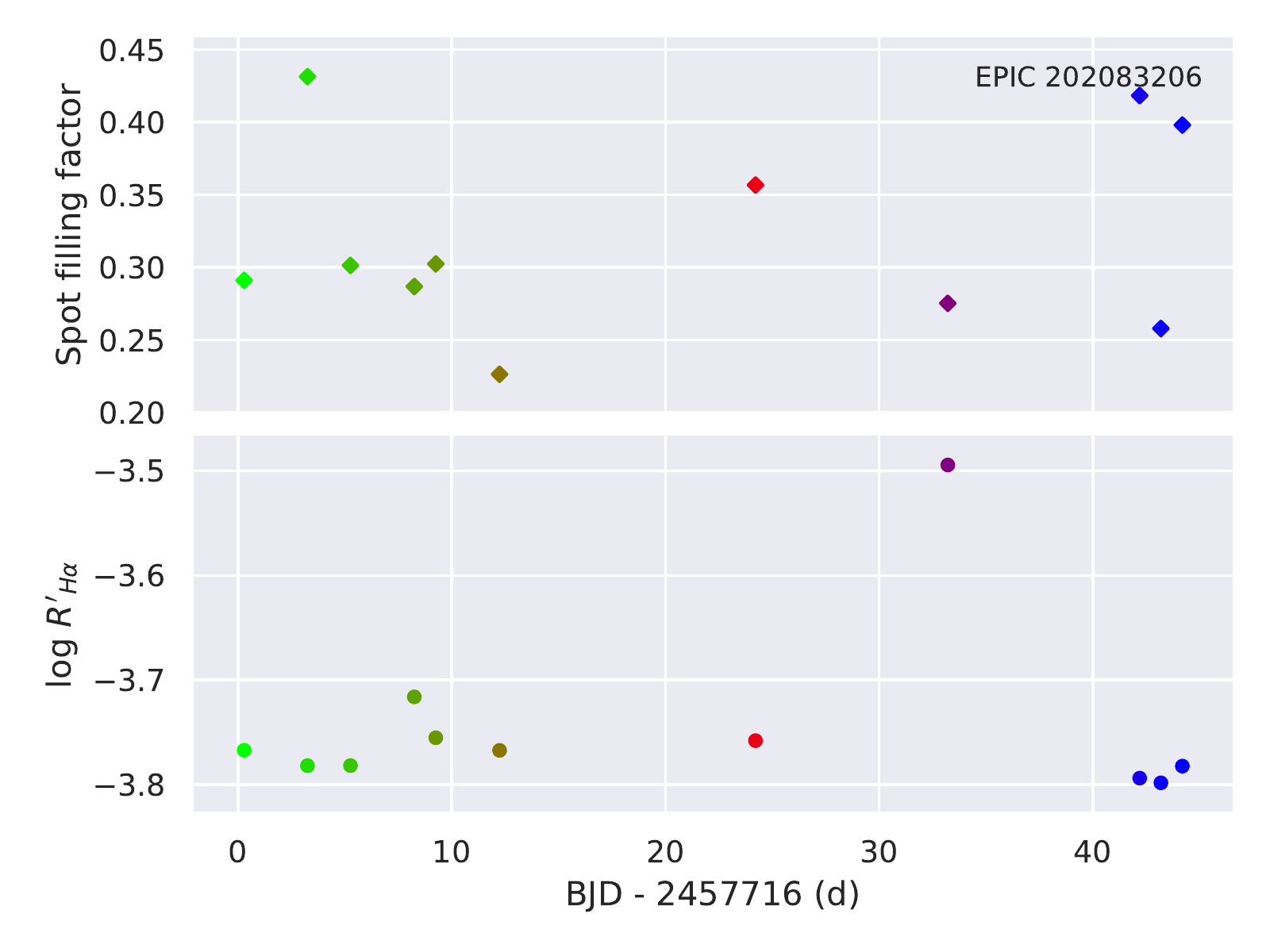}
\caption{Same as Fig.~\ref{fig:time_ewha_tio2n}, but for star EPIC 211892240, EPIC 202059188 and EPIC 202083206.}
\label{fig:time_ewha2}
\end{figure*}

\begin{figure*}
\centering
\includegraphics[width=\columnwidth]{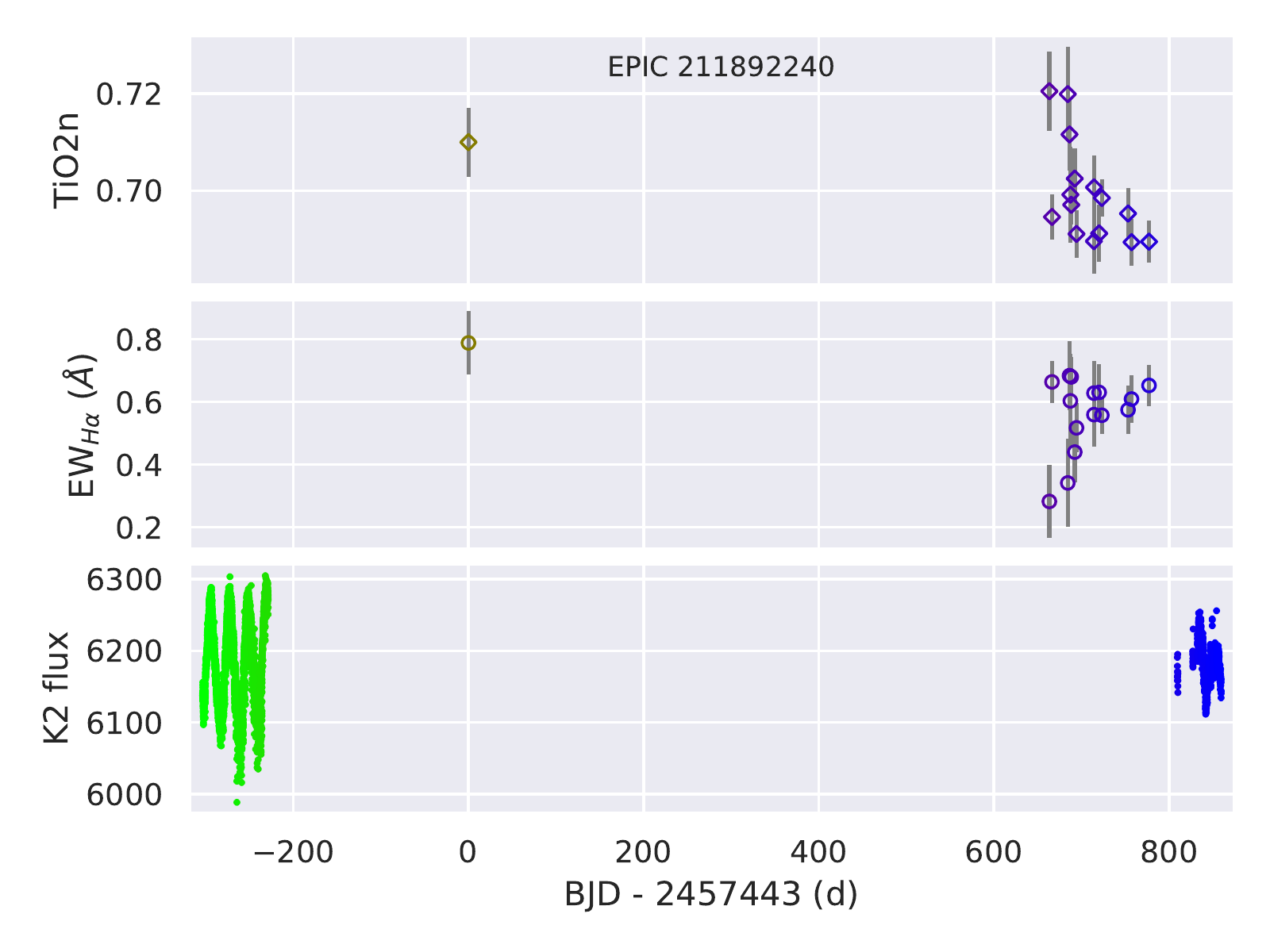}
\includegraphics[width=\columnwidth]{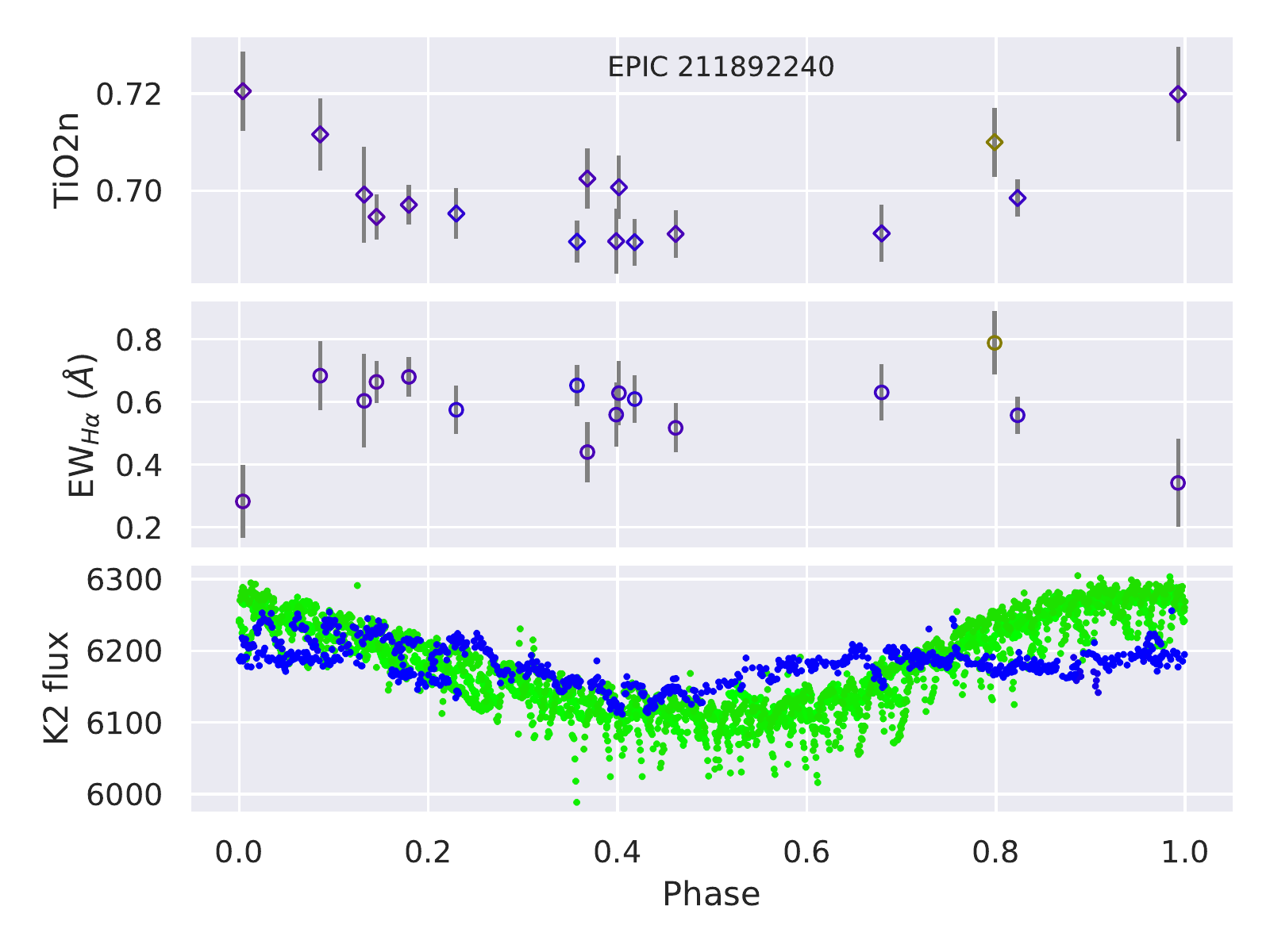}
\includegraphics[width=\columnwidth]{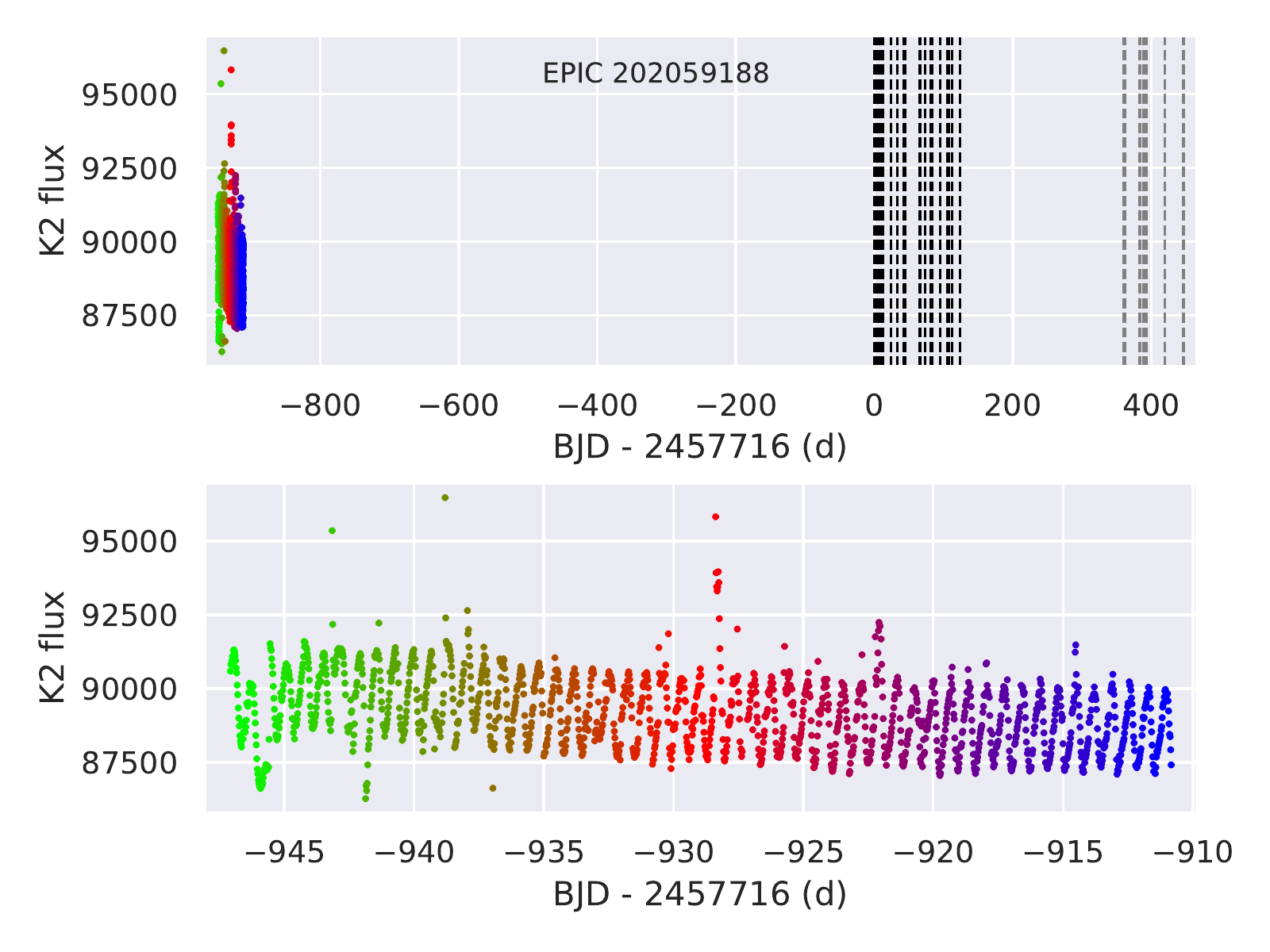}
\includegraphics[width=\columnwidth]{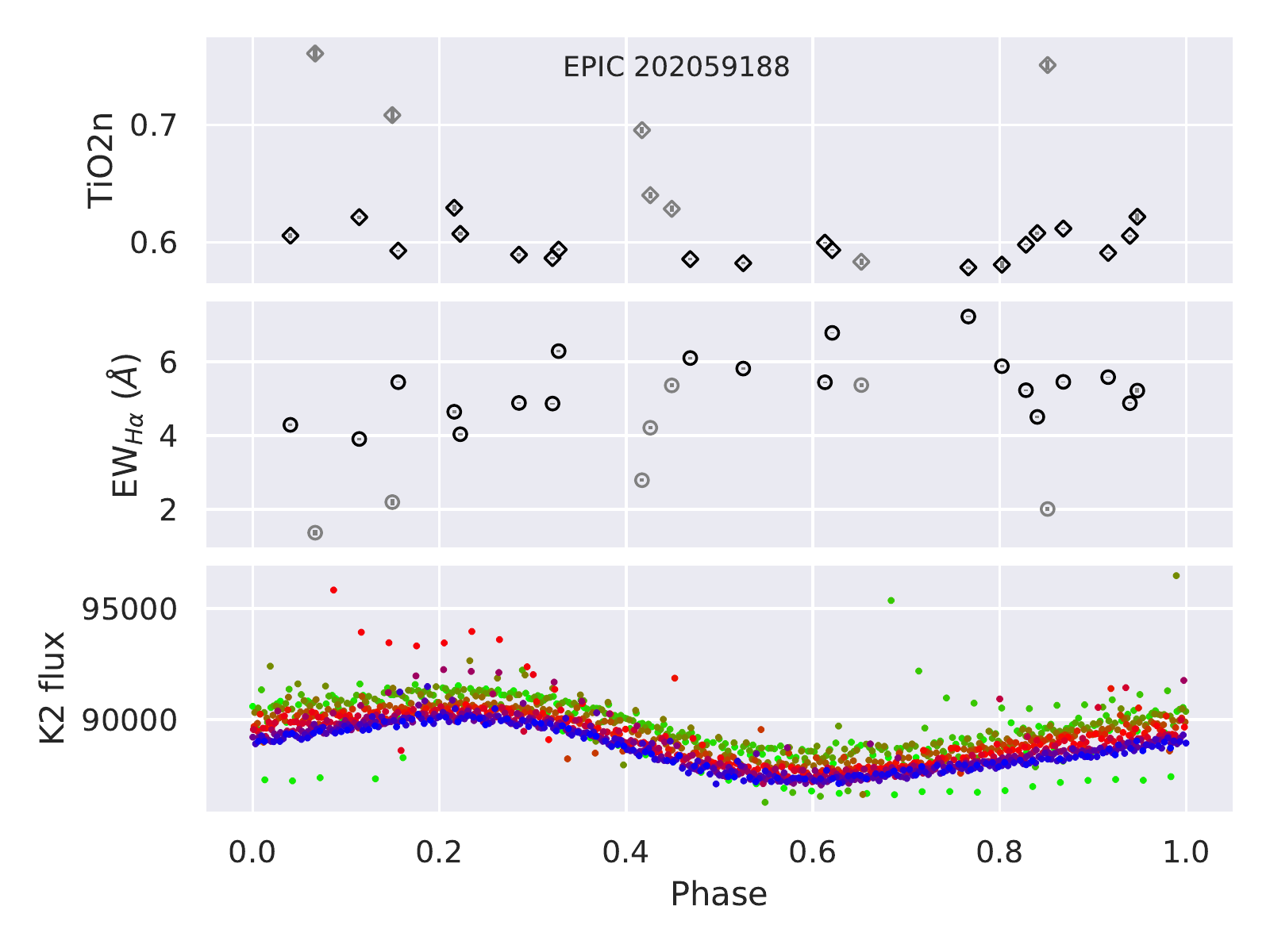}
\includegraphics[width=\columnwidth]{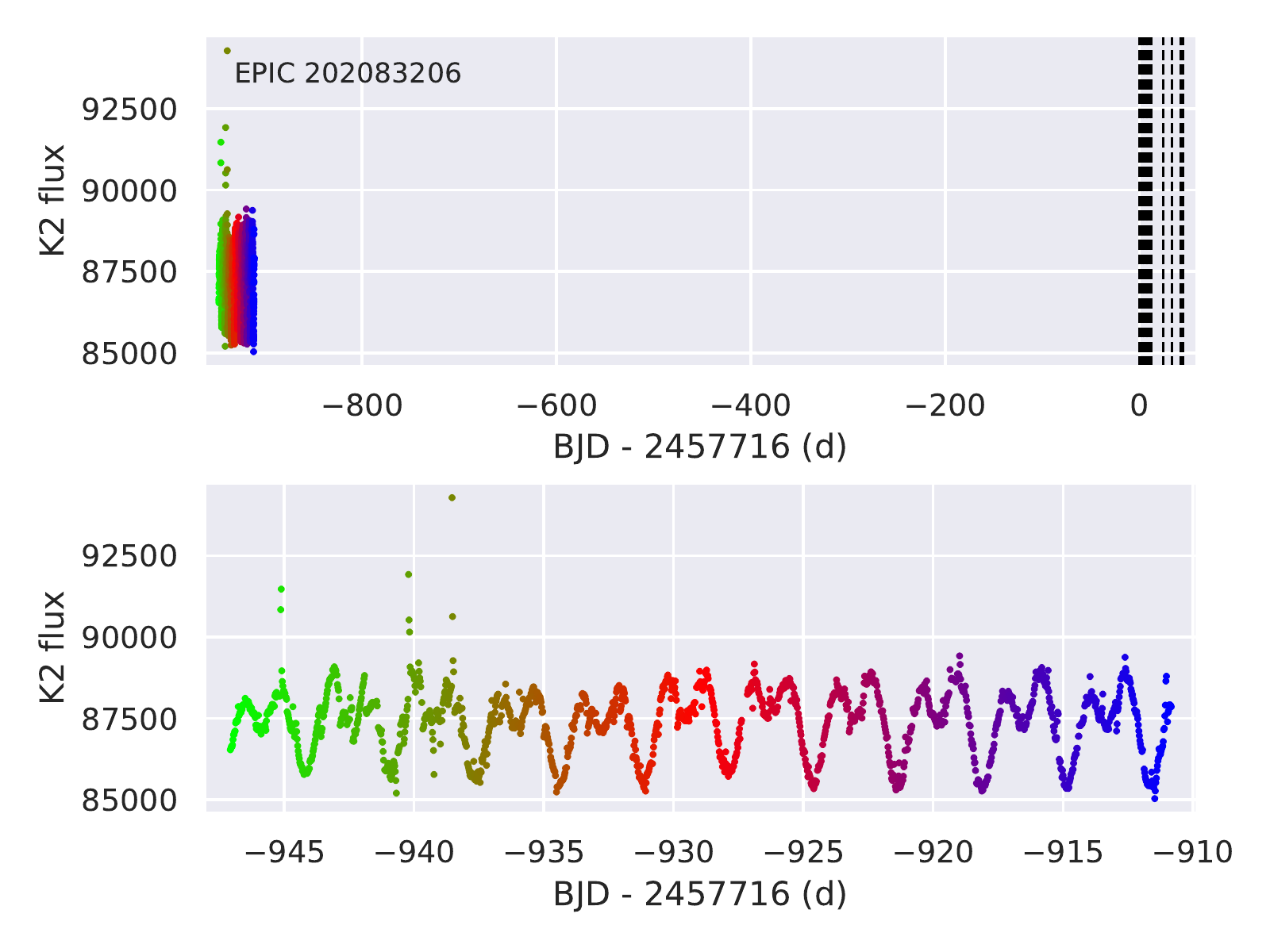}
\includegraphics[width=\columnwidth]{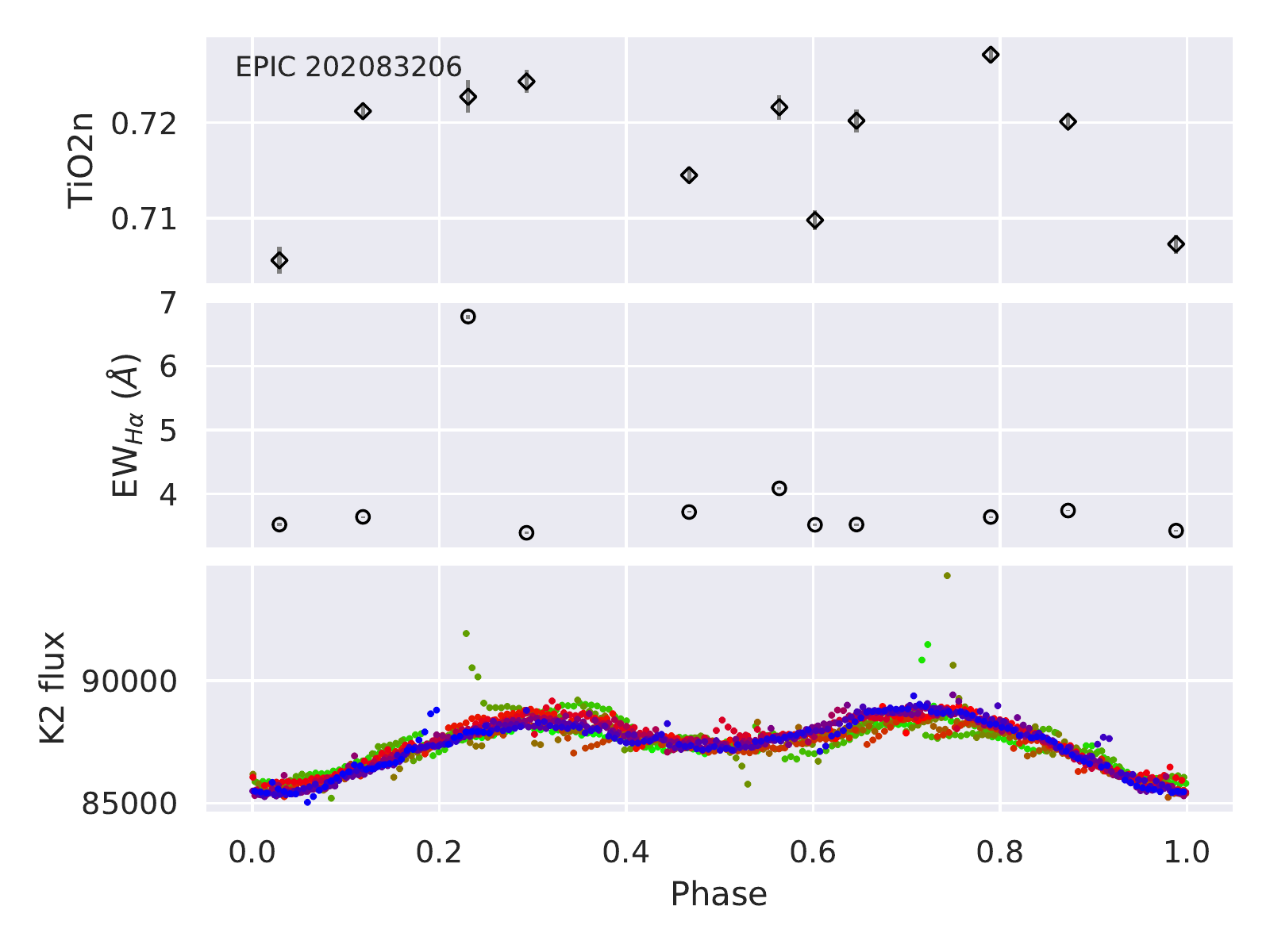}
\caption{Same as Fig.~\ref{fig:k2lc_1}, but for star EPIC 211892240, EPIC 202059188 and EPIC 202083206. Note that the vertical dashed lines (in left-middle and left-bottom plots) represent the LAMOST epochs. The observations in right plots were phase-folded with a period of $P=21.2513$, $P=0.6902$ and $P=3.2556$ days for EPI C211892240, EPIC 202059188 and EPIC 202083206, respectively. }
\label{fig:k2lc_2}
\end{figure*}

\begin{figure*}
\centering
\includegraphics[width=\columnwidth]{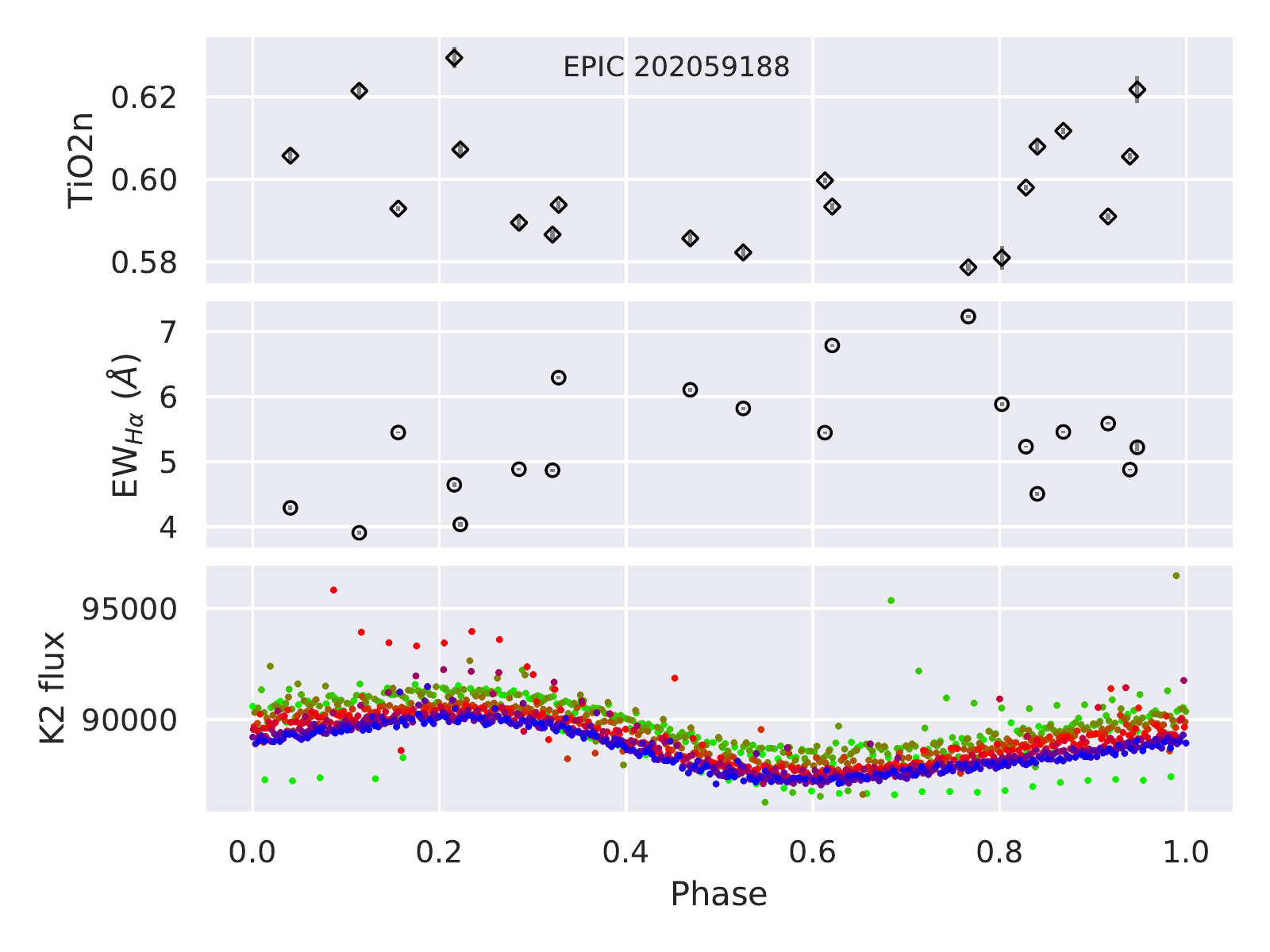}
\includegraphics[width=\columnwidth]{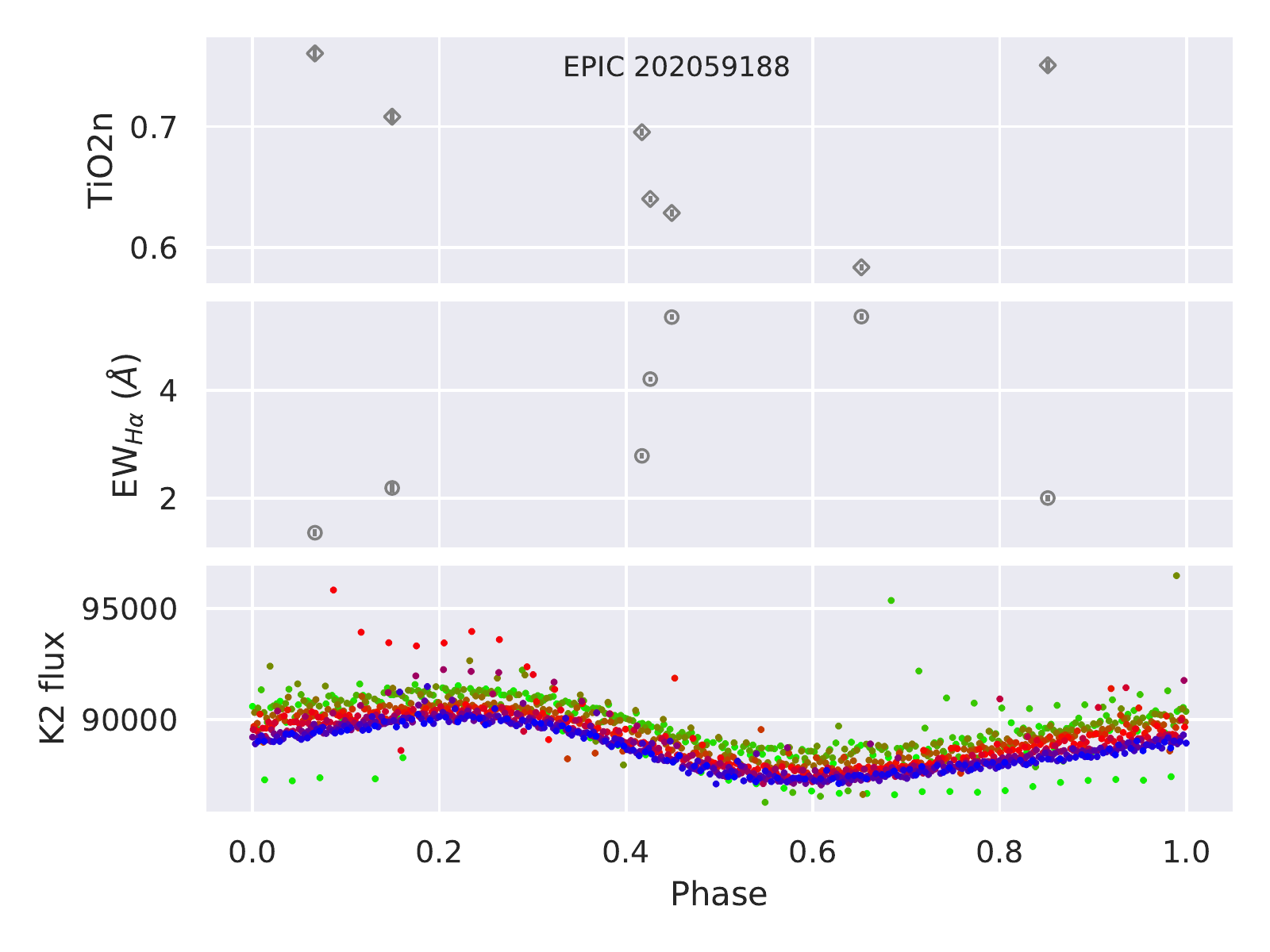}
\caption{Phase-folded LAMOST measurements and K2 brightnesses of EPIC 202059188, illustrating the correlation in phase between them. Left: LAMOST observations collected during from 2016-11-23 to 2017-03-27; Right: LAMOST observations collected during from 2017-11-18 to 2018-02-12.}
\label{fig:k2lc_3}
\end{figure*}

EPIC 202083206 ($\alpha_{\text{J2000}}=90.^\circ863670$, $\delta_{\text{J2000}}=24.^\circ037530$) is M2 dwarf  in Hyades, located at a distance about 62 pc \citep{gaia2018}. As shown in Fig.~\ref{fig:time_ewha2}, it exhibits strong H$\alpha$ emission with $\text{EW}_{\text{H}\alpha}\approx3.6$~\AA~and $\log R^{'}_{\text{H}\alpha}\approx-3.75$ dex (the outlier with $\text{EW}_{\text{H}\alpha}\approx7$~\AA~and $\log R^{'}_{\text{H}\alpha}\approx-3.5$ dex is probably due to a flare event). It is spotted with spot filling factors ranging from 20 to 45 percent during the LAMOST observing runs from November 2016 to January 2017. There is no clear correlation between TiO2n and H$\alpha$ emission ($C_{\text{TiO2n},\text{EW}_{\text{H}\alpha}}=+0.29$). This star was observed by the K2 mission in April and May 2014, as shown in Fig.~\ref{fig:k2lc_2}. By using the Lomb-Scargle periodograms mentioned above, we got a period of $P=3.2556$ days based on its K2 flux, very close to the period of $P=3.283667$ days derived by \citet{arms2015}. Though over two years apart, it’s clear that TiO2n is correlated with K2 flux in phase, e.g., its TiO2 absorption becomes stronger (equivalently more spotted) as it becomes faint as expected, however, we found no correlation between K2 flux and H$\alpha$ emission. The similar shape of K2 light-curves over two months indicates the presence of two stable spot groups with lifetimes over one month. Alternatively, perhaps it is a binary system, considering the two stable dips in its light-curve. Besides, this star probably suffers frequent and repeated flare events, e.g., near by phase of 0.2 and 0.7, as shown by both LAMOST spectra and K2 photometry in  Fig.~\ref{fig:k2lc_2}.


\bsp	
\label{lastpage}
\end{document}